\documentclass[aps,twocolumn,amsmath,amssymb,bbm,nofootinbib,preprintnumbers,showpacs]{revtex4}
\usepackage{epsfig}
\usepackage{axodraw}
\usepackage{mathrsfs,graphicx,bm,amsmath}

\newcommand{\sigex}{\tilde{\sigma}}

\newcommand{\rex}{\tilde{r}}

\newcommand{\hpn}{\tilde{h}_\phi}

\newcommand{\Tn}{\tilde{T}}
\newcommand{\qn}{\tilde{q}}

\newcommand{\nun}{\tilde{\nu}}
\newcommand{\hF}{^{(F)}}
\newcommand{\hB}{^{(B)}}
\newcommand{\lit}{^6\mathrm{Li}}
\newcommand{\kal}{^{40}\mathrm{K}}
\newcommand{\epm}{\epsilon_M}

\newcommand{\ef}{\epsilon_F}

\newcommand{\hpb}{\bar{h}_\phi}
\newcommand{\lpb}{\bar{\lambda}_\phi}

\newcommand{\rhob}{\bar{\rho}}

\newcommand{\rhon}{\tilde{\rho}}
\newcommand{\mpb}{\bar{m}_\phi^2}

\newcommand{\Apb}{\bar{A}_\phi}
\newcommand{\Apn}{\tilde{A}_\phi}
\newcommand{\ApR}{A_\phi}
\newcommand{\ZpR}{Z_\phi}

\newcommand{\nB}{\bar{n}_B}
\newcommand{\nF}{\bar{n}_F}
\newcommand{\nC}{\bar{n}_C}
\newcommand{\nM}{\bar{n}_M}
\newcommand{\OmB}{\bar{\Omega}_B}

\newcommand{\OmC}{\bar{\Omega}_C}
\newcommand{\OmM}{\bar{\Omega}_M}

\newcommand{\rhooR}{\rho_0}%\tilde{\rho_{0,R}
\newcommand{\rhoR}{\rho}%\tilde{\rho}_R

\newcommand{\lpR}{\lambda_\phi}

\newcommand{\nRF}{n_F}

\newcommand{\nRM}{n_M}

\begin{document}
%two columns:
%@twocolumnfalse\endcsname

\title{Universality in Phase Transitions for Ultracold Fermionic Atoms}

\author{S. Diehl}
\email{S.Diehl@thphys.uni-heidelberg.de}
\author{C. Wetterich}
\email{C.Wetterich@thpyhs.uni-heidelberg.de}

\address{
Institut f{\"u}r Theoretische Physik,
Philosophenweg 16, 69120 Heidelberg, Germany}

%\today

\begin{abstract}
We describe the gas of ultracold fermionic atoms by a functional integral for atom and molecule fields. The crossover
from Bose-Einstein condensation (BEC) to BCS-type superfluidity shows universal features in terms of a concentration parameter
for the ratio between scattering length and average interatomic distance. We discuss the relevance of the Yukawa coupling
between atoms and molecules, establish an exact narrow resonance limit and show that renormalized quantities are independent
of the Yukawa coupling for the broad resonance, BCS and BEC limits. Within our functional integral formalism we compute the
atom scattering in vacuum and the molecular binding energy. This connects the universal concentration parameter to the
magnetic field of a given experiment. Beyond mean field theory we include the fluctuations of the molecule field and the
renormalization effects for the atom-molecule coupling. We find excellent agreement with the observed fraction
of bare molecules in $\lit$ and qualitative agreement with the condensate fraction in $\lit$ and $\kal$. In addition to the
phase diagram and condensate fraction we compute the correlation length for molecules, the in-medium scattering length
for molecules and atoms and the sound velocity.
\end{abstract}

 %  \hfill HD-THEP-}

\maketitle
%two columns: ]

%%%%%%%%%%%%%%%%%%%%%%%%%%%%%%%%%%%%%%%%%%%%%%%%%%%%%%%%%%%%%%%%%%%%%%%%

%\input{Introduction}

\section{Introduction}
\label{sec:intro}

Ultracold fermionic atoms are developing into exciting experimental laboratories for the understanding of complex many
body quantum physics. Collective phenomena as the phase transition to low temperature superfluidity can be investigated
with controlled microscopic physics. Recent experimental progress \cite{Thomas02,Jin04,Ketterle04,ZGrimm04,Partridge05} in
the crossover region between a Bose-Einstein condensate (BEC) \cite{Einstein24,Einstein25} of molecules and the condensation
of correlated atom pairs similar to BCS-superconductivity \cite{ACooper56,BBCS57} reveals the universality of the
condensation phenomenon, as anticipated theoretically \cite{ALeggett80,BNozieres85,CMelo93,DStoof96,ECombescot99,FPethick00}.

At first sight the understanding of the crossover region in presence of a Feshbach resonance seems to depend on complex
atomic and molecular physics and therefore on many
parameters. The crucial role of molecular bound states involves the detailed knowledge of binding or resonance energies
as well as atomic scattering amplitudes in presence of a homogeneous magnetic field $B$. In particular, one needs the
effective coupling between the bound state and the atoms, usually encoded in the matrix element of the Hamiltonian between
``bound and open channels'' in the space of two-atom states. The many body physics depends in addition on the temperature
and density of the thermodynamic equilibrium state. Furthermore, the experimental observations are influenced by the
geometry of the trap which is used for the  confinement of the atom gas.

In this paper we underline the universal aspects of the equilibrium state of ultracold fermionic atoms. In the limit of
a homogeneous situation we argue that ``macroscopic'' observables can be expressed in terms of only three dimensionless
parameters:

(1) The ``concentration'' $c$ describes the ratio between the in-medium scattering length and the average distance between
two unbound atoms or molecules. For small negative $c$ the molecular resonance becomes unimportant - the gas is
approximated by fermionic atoms with a small pointlike interaction. The phase transition to superfluidity is of the BCS type
and we will call this region in parameter space ($|c|\ll 1$, $c<0$) the ``BCS regime''. In contrast, for small positive
$c$ the phase transition and the low temperature physics is well described by a gas of bosonic molecules. This limit will
be denoted by ``BEC regime'' - it shows the phenomenon of Bose-Einstein condensation very similar to a gas of bosonic
atoms. The perhaps most interesting region is the ``crossover regime'' between the BCS and BEC regimes for large $c$ or
$|c^{-1}|\lesssim 1$. All three regimes are continuously connected as $c^{-1}$ is varied from large negative to large
positive values. For low enough temperature $T$ a transition to a superfluid state with condensation of atom pairs or
molecules occurs for all $c$. The critical temperature $T_c$ for the phase transition to superfluidity clearly shows
(cf. fig. \ref{CrossoverTcAll}) the universal character of the crossover.

(2) The details of the crossover physics depends on a dimensionless Yukawa or Feshbach coupling $\hpn$ which describes the interaction
between atoms and molecules. This coupling depends on the density $\hpn \propto n^{-1/6}$. In the ``narrow resonance regime''
of small $\hpn$ the precise value of the coupling is not important for the location of the phase transition. A universal
``narrow resonance limit'' $\hpn \to 0$ exists for the ``symmetric phase'' for $T\geq T_c$. For
this limit we find a nontrivial \emph{exact} solution of the many body problem which is described by a suitably adapted
mean field theory. Small deviations from this universal limit lead for many (but not all) quantities to a controlled
expansion for small $\hpn$. For the ``broad resonance regime'' with large $\hpn$ the precise value of $\hpn$ does not affect the
observables that are connected to renormalized fields and parameters. In fact, for large $\hpn$ strong renormalization
effects lead to a new form of universality for large couplings as observed previously in relativistic systems with a
strong boson-fermion ``Yukawa coupling'' \cite{Jungnickel95}. Our general universality hypothesis
\footnote{The universality here differs from the universality argument given by Ho \cite{Ho104}. In particular, in the
crossover region the interaction is not pointlike and does not become scale-free for the limit of diverging scattering
length $\propto |c|$. The Yukawa coupling $\hpb$ introduces a further mass scale which manifests itself in the
dimensionless parameter $\hpn$. Ho's argument can be applied, however, in the limit $c^{-1} \to 0$, $\hpn \to \infty$.} states
states that the ``macroscopic quantities'' can be expressed in terms of only two couplings $c$ and $\hpn$.

(3) The third parameter is finally the temperature in units of the Fermi energy, $\tilde{T} = T/\epsilon_F$,
$\epsilon_F = k_F^2/2M$, $k_F = (3\pi^2 n )^{1/3}$ (with $M$ the mass of the atoms). Using the parameters $c$, $\hpn$ and
$\Tn$ we obtain results for dimensionless ratios and couplings for arbitrary densities $n$. For a given $c$ and $\Tn$ the
only implicit dependence on $n$ arises via the weak dependence $\hpn \propto n^{-1/6}$ and drops out in the limits of
``enhanced universality'' for broad and narrow resonances and the BEC or BEC regimes.

No further details of the ``microphysics'' are needed for the macroscopic quantities. In this sense the description becomes universal. In the language
of quantum field theory the parameters $c$ and $\hpn$ describe relevant couplings for the long distance physics.

At this point we may emphasize that the universal aspects discussed here
are not related to the ``universal critical behavior'' near a second order phase transition. For example, the divergence of
the correlation length for $T\to T_c$ is governed by a universal critical exponent - this has not yet been computed very
accurately here. The discussion of universality in the present paper rather concentrates on the parameters which are of relevance
and concerns ``non-universal quantities'' in the sense of critical phenomena.

Our results are based on a systematic formulation of the interacting many body system in terms of a functional integral.
A detailed derivation and discussion of this formalism, together with an extension to the inhomogeneous situation of a trap,
can be found in \cite{Diehl:2005ae}. We carefully address the issue of renormalization or, more precisely, the relation between ``microscopic parameters'' and
``macroscopic observables''. Expressed in terms of our renormalized parameters $c$, $\hpn$ and $\Tn$ all loop integrals
are ultraviolet convergent such that the actual value of the physical microscopic cutoff scale plays no role. Also the mass
of the atoms drops out.

As one of the important advantages of our functional integral formulation we can compute both the ``physics in the vacuum''
and the atom gas at nonvanishing density and temperature in the same framework. The vacuum physics corresponds to
appropriate low density limits and concerns properties of two-atom systems like the cross section and molecular binding
energy. These quantities are directly accessible to experiment. Combining the experimental information about the two-atom
system with knowledge of the density ($k_F$) we can relate the parameters $c$ and $\hpn$ to measurable quantities. In
particular, the concentration $c$ is related to the magnetic field, or, more precisely, to its ``detuning'' $\Delta B
=B - B_0$ from the vacuum Feshbach resonance at $B_0$.

The functional integral treats the fermionic fluctuations
of unbound atoms and the bosonic fluctuations of the molecule or di-atom field on equal footing. It is therefore
an ideal starting point for the inclusion of the molecule fluctuations. These fluctuations are important for the
quantitative understanding of the phase transition for a broad Feshbach resonance. For a large dimensionless
Yukawa coupling $\hpn$ the renormalization effects are crucial. In fact, the importance of the molecule fluctuations is
directly related to the size of $\hpn$ - for $\hpn\to 0$ a nontrivial exact ``narrow resonance limit'' can be established
where the molecule fluctuations can be neglected and mean field theory becomes valid. The present experiments in $\lit$ and
$\kal$ probe, however, broad resonances with large $\hpn$ for which the inclusion of the bosonic fluctuations cannot
be neglected.

We present a first computation of the correlation length for the molecule
field. It turns out to be larger than the average distance between two atoms for a rather broad range of temperature above
$T_c$, indicating the importance of collective effects. We also compute the effective scattering length for
molecule-molecule scattering. Again, it turns out to be large in the crossover region. Furthermore, we compute the
sound velocity in the superfluid phase.

Our functional integral approach permits a unified view of many previously obtained results. In fact, various other methods
have already revealed many aspects of the results of our investigation.
Strinati et. al. propose a purely fermionic description of the crossover problem based on a physically motivated choice of
diagrams in the frame of the operator formalism \cite{AAAAStrinati,BBStrinati,CCStrinati,YPieri,ZStrinati}.
Assuming that the chemical potential satisfies $\mu/T\to -\infty$ in the BEC limit, they can show by explicit calculation
\cite{YPieri} the appearance of a ``molecular'' and a condensate particle density in the BEC limit. A similar approximation
scheme is advocated in \cite{Drummond05}.

%Other approaches formulate the problem microscopically in terms of a Yukawa theory for a coupled system of fermionic
%unbound atoms and bosonic molecules. This is equivalent to a purely fermionic model with a particular momentum dependence
%of the four-fermion vertex on the level of the classical theory ***Achievements ***. All approaches of this kind try to
%tackle the issue of a strong
%microscopic coupling by a more sophisticated parameterization of the interaction close to the Feshbach resonance. Of
%course, this increases the number of parameters to be extracted from the experiment.
Other approaches formulate the problem microscopically in terms of a Yukawa theory for a coupled system of fermionic
unbound atoms and bosonic molecules
\cite{BBKokkelmans,CCKokkelmans,DDKokkelmans,EEGriffin,FFGriffin,AATimmermans,GGChen,HHChenReview,
WWStoofBos,XXStoofBos,YYStoof,ZZStoof}. In part, this is motivated by a functional integral
\cite{CCKokkelmans,WWStoofBos,XXStoofBos}. Approaches of this type are well suited for dealing with nonlocal interactions.
Another route of achieving this task is followed in \cite{Jensen04,Palo04}. A comparison of purely fermionic vs. bosonized
approach has recently been performed \cite{Goral05}. Working with the microscopic parameters appropriate for $\kal$, almost no
difference is found between the two approaches. This is consistent with our investigation, classifying $\kal$ in the
broad resonance regime where the two approaches become indeed equivalent.

A systematic study relating on the two body level the parameters of the microscopic Yukawa theory to binding energies and
two-atom cross sections in vacuum includes the UV-renormalization \cite{WWStoofBos,XXStoofBos,YYStoof,ZZStoof}. The
investigations of Feshbach resonances (see \cite{ZZStoof} for K) reveal a clear tendency that the Yukawa coupling extracted
``naively'' from scattering properties is much too large in
order to account for the large condensate fraction $\Omega_C$ advocated by observation, in particular for $\lit$ used
by many experimental groups \cite{Ketterle04,ZGrimm04,Thomas04,Bourdel04,Strecker04}. This has triggered recent work on
an effective Yukawa coupling and its momentum and energy dependence \cite{ZZStoof,Yi04,Pethick04}.
%\begin{itemize}
%\item Pairing gap theory Heiselberg \cite{Heiselberg00,Heiselberg04} first paper only attractive case, second in fermionic
%crossover scenario
%\end{itemize}

The plan of the paper is as follows. In sect. \ref{EffAtDens} we briefly discuss the formal ingredients of our functional
integral formalism. It is based on the effective action to be computed for our two-channel model of the crossover
system. A more extensive description of these issues can be found in \cite{Diehl:2005ae}. In sect. \ref{sec:renormalization}
we state our notion of universality more precisely. The aspects we consider differ from those analyzed by other authors,
e.g. \cite{Ho104}. We state universality here in a general field theoretical context, where for given external parameters
$n$ and $T$ only two relevant renormalized couplings describe the system, namely $c$ and $\hpn$. In sect.
\ref{sec:Enhanced} we briefly address enhanced universality in various limits. These limits are discussed in
more detail in sect. \ref{sec:EnhancedUniv} when the necessary formal framework has been provided.

In sect. \ref{DressedAndBareI} we turn to the important difference between bare (microscopic) and dressed molecules
(or ``macroscopic'' di-atom bound states). We compute the renormalization consant $Z_\phi$ that relates the bare and
dressed molecule fields. Sect. \ref{EvalBeyondMFT} turns to the effective potential for the molecule field. Here we state
our approximation scheme beyond mean field theory.

In sect. \ref{sec:corrlength} we compute the correlation length for the
molecule fluctuations. It diverges at the critical temperature and remains large for a rather wide range, typically
as long as $T$ is smaller than $1.5T_c$. A large correlation length is a clear sign for collective effects. We also
compute the in-medium scattering length for molecules and the sound velocity in the superfuid phase.

In sects. \ref{sec:lowdens},\ref{sec:concmag} we relate the universal couplings $c$, $\hpn$ to the parameters of
Feshbach resonances, in particular the detuning of the magnetic field $B-B_0$ and the response of the level
splitting to variations of the magnetic field, $\bar{\mu}$. Our functional integral formalism enables us to compute directly
the observable quantities of the two-atom system, like the molecular binding energy $\epsilon_M$ or the scattering cross
section between the unbound atoms. This is achieved in an appropriate limit where both the temperature $T$ and the density
$n$ vanish as shown in sect. \ref{sec:lowdens}. Sect. \ref{sec:concmag} establishes a direct link between the universal
concentration parameter $c$ and the magnetic field $B$ and determines the Yukawa coupling $\hpn$.

In sect. \ref{CompExp} we turn to our results for the crossover phase diagram for particular Feshbach resonances explored
by recent experiments. We first address the fraction of bare molecules $\OmB$ that has been measured by laser excitations
to a higher level \cite{Partridge05}. We find excellent agreement with the observations in a range covering five orders of
magnitude for $\OmB$ on both sides of the Feshbach resonance (cf. fig. \ref{BareExpPart}). The condensate fraction
$\Omega_C$ relates to dressed molecules. Here we find qualitative agreement with the experiments in $\kal$
\cite{Jin04} and $\lit$ \cite{Ketterle04}, whereas a quantitative comparison would need a more detailed relation between
the experimental observable and $\Omega_C$. In sect. \ref{sec:conclusions} we present our conclusions. Four appendices
deal with the renormalization effects for the Yukawa coupling (\ref{sec:YukRenorm}), the detailed form of the gap equation
(\ref{app:appb}), the dispersion relation for the molecules in vacuum (\ref{app:dispersion}) and the atom scattering
in vacuum (\ref{sec:scattvac}).

\section{Effective Action}
\label{EffAtDens}

%\subsection{Equal Treatment for unbound atoms and molecules}
The ultracold gas of fermionic atoms in the vicinity of a Feshbach resonance can be idealized by
two stable atomic states denoted by a two component spinor $\psi$. (For the example of $^6\mathrm{Li}$ these states
may be associated with the two lowest hyperfine states $|1\rangle$, $|2\rangle$.) The molecular state responsible for
the Feshbach resonance can be treated as a complex bosonic field $\hat{\phi}$. In our approximation this boson is stable for
negative binding energy and can decay into a pair of fermionic atoms for positive binding energy. Our functional integral
formalism is based on the euclidean microscopic action
\begin{eqnarray}\label{YukawaAction}
S_B\hspace{-0.15cm}&=&\hspace{-0.15cm}\int \hspace{-0.12cm} dx \Big[\psi^\dagger\big(\partial_{\tau}
-\frac{\triangle}{2M} -\sigma\big)\psi\nonumber\\
&&+\hat{\phi}^*\big(\partial_\tau -\frac{\triangle}{4M} + \bar{\nu}_\Lambda - 2\sigma\big)\hat{\phi} \nonumber\\
&&\hspace{-0.12cm}-\frac{\hpb}{2}\Big(\hat{\phi}^*\psi^T\epsilon\psi - \hat{\phi}\psi^\dagger\epsilon\psi^*\Big) +
\frac{\bar{\lambda}_\psi}{2}(\psi^\dagger\psi)^2\Big].
\end{eqnarray}
This is a simple model for fermions with Yukawa coupling to a scalar field $\hat{\phi}$ and a pointlike ``background ''
interaction (or ``four-fermion vertex'') $\bar{\lambda}_\psi$. The parameter $\sigma$ plays the role of an effective chemical
potential \footnote{We treat here only homogeneous situations such that the more complete formalism of \cite{Diehl:2005ae} is not
needed. We also have not bosonized the four-fermion vertex $\sim \bar{\lambda}_\psi$. In the language of \cite{Diehl:2005ae} this
corresponds to the formal limit $m^2\to \infty$.}. The different
prefactor for the coupling between $\sigma$ and the atoms resp. molecules accounts for the fact that the
molecules have atom number two. Similarly, their mass is twice the atom mass such that they have a
nonrelativistic kinetic energy $p^2/4M$. Further motivation and details of our functional integral approach can be found in
\cite{Diehl:2005ae}. Our units are $\hbar = c = k_B =1$.

The bare parameters $\bar{\nu}_\Lambda$, $\hpb$ and $\bar{\lambda}_\psi$ have to be fixed by appropriate
observable ``renormalized'' parameters. The quadratic term $\sim\hat{\phi}^* \hat{\phi}$ involves the ``bare'' binding
energy ($\bar{\nu}_\Lambda$) which typically depends on the magnetic field. The Yukawa coupling $\hpb$ accounts for the coupling
between the single atoms and molecules, $\epsilon_{\alpha\beta}=-\epsilon_{\beta\alpha}$, $\epsilon_{12}=1$. For
$\hpb\to 0$ the molecular states decouple, giving rise to a nontrivial exact ``narrow resonance limit'', cf. sect.
\ref{sec:decoupling}.

For a thermodynamic equilibrium situation the partition function is represented as a functional integral with weight factor
$e^{-S}$, with $S$ the Euclidean action
\begin{eqnarray}\label{7}
Z_B[j_\phi]&=&\int {\cal D}\psi {\cal D}\hat{\phi}\exp
\Big\{-S_B[\psi,\sigma,\hat{\phi}] \\\nonumber
&&\quad +\int dx\big[j_\phi^*(x)\phi(x) + j_\phi(x)\phi^*(x)\big]\Big\}.
\end{eqnarray}
Here $Z_B$ is defined as a functional of the source term $j_\phi(x)$ and we will set $j_\phi=0$ at the
end of the computations.

Our Yukawa type theory (\ref{YukawaAction}) is equivalent to a purely fermionic model. For this purpose we perform the
Gaussian functional integration over the $\hat{\phi}$ field. Expressed only in terms of fermions our model contains
now a four-fermion interaction with a momentum dependent coupling $\bar{\lambda}(Q_1,Q_2,Q_3,Q_4)$. The interaction has the
explicit form ($Q_4=Q_1+Q_2-Q_3$) \cite{CCKokkelmans}
\begin{eqnarray}\label{Mom4Fermion}
S_{int} = \frac{1}{2}\int\limits_{Q_1,Q_2,Q_3}\big(\psi^\dagger(-Q_1)\psi(Q_2)\big)\big(\psi^\dagger(Q_4)
\psi(-Q_3)\big)\nonumber\\
\Big\{ \bar{\lambda}_\psi - \frac{\hpb^2}{2\pi\mathrm{i}(n_1-n_4)T+(\vec{q}_1-\vec{q}_4)^2/4M + \bar{\nu}_\Lambda - 2\sigma}
\Big\}.\quad
\end{eqnarray}
In the pointlike limit the momentum dependence can be neglected and eq. (\ref{Mom4Fermion}) is replaced by the ``local
interaction approximation''
\begin{eqnarray}\label{BosonCond2}
\bar{\lambda}= \bar{\lambda}_\psi -\frac{\hpb^2}{\bar{\nu}_\Lambda - 2\sigma} .
\end{eqnarray}
Later we will argue that the pointlike limit can be established by $\hpb\to \infty$, $\bar{\nu} \propto\hpb^2$ (with
$\bar{\nu}$ the UV renormalized version of $\bar{\nu}_\Lambda$, cf. \cite{Diehl:2005ae} and sect. \ref{subsec:Renormalization}). This
settles the connection between broad Feshbach resonances and the pointlike limit, for which a purely fermionic description
as discussed by Strinati \emph{et al.} \cite{AAAAStrinati,BBStrinati,CCStrinati,YPieri,ZStrinati} is appropriate.

The action $S_B$ (\ref{YukawaAction}) has a global $U(1)$ symmetry. This implies the conservation of the particle number
$N=\int d^3x\, n$, where the total number density of atoms $n$ includes the unbound atoms and
those bound in molecules or forming condensates. It obeys
\begin{eqnarray}\label{TotDens}
n &=&  \nF + \nB = \langle\psi^\dagger \psi\rangle   + 2\langle \hat{\phi}^*\hat{\phi}\rangle\\\nonumber
  &=&  \langle\psi^\dagger \psi\rangle   + 2\langle \hat{\phi}^*\hat{\phi}\rangle_c + 2\langle\hat{\phi}^*\rangle
\langle\hat{\phi}\rangle\\\nonumber
&=& \nF + 2\nM + \nC.
\end{eqnarray}
In the second line, we have expressed the disconnected correlation function $\langle \hat{\phi}^*\hat{\phi}\rangle$ by the
connected Green's function $\langle \hat{\phi}^*\hat{\phi}\rangle_c$ and the expectation value $\langle\hat{\phi}\rangle$.
These terms can be interpreted \cite{Diehl:2005ae} as density contributions from open channel atoms ($\nF=\langle\psi^\dagger
\psi\rangle$) and closed channel atoms in uncondensed molecules ($\nM=\langle \hat{\phi}^*\hat{\phi}\rangle_c$) as well as a
contribution from closed channel atoms in the condensate ($\nC = 2\langle\hat{\phi}^*\rangle \langle\hat{\phi}\rangle$).
We emphasize that eq. (\ref{TotDens}) is no ``ad hoc'' assumption - it directly follows from the microscopic formulation of
the functional integral.

For a homogeneous setting the condensate is constant, $\langle\hat{\phi}(x)\rangle =\bar{\phi}_0$. It can be related \cite{Diehl:2005ae}
to the expectation value  of a composite di-atom operator
\begin{eqnarray}
\bar{\phi}_0 = \frac{\hpb}{2\bar{\nu}_\Lambda - 4\sigma}\langle \psi^T\epsilon\psi\rangle.
\end{eqnarray}
This demonstrates directly that our formulation in terms of a path integral makes no difference between a ``condensate of
molecules'' $\bar{\phi}$ and a ``condensate of atom pairs'' $\langle\psi^T\epsilon\psi\rangle$ - they are simply related by a
multiplicative constant. Hence, the universality of the condensation phenomenon itself - the fact that we do not
need to specify a precise microscopic mechanism as e.g. the formation of Cooper pairs - is particularly visible.

The field equations which determine $\bar{\phi}_0$ and $n$ can be computed from the effective action $\Gamma[ \sigma, \bar{\phi}]$,
which depends on $\sigma$ and the expectation value $\bar{\phi} = \langle\hat{\phi}\rangle$. As usual, it obtains by a
Legendre transform, eliminating the source $j_\phi(x)$ in favor of $\bar{\phi}(x)$,
\begin{eqnarray}
\Gamma[\bar{\phi}] = -\ln Z_B[j_\phi[\bar{\phi}] ] + \int dx(j_\phi^*\bar{\phi} + j_\phi\bar{\phi}^*).
\end{eqnarray}
For homogeneous $\sigma,
\bar{\phi}$, it is appropriate to consider the effective potential $U = (T/V) \Gamma$. To one loop order it is composed of
a classical piece and a fluctuation contribution,
\begin{eqnarray}
U &=& U^{(cl)} +U_1.
\end{eqnarray}
In our case, the classical contribution reads
\begin{eqnarray}\label{Uclass}
U^{(cl)} =% \frac{m^2}{2} \,\sigma^2
(\bar{\nu}_\Lambda -2\sigma) \bar{\phi}^* \bar{\phi}.
\end{eqnarray}
The one loop contribution is composed of a piece from fermion ($U_1\hF$) and boson ($U_1\hB$) fluctuations.
The explicit expression for the one loop part depends on the detailed approximation. It will be discussed in sect.
\ref{EvalBeyondMFT}, while we will first display some general features and results of our approach.

The homogeneous field equations read \cite{Diehl:2005ae}
%in presence of a nonvanishing source $J = \mu$ read
%\begin{eqnarray}\label{FESigma}
%\frac{\partial U_\sigma}{\partial\sigma} &=& m^2\mu = m^2\sigma - n,\\
%\frac{\partial U_\sigma}{\partial\phi}   &=& 0.
%\end{eqnarray}
%It is discussed in detail, including the subtle issues of counterterms, in \cite{Diehl:2005ae}. We redefine the
%effective potential
%\begin{eqnarray}\label{barU}
%U = U_\sigma - (m^2/2)\sigma^2
%\end{eqnarray}
%such that a homogeneous situation is characterized by
\begin{eqnarray}\label{FESigmaII}
-\frac{\partial U}{\partial\sigma} &=& n,\quad \frac{\partial U}{\partial\bar{\phi}} = 0.
\end{eqnarray}
The condensation of atoms pairs or molecules is signalled by a non-vanishing expectation value $\langle\hat{\phi}\rangle
=\bar{\phi}_0$, given by the location of the minimum of $U$. The associated spontaneous breaking of the global continuous
symmetry of phase rotations of $\psi$ and
$\hat{\phi}$ (related to the conservation of the number of atoms) induces a massless Goldstone boson. This is the origin of
superfluidity.

\section{Universality}
\label{sec:renormalization}

\subsection{Additive Renormalization}
\label{subsec:Renormalization}

The microscopic action (\ref{YukawaAction}) depends explicitly on three parameters $\bar{\nu}_\Lambda$, $\hpb$,
$\bar{\lambda}_\psi$. A fourth
parameter is introduced implicitly by the ultraviolet cutoff $\Lambda$ for the momentum integration in the fluctuation
effects. (Besides this, the results will depend on the thermodynamic variables $T$ and $\sigma$ - see \cite{Diehl:2005ae}
for the relation between the effective chemical potential $\sigma$ and the chemical potential $\mu$.) The renormalization
procedure eliminates the cutoff dependence and trades these ``bare'' parameters for renormalized ones which can
subsequently be related to observables of the concrete atomic system (cf. sect. \ref{sec:concmag}).

It turns out that it is sufficient to perform the ultraviolet renormalization in two instances: The first one concerns the
connected correlation functions (density contributions) $\langle\psi^\dagger \psi\rangle$ and $\langle \hat{\phi}^*
\hat{\phi}\rangle_c$. Carefully relating these field expectation values to their counterparts in the operator formalism shows
that the divergences $\propto \Lambda^3$ are due to a zero point shift of the density \cite{Diehl:2005ae}. In our setting the
contributions of the fermionic fluctuations are cancelled by the bosonic (molecule)
fluctuations and no renormalization is necessary in this respect. The second divergence for $\Lambda
\to \infty$ is associated to the bare detuning $\bar{\nu}_\Lambda$. It can be absorbed by using the renormalized detuning
\begin{eqnarray}\label{Watwees}
\bar{\nu} = \bar{\nu}_\Lambda - \frac{\hpb^2 M \Lambda}{2\pi^2}.
\end{eqnarray}
This prescription corresponds to a renormalization of the
effective atom interaction strength $\bar{\lambda}$ (\ref{BosonCond2}) \cite{Diehl:2005ae}.

Expressed in terms of $\bar{\nu}$ and $\hpb$ the
effective potential becomes very insensitive to the microscopic physics, i.e. the value of the cutoff $\Lambda$. Without
much loss of accuracy we can take the limit $\Lambda \to \infty$ for the computation of $U$.

\subsection{Dimensionless Parameters}
\label{sec:DimlessParam}

The action (\ref{YukawaAction}) still involves the atom mass $M$ which is much larger than all other characteristic mass
scales. By the choice of appropriate units the nonrelativistic physics becomes independent of $M$. We therefore introduce
a ``Fermi momentum'' $k_F$ and a ``Fermi energy'' $\epsilon_F = k_F^2/2M$. The inverse of the Fermi momentum should be
associated with the most important characteristic length scale in our problem. Roughly speaking, it corresponds to the
average distance $d$ between two unbound atoms or molecules. Nevertheless, the precise value of $k_F$ is not relevant - it
just sets the units and we could take a fiducial value $k_F= 1\mathrm{eV}$. For a homogenous situation it is convenient,
however, to define $k_F$ by the density,
\begin{eqnarray}
n = k_F^3/(3\pi^2).
\end{eqnarray}

Next we rescale coordinates and fields according to
\begin{eqnarray}\label{dimlessUnits1}
\tilde{\vec{x}} &=& k_F \vec{x}, \quad \tilde{\tau} = \epsilon_F \tau, \quad \tilde{T} = T/\epsilon_F,\,\,\qn = q/k_F,\\\nonumber
\quad \tilde{\psi}&=&
k_F^{-3/2}\psi, \quad \hat{\tilde{\phi}} = k_F^{-3/2}\hat{\phi}, \quad \tilde{\sigma}  = \sigma/\epsilon_F.
\end{eqnarray}
This yields the scaling form of the action
\begin{eqnarray}\label{DimlessYukawaAction}
S_B\hspace{-0.15cm}&=&\hspace{-0.15cm}\int \hspace{-0.12cm} d\tilde{x} \Big[\tilde{\psi}^\dagger\big(\tilde{\partial}_{\tau}
-\tilde{\triangle} -\tilde{\sigma}\big)\tilde{\psi}\nonumber\\
&&+\hat{\tilde{\phi}}^*\big(\tilde{\partial}_\tau -\frac{1}{2}\tilde{\triangle} + \tilde{\nu}_\Lambda
- 2\tilde{\sigma}\big)\hat{\tilde{\phi}} \nonumber\\
&&\hspace{-0.12cm}-\frac{\hpn}{2}\Big(\hat{\tilde{\phi}}^*\tilde{\psi}^T\epsilon\tilde{\psi}
- \hat{\tilde{\phi}}\tilde{\psi}^\dagger\epsilon\tilde{\psi}^*\Big)
+ \frac{\tilde{\lambda}_\psi}{2}(\tilde{\psi}^\dagger\tilde{\psi})^2\Big].
\end{eqnarray}
Accordingly, all quantities derived from the partition function can be brought to a scaling form and can only depend
on the dimensionless parameters ($\tilde{\nu}_\Lambda = \tilde{\nu} + \Lambda\hpn^2/(4\pi^2k_F)$)
\begin{eqnarray}\label{dimlessUnits2}
\tilde{\nu} = \bar{\nu}/\ef,  \quad \hpn= 2M k_F^{-1/2} \hpb, \,\, \tilde{\lambda}_\psi = 2Mk_F\bar{\lambda}_\psi.
\end{eqnarray}
This scaling form shows a first important aspect of universality. All computations can be performed at a fiducial $k_F=
1\mathrm{eV}$. Contact to a concrete physical situation is only made at the end by rescaling the results by using the
value of $k_F$ appropriate to the density of a given experiment. In our approximation for a homogeneous situation all
dimensionless physical observables can be expressed in terms of the dimensionless variables $\tilde{\nu}$, $\hpn$,
$\tilde{\lambda}_\psi$ and $\Tn$! The scale is then introduced by $n$.

The parameters in the action (\ref{DimlessYukawaAction}) are microscopic parameters $\tilde{\nu}_\Lambda,
\tilde{h}_{\phi,\Lambda}, \tilde{\lambda}_{\psi,\Lambda}$. Similar to the renormalization of $\tilde{\nu}$ also the
couplings $\hpn$ and $\tilde{\lambda}_\psi$ will be affected by renormalization effects. For a comparison with experiment
we have to relate the ``bare couplings'' $\tilde{h}_{\phi,\Lambda}$, $\tilde{\lambda}_{\psi,\Lambda}$ to renormalized
couplings, as for example the coupling $\tilde{h}_{\phi,0}$ which appears in the molecular binding energy in vacuum
of the scattering of two isolated atoms \footnote{In order not to overload the notation we often will not specify directly
the indices like $\tilde{h}_{\phi,\Lambda}$ or $\tilde{h}_{\phi,0}$. Where it becomes necessary we describe the status of
$\hpn$ more precisely}. We will argue below that in terms of the macroscopic relevant parameters only two parameters
remain which can be associated to the renormalized couplings $\nun$ and $\hpn$. The macroscopic value of
$\tilde{\lambda}_\psi$ becomes computable as a partial fixed point in terms of $\nun$ and $\hpn$, at least sufficiently
close to the resonance and for sufficiently large $\hpn$. Furthermore, for the broad resonance limit of a pointlike
interaction we may absorb $\tilde{\lambda}_\psi$ by a Fierz transformation into a redefinition of $\hpn^2$ such that
$\bar{\lambda}_\psi$ (\ref{BosonCond2}) remains invariant, i.e. $\hpb^2\to \hpb^2 - \bar{\lambda}_\psi\bar{\nu}_\Lambda$
(with $|2\sigma| \ll \bar{\nu}_\Lambda$). In sect. \ref{sec:concmag} we will relate $\nun$ and $\hpn$ to the parameters of
experiments with gases of $\lit$ and $\kal$.

\subsection{Concentration}

The space of model parameters and thermodynamic variables for the crossover problem is spanned by $\tilde{\nu}, \hpn^2$ and
$\Tn$. However, it
turns out that another set of parameters, $c, \hpn^2$ and $\Tn$, describes the system more efficiently, revealing further
aspects of universality more clearly. Using the relation between a four-fermion coupling $\bar{\lambda}$ and a scattering
length $a$ ,
\begin{eqnarray}\label{CoupScLength}
\bar{\lambda} = \frac{4\pi a}{M},
\end{eqnarray}
we define an effective in-medium resonant scattering length $\bar{a}$ or, in the scaling form,
\begin{eqnarray}\label{InMediumScattLength}
\frac{1}{c} = \frac{1}{\bar{a}k_F}  =- \frac{8\pi(\tilde{\nu} - 2\sigex)}{\hpn^2}.
\end{eqnarray}
We will refer to $c$ as ``concentration''. Indeed, with the inverse Fermi momentum setting the scale for the
average interparticle spacing $d$, the concentration $c$ is a measure for the ratio between scattering length $\bar{a}$
and average distance, $c\propto \bar{a}/d$. It is therefore of no surprise that in the case $c^{-1} <1$, where the
scattering length exceeds the typical interatomic distance, many-body effects play an essential role. Note that through
$\sigma$, this in-medium scattering length depends on the density. It will, in general, differ from the
``reduced scattering length'' $a_R = \bar{a}(\sigma = 0)$, which is defined by
\begin{eqnarray}\label{VacuumScattLength}
 \frac{1}{a_R k_F} = - \frac{8\pi\tilde{\nu}}{\hpn^2}.
\end{eqnarray}
For broad Feshbach resonances the concentration is related to the full scattering length $a(B)$ by $c=a(B) k_F$ (cf.
sect. \ref{sec:concmag}). Here the dependence of $a(B) = a_{res}(B) + a_{bg}$ on the magnetic field $B$ involves both a
resonant piece $a_{res}(B)$ that diverges at the Feshbach resonance $B\to B_0$, and an approximately $B$ - independent
background scattering length $a_{bg}$.

The impact of the choice of variables $c^{-1}, \hpn^2, \Tn$ is twofold. First, it absorbs a good deal of the $\hpn$ -
dependence in the final results. In the language of
critical phenomena, the dimensionless temperature $\Tn$ and the inverse concentration $c^{-1}$ are relevant parameters,
whereas $\hpn$ is marginal. A more systematic discussion of the role of $\hpn$ will be given in the next section. Second,
the choice of $c$ or $\bar{a}$ relates our formulation based on partial bosonization more closely to an interacting
fermion system, where the effective scattering length $\bar{a}$ is obviously a crucial parameter.

For a small concentration $|c|$ the gas is dilute in the sense that scattering can be treated as a
perturbation. In this range mean field theory is expected to work reasonably well. On the other hand, for large $|c|$ the
scattering length exceeds the average distance between two atoms and collective effects play a crucial role.

\begin{figure}
\begin{minipage}{\linewidth}
\begin{center}
\setlength{\unitlength}{1mm}
\begin{picture}(85,54)
      \put (0,0){
     \makebox(80,49){
     \begin{picture}(80,49)
       %\coordsyst{90}{54.9}{21}{14}
      \put(0,0){\epsfxsize80mm \epsffile{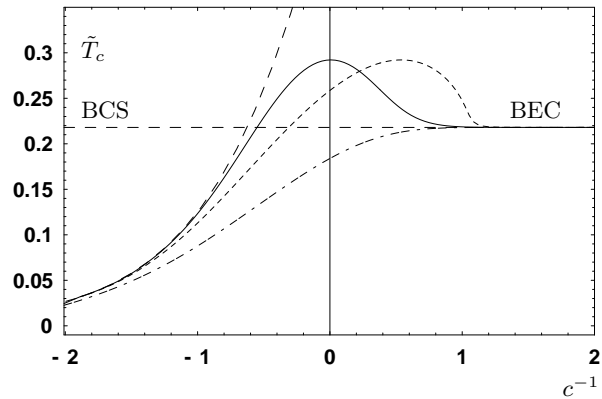}}
      \put(74,-3){$c^{-1}$}
      \put(10,42){$\Tn_c$ }
      \put(10,34){BCS}
      \put(67,34){BEC}
      \end{picture}
      }}
   \end{picture}
\end{center}
\vspace*{-1.25ex} \caption{Crossover phase diagram. The dependence of the critical temperature $\tilde{T}_c = T_c/\epsilon_F$
on the inverse concentration $c^{-1}=(\bar{a}k_F)^{-1}$ is shown for
the broad resonance limit $\hpn\to \infty$ (solid
and short-dashed line) and for the narrow resonance limit $\hpn \to 0$ (dashed-dotted line). We also indicate the standard
BEC (dashed horizontal line) and BCS (dashed rising line) limits which do not depend on the choice of the Yukawa coupling.
For the broad resonance limit we plot two different approximations.
}
\label{CrossoverTcAll}
\end{minipage}
\end{figure}

In summary we have now an effective low energy formulation where neither $M$ nor $\Lambda$ enter anymore. Everything is
expressed in terms of $k_F$ and three dimensionless parameters, namely $c$, $\Tn$ and $\hpn$.

\section{Enhanced Universality}
\label{sec:Enhanced}

Enhanced universality occurs in situations where one of the two parameters $c, \hpn$ becomes irrelevant (or fixed).
This concerns the limits of narrow and broad Feshbach resonances (``narrow resonance limit'', $\hpn\to 0$ and ``broad
resonance limit'', $\hpn \to \infty$), the BCS and BEC regimes, $c\to 0$, and the scaling limit, $c \to \infty$. For
a given $\Tn$ the crossover for narrow and broad Feshbach resonances can be described by a single parameter $c^{-1}$.
Results for the BCS, BEC and scaling limit can only depend on $\hpn$ and become parameter free for narrow or
broad resonances.

\begin{figure}[t!]
\begin{minipage}{\linewidth}
\begin{center}
\setlength{\unitlength}{1mm}
\begin{picture}(85,108)
      \put (0,0){
     \makebox(80,49){
     \begin{picture}(80,49)
      % \coordsyst{90}{54.9}{21}{14}
      \put(0,0){\epsfxsize80mm \epsffile{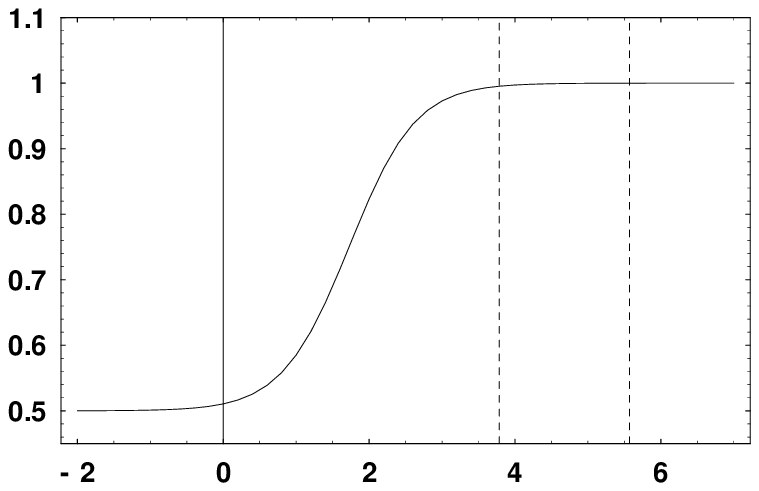}}
      \put(67,-2){$\log_{10}\hpn^2$}
      \put(11,40){$\ApR$ }
      \put(43,25){$\kal$ } \put(66,25){$\lit$ }
      \put(-3,1){(b)}
      \end{picture}
      }}
      \put (0,54){
    \makebox(80,49){
    \begin{picture}(80,49)
      % \coordsyst{90}{54.9}{21}{14}
      \put(0,0){\epsfxsize80mm \epsffile{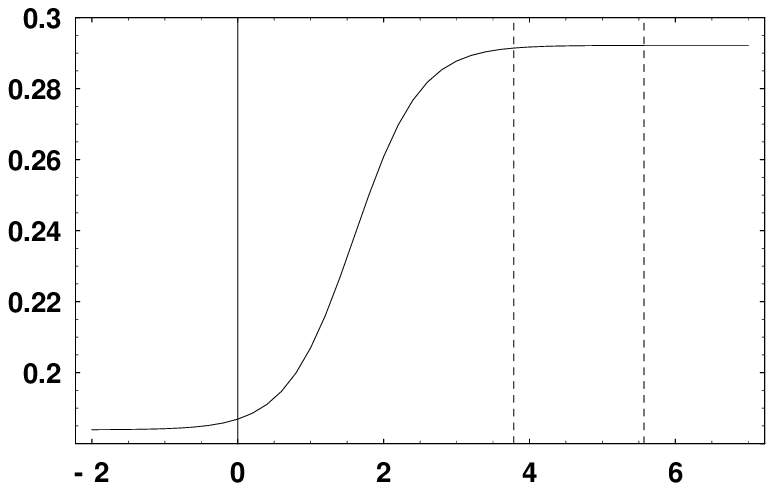}}
      \put(67,-3){$\log_{10}\hpn^2$}
      \put(46,25){$\kal$ } \put(68,25){$\lit$ }
      %\put(6,19){\footnotesize{[10$^{12}$ cm$^{-3}$]}}
      \put(12,40){$\Tn_c$ }
      \put(-3,1){(a)}
      \end{picture}
      }}
\end{picture}
\end{center}
\vspace*{-1.25ex} \caption{Enhanced universality for large and small $\hpn$. For $c^{-1} =0$ we plot the dependence on
$\hpn$ of (a) the dimensionless critical temperature $\Tn_c$ and (b) the gradient coefficient $\ApR$ for $T =T_c$ (cf. sect.
\ref{DressedAndBareI}).
For small $\hpn < 1$, a stable universal narrow resonance limit is approached. For large $\hpn$ we find a very pronounced
insensitivity of $\Tn_c$ and $\ApR$ to the precise value of $\hpn$ - note that we plot on a logarithmic scale. The
``crossover'' regime interpolates smoothly between the universal limits. }
\label{hpUniversality}
\end{minipage}
\end{figure}

We can qualitatively discuss the limits of enhanced universality taking the phase diagram in fig. \ref{CrossoverTcAll} as an example.
The values of $c^{-1}$ and $\hpn$ determine the phase diagram completely, i.e. the critical temperature depends only on
these two parameters, $T_c = T_c(c^{-1},\hpn)$. We display the critical temperature $T_c$ as a function of $c^{-1}$ and
compare the limits $\hpn\to 0$ (dashed) and $\hpn\to \infty$ (two approximations, solid and short dashed). (The underlying
approximation schemes are discussed in sect. \ref{EvalBeyondMFT}.) For intermediate values of
$\hpn$, a monotonic change between the two limits is found, cf. also fig.
\ref{hpUniversality}. Though the impact of the marginal parameter $\hpn$ is moderate, the precise value of $\hpn$
influences the details of the crossover and will be important for precision estimates for concrete physical systems, in
particular when narrower resonances than those presently investigated in $\lit$ or $\kal$ will be explored.

Fig. \ref{CrossoverTcAll} clearly reveals an $\hpn$ - independent approach to the limiting BCS and BEC regimes: for
$|c^{-1}| \to \infty$, $\hpn$ becomes an irrelevant parameter and the system can be described in terms of the
concentration only. For comparison, we have also plotted the extrapolated standard BCS (dashed, rising line) and BEC
(dashed, horizontal line) results.

The narrow resonance limit corresponds to an exact solution of the many-body problem. It requires $\hpn \to 0$ and
additionally a vanishing four-fermion background interaction, but is free of further approximations. The broad resonance
limit, instead, corresponds to a strongly interacting field theory, whose approximate solution still involves quantitative
uncertainties, in particular close to the critical temperature (cf. sect. \ref{EvalBeyondMFT}). This is reflected by the
two critical lines obtained in different approximations in the broad resonance limit. The difference between the solid
line and the short dashed line gives an idea of the remaining uncertainty in the treatment of the molecule fluctuations.
For the solid line our result at the resonance $\Tn_c = 0.292$ is in agreement with the
theoretical value obtained in \cite{Levin052,Thomas05} and compatible with the measurements reported in \cite{Thomas05}
$\Tn_c= 0.31\pm 0.04$. Including the molecule fluctuations in the gap equation (short dashed line) yields $\Tn_c(c^{-1}=0)
=0.255$. The maximum of $\Tn_c(c^{-1})$ has a similar value for both approximations.

We find a new form of crossover from the narrow to broad resonance limit in dependence on the Yukawa or Feshbach coupling
$\hpn$. This is shown in fig. \ref{hpUniversality}, where we plot $\Tn_c$ (fig. \ref{hpUniversality} (a)) as a function of
$\hpn$ for $c^{-1}=0$. (The curves for other values of large $|c|$ are similar.) One clearly sees a smooth
interpolation between the regimes. Within the broad resonance regime the precise value of $\hpn$ is not relevant
for $\Tn_c$ or, more generally, for all quantities not involving explicitly the microscopic constituents (``bare fields'').
Renormalization effects for $\hpn$ even as large as a factor of ten do not matter. In contrast, in the ``crossover regime''
of intermediate $\hpn\approx 10$ the physical results depend on the value of $\hpn$. Now renormalization effects must be
taken into account carefully for a reliable computation. Furthermore, we will learn in sect. \ref{DressedAndBareI} that
observations related to ``microscopic observables'' like the fraction of closed channel atoms (or microscopic molecules)
can depend strongly on $\hpn$ even in the broad resonance regime. We also show in fig. \ref{hpUniversality} (b) the
gradient coefficient $\ApR$ for $T=T_c$, $c^{-1}=0$ - this quantity will be defined in the next section.

More details on the different limits of enhanced universality will be given in sect. \ref{sec:EnhancedUniv} after the discussion of the
necessary
formalism in the next sections.

\section{Dressed and bare molecules: The wave function renormalization $\ZpR$}
\label{DressedAndBareI}

For an understanding of the crossover problem, it is crucial to distinguish bare and dressed molecules
\cite{Stoof05,IIChen05}. In our interpretation, the bare or ``closed channel'' molecules are associated to a true
nonlocality in the microscopic interaction (\ref{Mom4Fermion}). Their overall role for the thermodynamics should be
subdominant for broad resonances or almost pointlike microscopic interactions. The dressed molecules can instead be
effective degrees of freedom, generated dynamically by fluctuation effects. They correspond to a renormalized bosonic
field.

\subsection{Derivative Expansion}
For a homogeneous situations or weakly varying trap potentials a derivative expansion of the effective action becomes
appropriate (with $\tilde{\phi} = \langle \hat{\tilde{\phi}}\rangle = k_F^{-3/2}\bar{\phi}$ in presence of sources
$j_\phi$)
\begin{eqnarray}
\Gamma [\tilde{\phi}] = \int d\tilde{x}\big\{ \tilde{u}(\tilde{\phi}) + \Apn \tilde{\vec{\nabla}}\tilde{\phi}^*
\tilde{\vec{\nabla}}\tilde{\phi} + \ZpR \tilde{\phi}^*\tilde{\partial}_\tau\tilde{\phi} + ...\big\}.
\end{eqnarray}
By analytic continuation from euclidean time $\tau$ to Minkowski time (``real time'') we can derive a time evolution
equation for $\tilde{\phi}$ from the variational principle $\delta \Gamma /\delta\tilde{\phi}=0$. The coefficient of the
term with the time derivative, $\ZpR$, plays the role of a wave function renormalization. We therefore introduce
renormalized fields for bosonic quasiparticles or ``dressed molecules'' and appropriately renormalized couplings
\begin{eqnarray}
\phi = \ZpR^{1/2} \tilde{\phi}, \quad \rho = \phi^*\phi, \quad \ApR = \frac{\Apn}{\ZpR}, \quad \nu =
\frac{\tilde{\nu}}{\ZpR}
\end{eqnarray}
such that
\begin{eqnarray}\label{EffActTrunc}
\Gamma [\phi] = \int d\tilde{x}\big\{ \tilde{u}(\phi) + \ApR \tilde{\vec{\nabla}}\phi^*
\tilde{\vec{\nabla}}\phi + \phi^*\tilde{\partial}_\tau\phi + ...\big\}.
\end{eqnarray}

The effective potential can be decomposed into a classical contribution $\tilde{u}^{(cl)}$, a contribution from the fermion
fluctuations $\tilde{u}_1\hF$ and a contribution from the molecule fluctuations $\tilde{u}_1\hB$,
\begin{eqnarray}
\tilde{u} &=& \tilde{u}^{(cl)} + \tilde{u}_1\hF + \tilde{u}_1\hB,\\\nonumber
\tilde{u}^{(cl)} &=& (\tilde{\nu} - 2\sigex) \tilde{\phi}^*\tilde{\phi} = (\nu - 2\sigex/\ZpR) \rho.
\end{eqnarray}
The $\sigma$ - dependence of the different pieces is closely related to the different contributions to the total density of
atoms \cite{Diehl:2005ae}. Indeed, the pressure of the system obeys $p(T) =-(U_0(\sigma, T) - U_0(\sigma = T=0))$, where
$U_0(\sigma)$ denotes the value of the effective potential at its minimum (with respect to $\bar{\phi}$). A thermodynamic
relation associates the particle density with the derivative of $p$ with respect to the chemical potential.
\begin{figure}
\begin{minipage}{\linewidth}
\begin{center}
\setlength{\unitlength}{1mm}
\begin{picture}(85,52)
      \put (0,0){
     \makebox(80,49){
     \begin{picture}(80,49)
       %\coordsyst{90}{54.9}{21}{14}
      \put(0,0){\epsfxsize80mm \epsffile{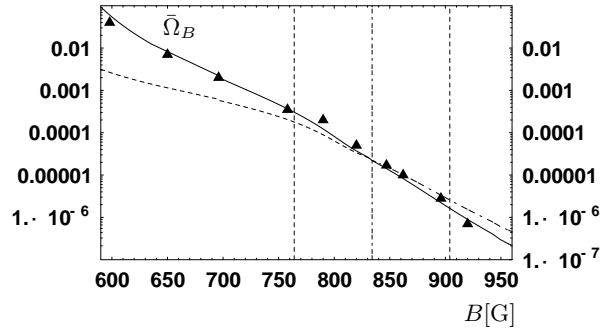}}%with the line vanishing at resonance: BareExp4.eps
      \put(61,2){$B [\mathrm{G}]$}
%      \put(-3,1){(b)}
      \put(21,40){$\OmB$ }
  %    \put(15,28){$\OmB$ }
      \end{picture}
      }}
\end{picture}
\end{center}
\vspace*{-1.25ex} \caption{Fraction of closed channel
molecules $\OmB = \OmC + \OmM$, compared to experimental values \cite{Partridge05}, for $T=0$ and $k_F= 0.493
\mathrm{eV}\hat{=}250\mathrm{nK}$. On the BEC side, we find that our results are quite insensitive to the precise choice
of $k_F$. The strongly interacting region $c^{-1}<1$ is indicated by vertical lines, where the center line denotes the
position of the resonance. For $B> B_0$ the solid and long dashed-dotted lines reflect the
uncertainty from the ``Fierz ambiguity'' discussed in app. \ref{sec:scattvac}. For $B<B_0$ the dashed line omits the
renormalization effects discussed in sect. \ref{sec:concmag} and app. \ref{sec:YukRenorm}.}
\label{BareExpPart}
\end{minipage}
\end{figure}

\subsection{Contributions to the atom density}
The wave function renormalization $\ZpR$ plays a crucial role for the precise understanding of the different contributions
to the total atom density, from condensed or uncondensed atoms, molecules or unbound atoms, open channel and closed
channel atoms. For example, the number density of uncondensed dressed molecules $n_M$ can be computed from the propagator
for the renormalized field $\phi$. It is related to the $\tilde{\sigma}$ - derivative of $\tilde{u}_1\hB$
\cite{Diehl:2005ae}
evaluated at fixed $\nu$ (not fixed $c^{-1}$!)
\begin{eqnarray}\label{YYA}
\frac{n_M}{k_F^3} = -\frac{1}{2}\frac{\partial \tilde{u}_1\hB}{\partial\sigex}\big|_{\phi_0} =
\frac{\Omega_M}{6\pi^2}.
\end{eqnarray}
Here $\phi_0$ corresponds to the minimum of $\tilde{u}$ and $\Omega_M = 2n_M/n$ counts the fraction of atoms bound
in the dressed molecules. Similarly, the number of bare or microscopic molecules $\bar{n}_M$ corresponds to the propagator
for $\tilde{\phi}$ such that
\begin{eqnarray}
n_M = \ZpR \bar{n}_M,\quad  \Omega_M = \ZpR \bar{\Omega}_M.
\end{eqnarray}
Now $\bar{\Omega}_M$ counts the fraction of closed channel atoms (not in the condensate) as defined by the microscopic
molecular Feshbach state.
For large $\ZpR$ the number of dressed molecules $n_M$ can be substantial even if the number of microscopic molecules
$\bar{n}_M$ is negligibly small. Of course, in this case the dressed molecules are dominantly composed of open channel
atoms and not of closed channel molecules.

Similarly, the condensate fraction $\Omega_C = n_C/n$ measures the fraction of atoms in the condensate
\begin{eqnarray}
\frac{\Omega_C}{3\pi^2}=\frac{n_C}{k_F^3} = 2\phi^*_0\phi_0 = 2\rho_0 =
-\ZpR\frac{\partial\tilde{u}^{(cl)}}{\partial\sigex} \big|_{\phi_0}.
\end{eqnarray}
It is related to $\bar{n}_C$ (\ref{TotDens}) by $n_C = \ZpR\bar{n}_C$. Thus the fraction of closed channel atoms in the
condensate obeys $\bar{\Omega}_C = \bar{n}_C/n = \Omega_C/\ZpR$ and is tiny for large $\ZpR$. The remaining number of
(dressed) unbound atoms obeys $n_F = n- n_M - n_C$. Correspondingly, $\Omega_F = n_F /n$ denotes the fraction of atoms that
are neither condensed nor bound in dressed molecules.

\subsection{Computation of $\ZpR$}
The wave function renormalization $\ZpR$ can be computed from the dependence of the molecule propagator on the (real)
frequency \cite{Diehl:2005ae}. We include in this paper only the contribution from fermion fluctuations. In this section
we employ the relation \cite{Diehl:2005ae}
\begin{eqnarray}\label{ZpR}
\ZpR = 1 - \frac{1}{2}\frac{\partial^3 \tilde{u}_1\hF}{\partial\sigex\partial\tilde{\phi}^*\partial\tilde{\phi}}
\end{eqnarray}
which relates $\ZpR$ to the dependence of the inverse molecule propagator on the effective chemical potential $\sigma$.
(In the app. \ref{app:dispersion} we discuss a definition of $\ZpR$ based on the explicit frequency dependence of the
inverse propagator.) In eq. (\ref{ZpR}) $\tilde{u}_1\hF$ is computed from a fermion loop in the background of constant
$\tilde{\phi}$ and $\sigex$. It reads
\begin{eqnarray}\label{USigmaPhi}
\tilde{u}_1^{(F)}(\sigex,\tilde{\phi})&=& - 2\Tn\int\frac{d^3\qn}{(2\pi)^3}\Big[\ln\big(\mathrm{e}^{\gamma_\phi - \gamma} +
               \mathrm{e}^{-\gamma_\phi-\gamma}\big)\nonumber\\
              &&\qquad\qquad\qquad - \frac{\rex }{4\Tn \qn^2}\Big]
%&=& -\frac{1}{2} \Tr \log P_F \\\nonumber
      %  &=& - 2T\int\frac{d^3q}{(2\pi)^3}\ln\cosh\gamma_\phi%\big(\mathrm{e}^{\gamma_\phi - \gamma} +
               %\mathrm{e}^{-\gamma_\phi-\gamma}\big)
               ,\\\nonumber
\end{eqnarray}
with \footnote{The appropriate choice of the Yukawa coupling $\hpn$ relating $\rex$ and $\tilde{\phi}$ is a subtle issue. We
follow here the concept of a systematic renormalization group improved perturbation theory \cite{Wetterich:1996eu} where
$\hpn$ is a $\sigex$ and $\Tn$ dependent renormalized coupling which is evaluated at the momentum scale which dominates
the integral (\ref{USigmaPhi}). This prescription differs from the Schwinger-Dyson equations for the fermion contributions
to the $\phi$ vertices which follow from the $\phi$ - derivatives of eq. (\ref{USigmaPhi}). In the SDE - approach one
factor of $\hpn$ would correspond to a microscopic or bare coupling $\tilde{h}_{\phi,\Lambda}$.}
\begin{eqnarray}
\gamma_\phi&=&\frac{1}{2\Tn}\Big(\Big(\qn^2 -\sigex\Big)^2 +\rex\Big)^{1/2},\\
\gamma &=& \gamma_\phi(\tilde{\phi}=0) = \frac{1}{2\Tn}\Big(\qn^2 -\sigex\Big), \nonumber\\
\rex &=& \hpn^2\tilde{\phi}^*\tilde{\phi} .%= \hpn^2 \rhon.
\label{RexRho}\nonumber
\end{eqnarray}
The relation (\ref{ZpR}) yields
\begin{eqnarray}\label{ZRFormula}
\ZpR &=& 1 + \frac{\hpn^2}{16\Tn^2}\int \frac{d^3\qn}{(2\pi)^3}\,\,\gamma \gamma_\phi^{-3}\big[\tanh\gamma_\phi -
        \gamma_\phi\cosh^{-2}\gamma_\phi\big]\nonumber\\
\end{eqnarray}
and agrees indeed with the definition from the frequency dependence of the propagator in app. \ref{app:dispersion}.

In the symmetric phase $\ZpR$ can be expressed in terms of the dimensionful parameters as
\begin{eqnarray}\label{FuncZphi}
\ZpR &=& 1 + \frac{1}{4\pi^2} M^{3/2} \hpb^2 T^{-1/2} H(s),\nonumber\\
H(s) &=& \int\limits_0^\infty dy \frac{y^2}{(y^2 - s)^2}\big\{\tanh(y^2 - s) - \frac{y^2-s}{\cosh^2(y^2 - s)}\big\},\nonumber\\
s &=& \frac{\sigma}{2T} =  \frac{\sigex}{2\Tn}
\end{eqnarray}
where $H(s)$ has the limit
\begin{eqnarray}\label{FuncHs}
H(s\to -\infty) &\to& \frac{\pi}{4}(-s)^{-1/2}.
%H(s\to \infty) &\to& ??s^{3/2},\nonumber\\
%H(s\to 0) &\to& ??
\end{eqnarray}
The limit $T\to 0, \sigma\to 0$ is infrared divergent. This divergence is cured by a nonzero momentum $\vec{q}$
in the propagator: For $T=\sigma =0$ the correction $\propto \hpb^2(\vec{q}^2)^{-1/2}$ diverges only for $\vec{q}^2\to 0$.
In the vacuum the physics of the ``dressed molecules'' has therefore to be handled with care. We will see in sect.
\ref{sec:lowdens} that the case $\sigma \to 0$ applies to a situation where the molecules correspond to an unstable resonant
state with positive binding energy. In contrast,
stable molecules correspond to a nonzero negative value of $\sigma$ without infrared divergence in $\ZpR$. Furthermore,
for nonzero density a condensate occurs for $T=0$ and $\ZpR$ is regularized by the nonvanishing value of $\rex$ in eq.
(\ref{ZRFormula}).

\subsection{Renormalized Yukawa coupling}
We observe that for large $\hpn$ the wave function renormalization diverges $\ZpR \propto \hpn^2$. We will find that the
total fraction of dressed molecules $\Omega_B = \Omega_M + \Omega_C$ is a quantity of the order one in the crossover
region. We conclude that in the broad resonance limit the total number of bare molecules $\bar{\Omega}_B =\bar{\Omega}_M
+\bar{\Omega}_C = \Omega_B/\ZpR\propto \hpn^{-2}$ becomes negligibly small. As argued in sect. \ref{sec:Enhanced} the
microscopic molecules can be omitted for the thermodynamics and the broad resonance limit is well described by a pointlike
interaction for the fermionic atoms. Nevertheless, the renormalized bosonic field $\phi$ remains a very useful concept.
We will see that in the limit $\hpn\to \infty$ the appropriately renormalized couplings remain finite and become therefore
independent of the precise value of $\hpn$. This holds, in particular, for the renormalized Yukawa coupling
\footnote{The renormalization of the Yukawa coupling is done in two steps. A first renormalization step computes a
renormalized coupling $\hpn$, as extracted from binding energy and scattering length in vacuum, e.g. $\tilde{h}_{\phi,0}$
in sect. \ref{sec:concmag}. A second step (\ref{SecondRenorm}) takes the wave function renormalization $\ZpR$ into
account.}
\begin{eqnarray}\label{SecondRenorm}
h_\phi = \ZpR^{-1/2} \hpn
\end{eqnarray}
and the gap
\begin{eqnarray}\label{list2}
\rex = \hpn^2\tilde{\phi}^*\tilde{\phi} = (\hpn^2 /\ZpR)\phi^*\phi = h_\phi^2\rho.
\end{eqnarray}
The fact that these quantities are effectively $\hpn$ - independent has a simple but
important implication regarding the renormalization effects for $\hpn$ itself: Even if the actual value of $\hpn$ is changed
by, say, an order of magnitude, this will not affect these parameters. All quantities that can be expressed in terms of
the renormalized couplings become very insensitive to the value of $\hpn$, as seen in fig. \ref{hpUniversality}.

\subsection{Bosonic quasiparticles}
We have computed the gradient coefficient $\Apn$ in \cite{Diehl:2005ae}. In fig. \ref{hpUniversality} (b) we have
displayed the renormalized gradient coefficient $\ApR$ at the resonance, $c^{-1} =0$. It switches from 0.5 for small
$\hpn$ to a value close to one for large $\hpn$. On the other hand, deep in the BEC limit we find $\ApR = 0.5$ both for
small and large $\hpn$. Indeed, the BEC limit is well described by pointlike bosonic quasiparticles, even if the
microscopic molecules become irrelevant for $\hpn\to \infty$. The limit
\begin{eqnarray}
\lim\limits_{\hpn\to \infty} \lim\limits_{c\to 0} \ApR = \frac{1}{2}
\end{eqnarray}
states that the dressed molecules always have mass $2M$. Furthermore, in the BEC limit the number density of atoms
($\hat{\phi}_R= \ZpR^{1/2}\hat{\phi}$)
\begin{eqnarray}
n= 2 n_M + n_C = 2(\langle \hat{\phi}^*_R\hat{\phi}_R\rangle_c + \langle\hat{\phi}^*_R\rangle\langle\hat{\phi}_R\rangle)
\end{eqnarray}
turns out to be twice the number of dressed molecules (including the ones in the condensate) according to the standard
Bogoliubov formula. (For more details, we
refer to \cite{Diehl:2005ae}. A similar result has been obtained by Strinati \emph{et al.} in a purely fermionic setting
\cite{BBStrinati,YPieri}.) In the BEC limit all quantities that can be expressed in terms of the renormalized couplings
can equally well be described by a simple model involving only bosonic effective degrees of freedom. This holds both for
the case where microscopic molecules dominate and for the case of a purely pointlike interaction for fermionic atoms
without microscopic molecules. In this case an effective bosonic theory looses memory of its microscopic origin.

\subsection{Fraction of microscopic molecules}
Our calculations relate the ``macroscopic observables'' and the renormalized parameters to the microphysical interactions.
In view of the large universality for the experimentally tested broad resonances one may wonder if these calculations can
be checked experimentally or if only universal relations will remain at the end. A first possible check of the detailed
relation between the microphysics and macrophysics concerns the relation of the universal concentration parameter $c$ with
the detailed experimental setting, in particular its dependence on the magnetic field $B$. This will be discussed in sect.
\ref{sec:concmag}. A second access is provided by tests that directly observe ``microscopic quantities'' as, for example,
the fraction of microscopic molecules.

Indeed, even for a broad Feshbach resonance the microscopic molecules exist and one can measure their number.
The laser probe in the experiment by Partridge \emph{et al.} \cite{Partridge05} triggers resonant transitions from closed
channel microscopic molecules to open channel atoms. It therefore directly couples to the total
number of microscopic molecules $\langle\hat{\phi}^*\hat{\phi}\rangle$. The quantity $Z$ in \cite{Partridge05} can be
identified with the fraction of bare molecules
\begin{eqnarray}\label{EqOmegaB}
\bar{\Omega}_B = \bar{\Omega}_M + \bar{\Omega}_C = (\Omega_M + \Omega_C)/\ZpR.
\end{eqnarray}
We have computed this quantity as a function of $c^{-1}$ and relate $c^{-1}$ to the magnetic field in sect.
\ref{sec:concmag}. The result is shown in fig. \ref{BareExpPart}. The comparison with the experimental values is very
good. Since $\Omega_M + \Omega_C$ is of the order one the dominant effect actually checks our computations of $\ZpR$.
Obviously, $\bar{\Omega}_B$ depends explicitly on the Yukawa coupling, $\bar{\Omega}_B \propto \hpn^{-2}$, in
distinction to the universal quantities for a broad Feshbach resonance. For $T=0$ and $n=0$ the Yukawa coupling
$\tilde{h}_{\phi,0}$ can be extracted from properties of single atoms and molecules (cf. sect. \ref{sec:concmag}) and one
finds for $\lit$ $\tilde{h}_{\phi,0}^{(\mathrm{Li})}= 610$ ($k_F =1\mathrm{eV}$). However, even for $T=0$
(as appropriate for the experiment \cite{Partridge05}) the renormalization effects induce an effective dependence of $\hpn$
on $c^{-1}$ as shown in app. \ref{sec:YukRenorm}. Omitting these effects gives substantial disagreement with the measured values, as
shown by the dashed curve in fig. \ref{BareExpPart} for $B< B_0$. We consider fig. \ref{BareExpPart} as an important
quantitative check of our method \footnote{Another quantitative check concerns the agreement or our results for $T=0$
\cite{Diehl:2005ae} with Quantum Monte Carlo simulations \cite{Carlson03,Giorgini04}.} that we will describe in more
detail below.

\section{Effective potential}
\label{EvalBeyondMFT}

In our approximation all relevant information is contained in the various contributions to the effective potential
$\tilde{u}^{(cl)}, \tilde{u}_1\hF, \tilde{u}_1\hB$ and in $\ApR, \ZpR$. These quantities depend on $\Tn, \sigex, c^{-1}$ and
$\hpn$ where $\sigex$ can be fixed by the first equation (\ref{FESigmaII})
\begin{eqnarray}\label{FieldEqSigex}
- \frac{\partial \tilde{u}}{\partial \sigex} = \frac{1}{3\pi^2}.
\end{eqnarray}
With $\rex = h_\phi^2\rho$, $(8\pi c)^{-1} = -\hpn^{-2}(\tilde{\nu}- 2\sigex)$ and
\begin{eqnarray}
\tilde{u} &=& -\frac{\rex}{8\pi c} + \tilde{u}_1^{(F)}(\rex,\sigex)+ \tilde{u}_1^{(B)}(\rho)\label{FermPot}
\end{eqnarray}
we see that for fixed $c$ the explicit dependence of $\tilde{u}(\rex)$ on $\hpn$ enters only through the contribution
from the molecule fluctuations $\tilde{u}_1\hB$. The mean field theory (MFT) only includes the fermionic fluctuations and
omits $\tilde{u}_1\hB$. According to eq. (\ref{YYA}) standard MFT also omits the density of unbound dressed molecules, i.e.
$\Omega_M=0$. With
\begin{eqnarray}\label{EqOmegaC}
\Omega_{F,0} &=& - 3\pi^2\frac{\partial\tilde{u}_1\hF}{\partial \sigex} ,\nonumber\\
\bar{\Omega}_C &=& -3\pi^2\frac{\partial\tilde{u}^{(cl)}}{\partial \sigex} = \frac{6\pi^2 \rho_0}{\ZpR}
=\frac{6\pi^2\rex_0}{\hpn^2}
\end{eqnarray}
eq. (\ref{FieldEqSigex}) is equivalent to $\Omega_{F,0} + \bar{\Omega}_C = 1$. For small (or zero) $\bar{\Omega}_C$ the
dependence on $\hpn$ becomes negligible. The MFT equations simply reduce to
\begin{eqnarray}\label{NewFieldEqs}
- 3\pi^2\frac{\partial\tilde{u}_1\hF}{\partial \sigex} = 1-\bar{\Omega}_C, \quad \frac{\partial\tilde{u}_1\hF}{\partial \rex} =
\frac{1}{8\pi c},
\end{eqnarray}
with $\tilde{u}_1\hF$ given by eq. (\ref{USigmaPhi}).

A first step beyond MFT includes the contribution of $\tilde{u}_1\hB$ in eq. (\ref{FieldEqSigex}) and omits it in the
field equation $\partial \tilde{u} /\partial \phi$. This changes the first equation (\ref{NewFieldEqs}) to
\begin{eqnarray}
\Omega_{F,0} = -3\pi^2\frac{\partial\tilde{u}_1\hF}{\partial \sigex} = 1 - \Omega_M - \bar{\Omega}_C.
\end{eqnarray}
The presence of a substantial $\Omega_M$ on the BEC side reduces $\Omega_{F,0}$ and therefore changes the value of $\sigex$.
This effect is already sufficient to produce a crossover from the BCS to the BEC regime. We therefore need an estimate of
$\Omega_M$.

According to eq. (\ref{TotDens}) the bare molecule density is related to the exact propagator $\bar{G}_\phi$ of the
bosonic field $\hat{\phi}$
\begin{eqnarray}\label{ExactDens}
\nM = \frac{T}{2}\mathrm{tr}\sum\limits_n \int \frac{d^3q}{(2\pi^3)} \bar{G}_\phi(q,n) - \hat{n}_B.
\end{eqnarray}
with $\hat{n}_B$ an additive renormalization \cite{Diehl:2005ae}. Here we work in a basis with dimensionless renormalized real fields
$\phi_1,\phi_2$ with $\phi=(\phi_1+\mathrm{i}\phi_2)/ \sqrt{2}$ and $\rho = \phi^* \phi= (\phi_1^2+\phi_2^2)/2$.
Therefore $\bar{G}_\phi = \bar{\mathcal{P}}_\phi^{-1}$ is a $2\times 2$ matrix over which tr takes the trace. In terms of
renormalized dimensionless quantities, $\mathcal{P}_\phi = \bar{\mathcal{P}}_\phi/(\ZpR\epsilon_F)$, this yields for the
density of dressed molecules
\begin{eqnarray}\label{DressedDensDef}
\frac{n_M}{k_F^3} = \frac{\Tn}{2}\mathrm{tr}\sum\limits_n \int \frac{d^3\qn}{(2\pi^3)}\mathcal{P}_\phi^{-1} -
\frac{\hat{n}_B \ZpR}{k_F^3}.
\end{eqnarray}
This relation is exact. It involves the exact inverse propagator $\mathcal{P}_\phi$ which is directly related to the second
functional derivative of the effective action (\ref{EffActTrunc}) with respect to $\phi$,
\begin{eqnarray}
\big(\mathcal{P}_\phi\big)_{ab} (\tilde{x}, \tilde{x}') = \frac{\delta^2\Gamma}{\delta \phi^*_a(\tilde{x})
\delta\phi_b(\tilde{x}')}.
\end{eqnarray}
\begin{figure}
\begin{minipage}{\linewidth}
\begin{center}
\setlength{\unitlength}{1mm}
\begin{picture}(85,54)
     \put (0,0){
    \makebox(80,49){
    \begin{picture}(80,49)
      % \coordsyst{90}{54.9}{21}{14}
      \put(0,0){\epsfxsize80mm \epsffile{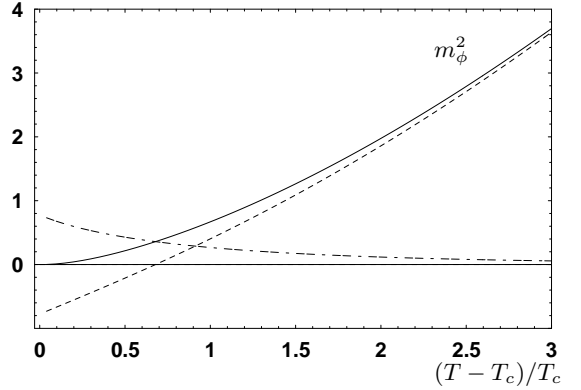}}%CorrLengthComp2.eps
      %earlier CorrLengthSmall2.eps
      \put(60,-2){$(T-T_c)/T_c$}
      %\put(6,19){\footnotesize{[10$^{12}$ cm$^{-3}$]}}
      \put(60,41){$m_\phi^2$ }
    %\put(-3,1){(a)}
    \end{picture}
      }}
      \end{picture}
\end{center}
\vspace*{-1.25ex} \caption{Temperature dependence of the bosonic mass term $m_\phi^2$ for $c^{-1} =0, \hpn\to \infty$. The
vanishing of the mass term $m_\phi^2$ is approached continuously (solid line). We also plot the separate contributions from
``mean field'' (classical plus fermions, dashed) and boson fluctuations (dashed-dotted). For $T>T_c$ the latter become
unimportant away from the phase transition. }
\label{MassesSYM}
\end{minipage}
\end{figure}

We will employ a Taylor expansion of the effective potential $\tilde{u}$ around its minimum
\begin{eqnarray}
\tilde{u} &=&\left\{
\begin{array}{c}
  { m_\phi^2\rho + \frac{1}{2}\lambda_\phi\rho^2 + ...\quad \qquad\quad \text{SYM}}  \\
  {  \frac{\lambda_\phi}{2}(\rho-\rho_0)^2 + ...\qquad \,\,\qquad\text{SSB}}
\end{array}\right.
\end{eqnarray}
where SYM and SSB denote the symmetric and superfluid (spontaneously symmetry-broken) phases. Without loss of generality
we also choose $\phi_0$ real such that $\phi_1$ corresponds to the ``radial mode'' and $\phi_2$
to the ``Goldstone mode''. In the approximation (\ref{EffActTrunc})  this yields \cite{Diehl:2005ae}
\begin{eqnarray}\label{FBosonProp}
\mathcal{P}_\phi  = \left(
\begin{array}{rr}
  {\ApR \qn^2 +m_\phi^2 + 2\lpR \rhoR_0,} & {-\tilde{\omega}_n} \\
  {\tilde{\omega}_n \qquad \qquad,} & {\ApR \qn^2 + m_\phi^2}
\end{array}\right)
\end{eqnarray}
and we note the appearance of the Matsubara frequencies in the off diagonal entries. Here we use a formulation that
can be used in both the normal (SYM) and the superfluid (SSB) phase. We will specialize to these cases in
the following subsections. The mass matrix
\begin{eqnarray}\label{Mass2}
\big(m^2_\phi\big)_{ab} =\frac{\partial^2\tilde{u}}{\partial\phi_a\partial\phi_b} =
\left(
\begin{array}{rr}
  {m_\phi^2 + 2\lpR \rhoR_0 ,}& {0} \\
  {0 \qquad ,} & {m_\phi^2}
\end{array}
\right)
\end{eqnarray}
involves both the fermionic and bosonic fluctuation corrections to $\tilde{u}$. At the phase transition $(m_\phi^2)_{ab}$
vanishes and we obtain
\begin{eqnarray}\label{SYMDens}
\nM=\frac{\Gamma (3/2)\zeta (3/2)}{4\pi^2}(4MT)^{3/2}.
\end{eqnarray}

The field equation for the renormalized field expectation value $\phi$ can be expressed as
\begin{eqnarray}\label{Rho0Cond}
m_\phi^2 \cdot \phi =0, \quad m_\phi^2 = \frac{\partial \tilde{u}}{\partial\rho}\Big|_{\rho_0}.
\end{eqnarray}
In the superfluid phase ($\phi_0 \neq 0, \rho_0\neq 0$) one infers $m_\phi^2=0$. Therefore the mass matrix (\ref{Mass2})
has a zero eigenvalue and $\phi_2$ can be identified with the massless Goldstone boson. The presence of a ``massless mode''
with infinite correlation length is the central ingredient for superfluidity. For the symmetric phase the two eigenvalues
of $(m_\phi^2)_{ab}$ are equal since $\rho_0=0$.

In principle, the inverse propagator (\ref{FBosonProp}) involves a momentum dependent vertex $\lambda_\phi(q)$. Including
the molecule fluctuations $\lambda_\phi (q)$ vanishes for $q=0$. As advocated in
\cite{Diehl:2005ae} we take this difficulty into
account by using in the density equation (\ref{DressedDensDef}) only the fermionic contribution $\lambda_\phi \to
\lambda_\phi\hF = \partial^2\tilde{u}_1\hF/\partial \rho^2|_{\rho_0}$. To reduce the error connected to this approximation
one may either compute the momentum dependence of the four-boson vertex or, perhaps more promising, use functional
renormalization group methods \cite{AAWetti,CWRG,Tetradis}. The approximation $\lambda_\phi = \lambda_\phi\hF$ is, of course, consistent with
our ``intermediate approximation'' where the field equation for $\phi$ neglects the bosonic fluctuation contributions.

A more consistent inclusion of the molecule fluctuations also takes them into account in the field equation for $\phi$,
modifying also the second equation (\ref{NewFieldEqs}). We need to evaluate the bosonic mass term
\begin{eqnarray}\label{MassDef}
m_\phi^2&=& \frac{\partial\tilde{u}}{\partial\rho} = (\nun - 2\sigex)/\ZpR +
\frac{\partial\tilde{u}_1\hF}{\partial\rho}  + \frac{\partial\tilde{u}_1\hB}{\partial\rho}\nonumber\\
&=& m_\phi^{(F)\, 2} + \frac{\partial\tilde{u}_1\hB}{\partial\rho}.
\end{eqnarray}
While $\tilde{u}_1\hF$ is given by eq. (\ref{USigmaPhi}) we still need to determine $\partial\tilde{u}_1\hB/\partial
\rho$. For this purpose we compute the lowest order ``gap'' or Schwinger-Dyson equation for the molecule propagator.
In the context of a perturbation expansion the gap equation amounts to a self-consistent resummation of a series of
diagrams involving bosonic fluctuations beyond the Gaussian (one loop) level. The explicit form of the gap equations for the
normal and superfluid phases as well as the phase boundary is displayed in app. \ref{app:appb}. Details of their
derivation and limitations of validity can be found in \cite{Diehl:2005ae}.

\begin{figure}
\begin{minipage}{\linewidth}
\begin{center}
\setlength{\unitlength}{1mm}
\begin{picture}(85,54)
     \put (0,0){
    \makebox(80,49){
    \begin{picture}(80,49)
      % \coordsyst{90}{54.9}{21}{14}
      \put(0,0){\epsfxsize80mm \epsffile{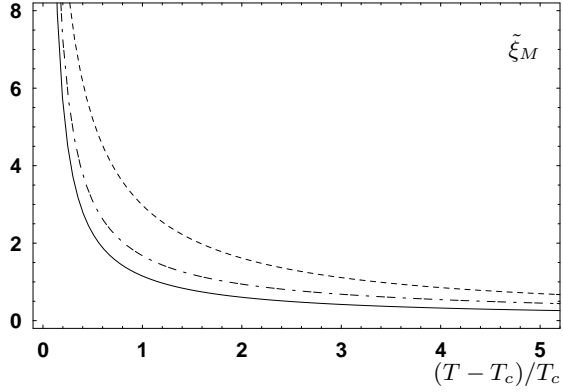}}%CorrLengthComp2.eps
      %earlier CorrLengthSmall2.eps
      \put(60,-2){$(T-T_c)/T_c$}
      %\put(6,19){\footnotesize{[10$^{12}$ cm$^{-3}$]}}
      \put(70,41){$\tilde{\xi}_M$ }
    %\put(-3,1){(a)}
    \end{picture}
      }}
      \end{picture}
\end{center}
\vspace*{-1.25ex} \caption{Molecular correlation length $\tilde{\xi}_M$ as a function of the reduced temperature
$(T-T_c)/T_c$. We consider three regimes: crossover ($c^{-1}=0$, solid), BCS ($c^{-1}=-1.5$, dashed), BEC ($c^{-1}=1.5$,
dashed-dotted) and the broad resonance limit $\hpn\to \infty$.}
\label{CorrLengthPlots}
\end{minipage}
\end{figure}

\section{Correlation length and scattering length for molecules}
\label{sec:corrlength}

In the symmetric phase the static effective molecule propagator reads (for zero Matsubara frequency)
\begin{eqnarray}
\bar{G}_\phi(q) = \bar{P}_\phi^{-1} = \frac{1}{\Apb q^2+ \bar{m}_\phi^2}.
\end{eqnarray}
We approximate here $\Apb$ as a constant. The Fourier transform of $\bar{G}_\phi$ decays for large distances $r$ exponentially
\begin{eqnarray}
\bar{G}_\phi(r) \sim \exp (-\xi_M r)
\end{eqnarray}
with correlation length
\begin{eqnarray}\label{TheCorrLength}
\xi_M =\bar{m}_\phi^{-1}\Apb^{1/2}.
\end{eqnarray}
(More precisely, $\xi_M^{-2}$ corresponds to the location of the pole of the propagator for negative $q^2$ and therefore
$\Apb$ should be evaluated at the location of the pole instead of $q^2 = 0$ as approximated here.) At the critical
temperature $T_c$ the correlation length $\xi$ diverges since $\bar{m}_\phi=0$.
\begin{figure}
\begin{minipage}{\linewidth}
\begin{center}
\setlength{\unitlength}{1mm}
\begin{picture}(85,54)
    \put (0,0){
    \makebox(80,49){
    \begin{picture}(80,49)
      % \coordsyst{90}{54.9}{21}{14}
      \put(0,0){\epsfxsize80mm \epsffile{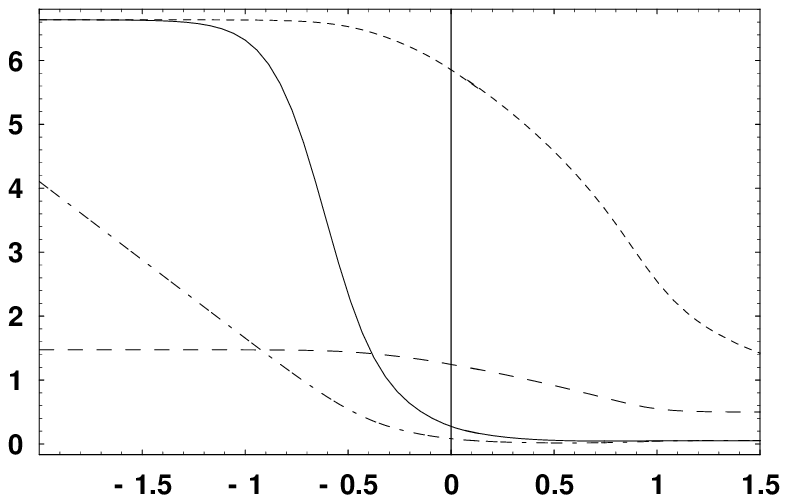}}
      \put(72,-2){$c^{-1}$}
      \put(-3,1){(a)}
      \put(33,35){$a_M k_F$ }
      \put(59,35){$a_M\hF k_F$ }
      \put(13,28){$m_\phi^2$ }
      \put(13,17){$\ApR$ }
      \put(10,10){BCS}
      \put(69,11){BEC}
      \end{picture}
      }}
\end{picture}
\end{center}
\vspace*{-1.25ex} \caption{
Effective potential in the crossover region, $\Tn =0.33, \hpn\to \infty$ . The solid line expresses the quartic coupling
$\lambda_\phi$ in terms of the renormalized molecule scattering length obtained with $a_M k_F = \lpR/(4\pi)$. The short
dashed line neglects the molecule fluctuations and shows $a_M\hF k_F = \lpR^{(F)}/(4\pi)$.  Large renormalization effects
are observed in the crossover regime. In the BEC limit, the bosons are only weakly interacting. In addition, we plot the
renormalized mass term for the molecule field $m_\phi^2$ (dashed-dotted) and the renormalized gradient coefficient $\ApR$
(long dashed).}
\label{BosonCoupling}
\end{minipage}
\end{figure}

In the superfluid phase the propagators for the radial and Goldstone modes differ. The correlation length for the Goldstone
fluctuations ($\phi_2$) is infinite according to eq. (\ref{Rho0Cond}). In the ``radial direction'' ($\phi_1$) the effective
mass term for the fluctuations around the minimum reads $2\lpb\rhob_0$.
It vanishes (at zero momentum) since our approximation based on Schwinger-Dyson equations \cite{Diehl:2005ae} yields $\lpb =0$ for
the deep infrared value of the molecule coupling. Hence we obtain in our approximation a diverging correlation length
for the radial mode. Nevertheless, the pole is actually not located at $q^2=0$ - using a momentum dependent
$\lpb(q^2)$ in $\mpb$ or taking this effect into account in the form of $\Apb(q^2)$ would be a better
approximation. In this paper we will not discuss further the effects that render the radial correlation length finite for
$T <T_c$.

The dimensionless molecular correlation length in the symmetric phase
\begin{eqnarray}
\tilde{\xi}_M  = \xi_M k_F = \ApR^{1/2}/m_\phi
\end{eqnarray}
is expressed in terms of the renormalized quantities $\ApR$ and $m_\phi^2$
\begin{eqnarray}\label{result145}
\tilde{\xi}_M^{-2}&=& \ApR^{-1}\Big\{\nu - 2\frac{\sigex}{\ZpR} +\Delta m_\phi^{(F)\,2}+ \frac{\lambda_\phi\hF}{3\pi^2}
\Omega_M \Big\}.
\end{eqnarray}
For the computation of $\ApR$ we refer to \cite{Diehl:2005ae}. We plot the renormalized boson mass, displayed in curly
brackets in eq. (\ref{result145}), in fig. \ref{MassesSYM}. Additionally, we indicate the separate contributions from
fermion (dashed) and boson fluctuations (dashed-dotted). Boson fluctuations are important close to the critical temperature.

The condition $\tilde{\xi}_M^{-2} = 0$ defines the critical temperature where the molecule correlation length diverges.
Fig. \ref{CorrLengthPlots} displays $\tilde{\xi}_M$ in the BEC, crossover and BCS regime. We observe
that the molecular correlation length remains larger than the average distance between atoms or molecules $\sim k_F^{-1}$
even rather far away from the critical temperature. This feature is quite independent of the BEC, BCS and crossover regime.
It indicates the importance of collective phenomena.

For $\tilde{\xi}_M\gg 1$ the correlation length exceeds by far the average distance between two atoms. We do not expect
the loop expansion to remain accurate in this limit. This is the typical range for the universal critical behavior
near a second order phase transition which is known to be poorly described by MFT or a loop expansion. A proper
renormalization group framework is needed for $\tilde{\xi}_M\gg 1$. In contrast, for $\tilde{\xi}_M\ll 1$ the molecules are
essentially uncorrelated.

Let us define an effective molecule scattering length $a_M$ for the dressed molecules in analogy to eq. (\ref{CoupScLength})
(with ``classical'' boson mass $2M$ for composite particles)
\begin{eqnarray}
\frac{\bar{\lambda}_\phi}{Z_\phi^2} = \frac{4\pi a_M}{2M}, \quad \bar{\lambda}_\phi = \frac{Z_\phi^2}{2Mk_F}\lambda_\phi
\end{eqnarray}
such that
\begin{eqnarray}
a_Mk_F = \frac{\lambda_\phi}{4\pi},\quad a_M\hF k_F = \frac{\lambda_\phi\hF}{4\pi}.\nonumber
\end{eqnarray}
A second possible criterion for the validity of the loop expansion is the smallness of the molecular self interaction
$\lambda_\phi$. Typically, a one loop expression becomes questionable for $a_M k_F \gg 1$.

\begin{figure}
\begin{minipage}{\linewidth}
\begin{center}
\setlength{\unitlength}{1mm}
\begin{picture}(85,54)
      \put (0,0){
     \makebox(80,49){
     \begin{picture}(80,49)
       %\coordsyst{90}{54.9}{21}{14}
      \put(0,0){\epsfxsize80mm \epsffile{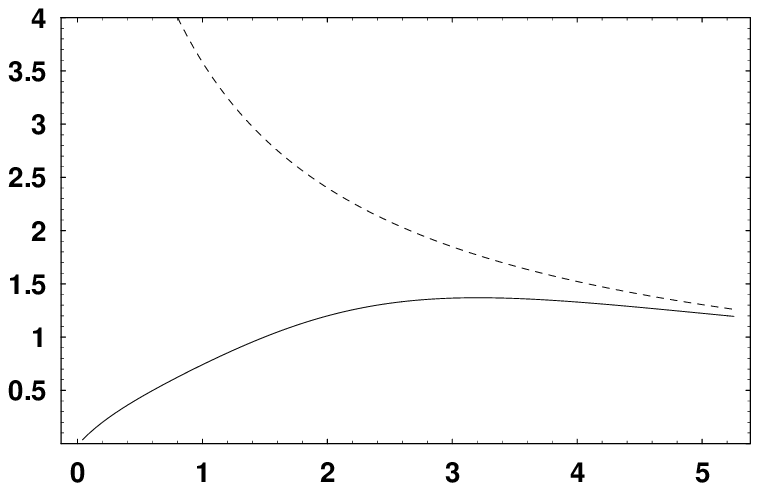}}%CorrLengthR: with second line for different fluct int
      \put(65,-2){$(T-T_c)/T_c$}
   %   \put(-3,1){(b)}
      \put(12,42){$a_M\hF k_F$ }
      \put(11,19){$a_M k_F$ }
      \end{picture}
      }}
\end{picture}
\end{center}
\vspace*{-1.25ex} \caption{Scattering length for molecules $a_M k_F$ as a function of the reduced temperature $(T-T_c)/T_c$
in the crossover regime ($c^{-1}=0, \hpn\to \infty, \Tn_c =0.255$). Boson fluctuations renormalize the bosonic scattering
length down to zero in the vicinity of the critical temperature. This may be compared to the increase of the
fermion fluctuation induced part $a_M\hF k_F$.}
\label{BosonCoupling2}
\end{minipage}
\end{figure}

We plot $a_M k_F$ and $a_M\hF k_F$ in fig. \ref{BosonCoupling} as a function of $c^{-1}$ and in fig. \ref{BosonCoupling2}
as a function of temperature. Both plots are for the broad resonance limit. While $a_M\hF k_F$ grows large near $T_c$ the
``full'' molecule scattering length $a_M k_F$ goes to zero.
The reader should be warned, however, that for $T\to T_c$ the momentum dependence of $\lambda_\phi(q)$ becomes crucial -
for $T=T_c$ one has $\lambda_\phi(q)\propto \sqrt{q^2}$ such that $\lambda_\phi$ vanishes only for $q=0$. At $T_c$ the
interaction between the dressed molecules cannot be approximated by a pointlike interaction.

Finally, we comment on the ratio of fermionic and molecular scattering length in the two-body limit. The diagrammatic
approach of
\cite{AAAAStrinati} yields $a_M/a \approx 0.75$. In \cite{Petrov04}, the ratio was extracted from an analysis of the
quantum mechanical four-body problem in the BEC limit, establishing the result $a_M/a \approx 0.6$. This was reproduced
in a Quantum Monte Carlo simulation in \cite{Giorgini04}.

We discuss a procedure to project on the low density limit in sect. \ref{sec:lowdens}. For the fermionic contribution to
the molecular scattering length, this yields in the BEC regime %(cf. eq. (\ref{LambdaPhi}))
\begin{eqnarray}
a_M\hF  %\frac{M\bar{\lambda}_\phi}{2\pi \ZpR^2} = \frac{M}{2\pi \ZpR^2} \frac{\partial^2 U}{\partial \bar{\rho}^2}(\bar{\rho}
 %=0, \bar{m}_\phi^2=0, \sigma=\sigma_A)
 = 2 a.
\end{eqnarray}
This is the Born approximation. In the low density limit, the one-loop expression for the bosonic fluctuations reduces
$a_M$ but is plagued by infrared divergences similar to the ones encountered above in the superfluid phase. A resolution
of this issue is expected from the consideration for the momentum dependence of the couplings or from the use of functional
renormalization group equations.

\begin{figure}
\begin{minipage}{\linewidth}
\begin{center}
\setlength{\unitlength}{1mm}
\begin{picture}(85,54)
      \put (0,0){
     \makebox(80,49){
     \begin{picture}(80,49)
       %\coordsyst{90}{54.9}{21}{14}
      \put(0,0){\epsfxsize85mm \epsffile{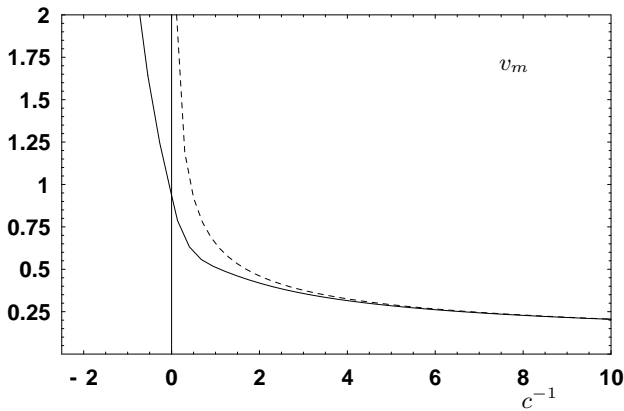}}
      \put(70,-2){$c^{-1}$}
      \put(67,43){$v_m$}
      \end{picture}
      }}
\end{picture}
\end{center}
\vspace*{-1.25ex} \caption{Dimensionless speed of sound $\tilde{v}_s$ for the molecules at $T=0$. In the
extreme BEC regime it approaches $\tilde{v}_s %=c^{-1}/2\sqrt{\rex}
\to 2\sqrt{c/(3\pi)}$. This limit (plotted dashed) is obtained by neglecting the
contribution from the boson density in the equation of state.}
\label{SpeedOfSound}
\end{minipage}
\end{figure}

As a last application of our formalism we compute the speed of sound as extracted from the low energy dispersion relation
$\omega = v_m |q|$ \cite{Diehl:2005ae}. In the dimensionless formulation $\tilde{v_s}=2M v_s/k_F$ it takes the form
\begin{eqnarray}
\tilde{v}_s = \sqrt{2\lambda_\phi\rho_0 \ApR}.
\end{eqnarray}
For the broad resonance limit the result is plotted at $T=0$ throughout the crossover in fig. \ref{SpeedOfSound}.
%A The contribution from the boson fluctuations is more
%complex. At nonzero temperature the condition $\bar{m}_\phi^2 =0$ obtained from the $k_F\to 0$ limit on the BEC side
%implies and infrared divergence of the corresponding integral, displayed in eq. (\ref{LambdaPhi}). This would imply
%$a_M =0$ as in the many-body context. On the other hand, drawing the $T\to 0$ limit first, we would get no
%contribution from from the boson terms such that $a_M =2a_R$. In sum, it is unclear how to draw the double limit
%$T\to 0$, $k_F\to 0$ for the bosonic contribution at present. A more sophisticated method for the resolution of this issue
%might be available by the use of functional renormalization group equations.
%VERY OLD
%Let us finally briefly comment on the infrared freedom, i.e. the fact that the ``many-body'' bosonic scattering length
%vanishes. At first sight this result is quite counterintuitive, since naively, it states that the dressed molecules do not
%interact. If it were so, this would immediately show up in an experiment, e.g. in the ballistic expansion of a trapped cloud.
%This is neither expected nor observed. However, as we have argued in \cite{Diehl:2005ae}, the infrared freedom does not mean that
%the gas is completely free. On smaller length scales or higher momenta, $\lambda_\phi(q)$ does not vanish, though a
%momentum independent approximation scheme yields $\lambda_\phi(q \to 0)=0$. In that sense, the ``many-body quantity''
%$a_M k_F$ is not the right observable to characterize the microscopic interaction between the dressed molecules.

\section{Limits of Enhanced Universality}
\label{sec:EnhancedUniv}

In this section we discuss in more detail the status of the different limits of enhanced universality presented in sect.
\ref{sec:Enhanced}.

\subsection{Exact Narrow Resonance Limit}
\label{sec:decoupling}

A nontrivial exact limit exists for which $\tilde{h}_\phi\to 0$ and $\tilde{\lambda}_\psi\to 0$ while $c$
and $\Tn$ are kept fixed. It applies to the symmetric phase including the location of the critical line. This exact limit
remains valid for arbitrary concentration $c$, even if the scattering length $a$ is arbitrarily large. As we have
discussed before, the molecules and fermionic atoms decouple in the limit $\hpn\to 0$ such that the Gaussian (one loop)
approximation becomes exact. Nevertheless, our limit can describe the full BCS-BEC crossover as visible from
fig. \ref{CrossoverTcAll}. In practice, the applicability of this limit corresponds
to a narrow resonance, much narrower than the ones currently investigated for lithium or potassium.

We learn from eq. (\ref{InMediumScattLength}) that our limit corresponds to $\bar{\nu} - 2\sigma \propto \hpb^2 \to 0$. Due
to the simultaneous vanishing of $\bar{\nu} -2\sigma$ and $\hpb$ our limit is not the trivial limit of a mixture of a
noninteracting atom gas plus a noninteracting molecule gas. Nevertheless, we will see that in the narrow resonance limit
an appropriate mean field theory becomes exact and can describe the high temperature phase including the approach to the
critical temperature of the phase transition. (This mean field theory differs from standard BCS mean field theory results,
cf. fig. \ref{CrossoverTcAll}.) The existence of this limit guarantees that mean field theory remains valid as long as
$\hpn$ remains small, say $\hpn <1$. This is confirmed in fig. \ref{hpUniversality}.

In order to establish the exact results for this limit we first perform in eq. (\ref{7}) the functional integral for
the fermions $\psi$. For $\bar{\lambda}_\psi = 0$ this can be done exactly. The Gaussian integral yields an intermediate
action $\bar{S}[\hat{\phi}]$ depending only on $\hat{\phi}$, with
\begin{eqnarray}\label{Intermediate}
Z = \int \mathcal{D} \hat{\phi} e^{-\bar{S}[\hat{\phi}]}.
\end{eqnarray}
Let us discuss the properties of $\bar{S}[\hat{\phi}]$ in the limit $\hpn\to 0$. The effective potential $\tilde{u}$ in $\bar{S}$
corresponds to the mean field potential and we note that $\tilde{u} = -\rex/(8\pi c(\sigex)) + \tilde{u}_1^{(F)}(\rex,
\sigex)$ depends on the gap parameter $\rex =\hpn^2\hat{\tilde{\phi}}^*\hat{\tilde{\phi}}$ but not explicitly on $\hpn$
(For the explicit mean field formulae cf. sect. \ref{DressedAndBareI}, eq. (\ref{USigmaPhi}), and app. \ref{app:appb}.) The
loop correction to the inverse boson propagator $\bar{\mathcal{P}}_\phi(Q)$ vanishes for $\hpn \to 0$ due to the overall
factor $\hpn^2$. In consequence, the inverse bosonic propagator for $\hat{\phi}$ is precisely given by the classical part
$\mathcal{P}_\phi^{(cl)} = 2\pi \mathrm{i} n T + q^2/4M + \bar{\nu} - 2 \sigma$.
This is the inverse propagator for free bosons with mass $2M$. All terms in $\bar{S}[\hat{\phi}]$ with more than two
powers of $\hat{\phi}$ also involve powers of $\hpn$ and therefore vanish in our limit. In consequence, the functional
integral for $\hat{\phi}$ becomes Gaussian in the limit $\hpn=0$ and we can solve the functional integral
(\ref{Intermediate}) exactly. In our limit it should be evaluated for $\sigma = \bar{\nu} /2$. In this limit the number
density of microscopic molecules $\nM$ therefore corresponds to the density of free nonrelativistic bosons of mass $2M$
with vanishing chemical potential (\ref{SYMDens}), as given by the canonical partition function.

In the narrow resonance limit the mean field approximation for the effective potential becomes exact. This yields exact estimates
for the critical temperature and all quantities of the symmetric phase for $T \geq T_c$. In contrast, the low temperature
phase has only a partially meaningful limit for $\hpn\to 0$. For nonzero $\rex$ the bare condensate fraction $\bar{\Omega}_C =
6\pi^2\rex/\hpn^2$ (cf. sect. \ref{EvalBeyondMFT}) diverges. The constraint $\OmC\leq 1$ then implies that $\rex$ has to
vanish $\propto\hpn^2$. In summary, the limit $\hpn\to 0$ corresponds to a universal phase diagram in the narrow resonance
limit, as given by the
lowest curve in fig. \ref{CrossoverTcAll}.%,\ref{CrossoverPhaseDiagramSigmaAll},\ref{CrossoverPhaseDiagramFractionsAll}
For small $\hpn$ this universal curve is smoothly approached (fig. \ref{hpUniversality}). All corrections to the
``narrow resonance universality'' are proportional to the dimensionless quantity $\hpn^2$. Obviously, for
$\tilde{h}_\phi \to 0$ all exact results depend only on the parameters $c$ and $\Tn$, whereby every point in the
$(c,\Tn)$-plane may be realized by different combinations of $a$ (or $\bar{\nu}$), $n$ (or $\sigma$) and $T$.

At this point a comment on the interpretation of the narrow resonance limit is in order. The choice of the classical values
for the coefficients in the boson propagator on the microscopic level corresponds to an intrinsically nonlocal situation.
Their actual values are found by simple physical considerations, i.e. $\Apb = 1/4M$ as the gradient coefficient for
bosons of mass $2M$ and chemical potential $-2\sigma$ for particle number two. However, for realistic narrow Feshbach
resonances additional physical features, as a nonvanishing $\bar{\lambda}_\psi$, may become important.

\subsection{Broad Resonance limit}
\label{sec:broadres}

The broad resonance limit obtains for $\hpn \to \infty$ while keeping $c$ fixed. We will see that it corresponds to a model
for fermionic atoms with local interaction and without explicit molecule degrees of freedom. For $\hpn\to \infty$ all
quantities depend only on $c$ and $\Tn$.
The broad resonance limit therefore shows a particularly high degree of universality. For fixed $c$ and finite $\sigex$
the limit $\hpn \to \infty$ is accompanied by $\nun \to \infty$ according to eq. (\ref{InMediumScattLength})
\begin{eqnarray}
c = - \frac{\hpn^2}{8\pi \nun}.
\end{eqnarray}
This relation is independent of $\sigex$. For finite $\sigex$ the in-medium scattering length $\bar{a}$ coincides with the
reduced scattering length $a_R$ (\ref{VacuumScattLength}). Going back to the original definition of the action for the
functional integral (\ref{7}) the broad resonance limit corresponds to the double limit $|\bar{\nu}| \to \infty$,
$\hpb \to \infty$. In this limit, the ``classical'' terms $\vec{\nabla}\hat{\phi}^*\vec{\nabla} \hat{\phi}$ and
$\hat{\phi}^*\partial_\tau\hat{\phi}$ become subdominant and can be neglected. In the fermionic language (\ref{Mom4Fermion})
our model therefore reduces to a purely
local four-fermion interaction. Our partially bosonized description has to match the purely fermionic description
with a pointlike interaction term. In other words, all physical results can only depend on the effective coupling
(cf. \ref{BosonCond2})
\begin{eqnarray}
\bar{\lambda} = \bar{\lambda}_\psi -\frac{\hpb^2}{\bar{\nu}_\Lambda}.
\end{eqnarray}
Any individual dependence on $\hpb$, $\bar{\nu}_\Lambda$ or $\bar{\lambda}_\psi$ (at fixed $\bar{\lambda}$) can only be an
artefact of an insufficient approximation and can therefore be used to test the validity of various approximation or
truncation schemes.

As the scale set by $\hpb$ drops out in this limit, we can express the crossover in terms of a single parameter, the inverse
dimensionless scattering length $c^{-1}$. In the superfluid phase we can use for the order parameter the squared
dimensionless fermionic mass gap $\rex = \hpn^2\tilde{\phi}^*\tilde{\phi}$. These are precisely the generic parameters
used in a purely fermionic description. They characterize the system uniquely in the
strict limit $\hpn\to \infty$, but also remain very efficient for $\hpn\gg 1$. Corrections will be $\mathcal{O}(\hpn^{-2})$
or less.

The systems currently investigated experimentally are $\lit$ and $\kal$. Both range in the broad resonance regime, as can
be seen from fig. \ref{hpUniversality} and discussed in sect. \ref{sec:concmag}. This clearly motivates the theoretical
investigation of this regime.

In contrast to the narrow resonance limit ($\hpn\to 0$, $\bar{\lambda}_\psi\to 0$) the broad resonance limit cannot be
solved exactly. It still corresponds to an interacting fermion model, as discussed in detail by Strinati \emph{et al.}
\cite{AAAAStrinati,BBStrinati,CCStrinati,YPieri,ZStrinati}. Nevertheless, it allows us to make a detailed matching with
a purely fermionic description and to compare all physical results directly by computing the concentration $c$ in both
approaches. Furthermore, the broad resonance limit can be used as a starting point for a systematic investigation of
corrections beyond a local four-fermion interaction.

Our present calculations in the broad resonance limit still involve quantitative uncertainties related to our approximation
scheme. This differs from the exactly solvable narrow resonance limit. On the other hand, due to the ``loss of memory''
concerning the details of the microscopic interaction vertex (\ref{Mom4Fermion}), this limit has a higher degree of
robustness as compared to the narrow resonance limit, where the details of the microscopic interactions must be known.

Finally, we would like to point out that a further ``crossover problem'', i.e. the crossover from small to large $\hpn$,
emerges from the above discussion. We leave this exciting field, physically describing the crossover from nonlocal
to pointlike interactions, for future work. Our interpretation of the narrow and broad resonance limits is confirmed
in the numerical study \cite{Jensen04} comparing different types of interaction potentials.

\subsection{BCS and BEC regimes}
\label{sec:UnivBCSBEC}

The BCS and BEC regimes correspond to the limits $c\to 0_-$ and $c\to 0_+$. In the limit of small $|c|$, the marginal
parameter $\hpn$ turns irrelevant. Formally, this can be seen form the microscopic
four-fermion vertex (\ref{Mom4Fermion}): In these limits, the absolute value of $|\bar{\nu}-2\sigma|$ becomes large compared
to the momentum dependent terms. Indeed, the temperature and momentum dependent terms in the denominator of eq.
(\ref{Mom4Fermion}) are typically of the order $\epsilon_F = k_F^2/(2M)$ whereas $|\bar{\nu}-2\sigma|\sim \epsilon_F \hpn^2/|c|$.
For $|c| \to 0$ the interaction becomes therefore effectively pointlike for any given nonzero value of $\hpn^2$. In
consequence, the results can only depend on $\bar{\lambda}$ or $c$, but not on $\hpn$ separately.

Though an intuitively expected result, this kind of universality is interesting from a conceptual point of view. It is
associated to a loss of memory concerning field degrees of freedom: On the BCS side, the molecules could have been omitted,
and the BCS picture of a weakly interacting Fermi gas exhibiting the formation of Cooper pairs becomes valid. On the BEC
side, we end up with an effective theory for the ``dressed'' molecules. They behave exactly like weakly interacting
pointlike fundamental bosons (more details will be presented in the next section), whereas atom degrees of freedom are
completely negligible for the macroscopic features.

The reason for this behavior can be seen directly from the functional
integral (\ref{7}) and the microscopic action (\ref{YukawaAction}): Introducing the variables $c^{-1}$ and $\hat{\varphi}
= \hpb\hat{\phi}$ ($\langle\hat{\varphi}^*\rangle\langle\hat{\varphi}\rangle = r$), which is an exhaustive set to describe
these limits, we observe the following: For $c^{-1} \to -\infty$ (BCS limit), the classical contribution to
$\partial\tilde{u}/\partial\rex$ (\ref{FermPot}) acts as a large positive mass term $\propto -c^{-1}$ for the bosons, and
the correlation functions associated to $\hat{\varphi}$ are heavily suppressed. On the other hand, for $c^{-1} \to
+\infty$ (BEC limit), the
solution of the crossover problem implies for the effective chemical potential $\sigex \to -\infty$, and $-\sigma$
constitutes a mass term for the fermions, similarly suppressing their propagation. Though the interaction between fermions
and bosons remains strong through $\hat{\varphi}\psi^T\epsilon\psi^\dagger$, the overall role of one of the two different
degrees of freedom becomes unimportant in the two respective limits.

It is instructive to discuss the BEC limit from a somewhat different formal perspective. Let us consider the model defined
by $S_B$ (\ref{DimlessYukawaAction}) and take the limit $\sigex \to -\infty$. The negative effective chemical potential acts
as an infrared cutoff \footnote{The infrared cutoff $k$ should not be confused with the Fermi momentum $k_F$.},
$-\sigex =k^2$, both for the fermion and boson propagator. We may now integrate out the fermions. In the limit
$k\to \infty$ the contribution from the fermion fluctuations to the effective action $\bar{S}[\hat{\phi}]$ (similar to
eq. (\ref{Intermediate}) discussed for the narrow resonance limit) vanishes. Thus $\bar{S}$ reduces to the action for
free bosonic molecules, $\bar{S} = \int_{\tilde{x}} \hat{\tilde{\phi}}^*(\tilde{\partial}_\tau - \tilde{\triangle}/2 +
\tilde{\nu}_\Lambda - 2\sigex)\hat{\tilde{\phi}}$. This free theory is easily solved and exhibits the standard BEC for
$T<T_c$.

More in detail, we observe for $\tilde{u}_1\hF$ (\ref{USigmaPhi}) that both $\gamma_\phi$ and $\gamma$ diverge for $k\to
\infty$, resulting in
\begin{eqnarray}
&&\tilde{u}_1^{(F)}(\sigex,\rex)= - 2\Tn\int\frac{d^3\qn}{(2\pi)^3}\big[\gamma_\phi - \gamma -
 \frac{\rex }{4\Tn \qn^2}\big]\\\nonumber
 &=& - \int\frac{d^3\qn}{(2\pi)^3}\big\{\frac{\rex}{2} \big(\frac{1}{\qn^2 + k^2} - \frac{1}{q^2}\big)
       + \mathcal{O}\big(\rex^2/(\qn^2 + k^2)^3\big)\big\}\\\nonumber
 &=& \frac{\rex k}{8\pi}.
\end{eqnarray}
For large $k$ only the remnant of the additive renormalization of $\bar{\nu}$ remains \footnote{In
the formal limit $k\to \infty$ at fixed momentum cutoff $\Lambda$ no renormalization of $\bar{\nu}$ occurs.}. Similarly,
for $\ZpR$ and $\tilde{A}_\phi$ one obtains
\begin{eqnarray}
\ZpR = 1 + \frac{\hpn^2}{32\pi k}, \quad \tilde{A}_\phi = \frac{1}{2} + \frac{\hpn^2}{64\pi k}.
\end{eqnarray}
Both $\ZpR$ and $\Apb$ take their classical values for $k\to \infty$.

In consequence, the limit $-\sigex \to \infty$ is exactly solved. Physically, this describes, however, only the very extreme
BEC limit. Still, for large but finite $-\sigex$ we recover an effective bosonic theory that can be solved by by perturbative
methods.

\subsection{Scaling limit}

A further interesting form of universality was pointed out by Ho \cite{Ho104} in the limit $c^{-1}\to 0$ \footnote{A similar
argument was proposed even earlier in the context of bosonic systems in \cite{Bulgac02}.}. It is argued that if the
scattering length drops out, then the density (or $k_F$) remains the only relevant scale (besides $T$),
and the thermodynamics of the system should be governed by simple scaling laws.

In this form the argument is not complete since it assumes in addition a strictly pointlike interaction. Deviations from
the pointlike structure, as reflected in the Feshbach coupling $\hpb$, introduce new scales and therefore new
dimensionless parameters as $\hpn$. This is clearly seen by the $\hpn$ dependence of the quantities $\Tn_c$ and $\ApR$
(effective gradient coefficient for the dressed bosons) in fig. \ref{hpUniversality}, which indeed refers to the resonance
$c^{-1}=0$. For example, a given $\hpb$ corresponds to a particular density $k_F^{(cr)}$ for the crossover from a broad to
a narrow resonance. (We may define $k_F^{(cr)}$ by the condition $\hpn(\hpb, k_F^{(cr)})  =10$.) In this sense the scaling
limit does not exhibit complete universality since it depends on an additional parameter.

Nevertheless, for a broad resonance the actual value of $\hpn$ is irrelevant for the dressed quantities
(as long as we are in a range of $k_F$ sufficiently away from $k_F^{(cr)}$).
Thus Ho's argument becomes valid in the double limit \footnote{In our approach this also holds for the narrow resonance
limit $c^{-1}\to 0, \hpn\to 0,\bar{\lambda}_\psi \to 0$. However, in practice $\bar{\lambda}_\psi$ or similar terms will not
vanish and induce scaling violations.} $c^{-1}\to 0, \hpn\to \infty$: For dressed quantities all dimensionless numbers and
ratios can be predicted completely independently of the microphysical details of the system!

Our results for a broad Feshbach resonance extend this argument also away from the location of the resonance. Universality
holds not only for one particular value of $B$ at the Feshbach resonance. For arbitrary $B$ in the whole crossover
region a single parameter $c^{-1}$ describes all relevant macroscopic properties of the dressed quantities. The concrete
microphysics of a system is only needed to relate $c$ to $B$, i.e. it is only reflected in one function $c(B)$ which varies
from one system to another.

\section{Low density limits}
\label{sec:lowdens}

So far we have developed a rather complete picture of the phase diagram for ultracold fermionic atoms in terms of the
parameters $c$ and $\hpn$. In order to make contact with experiment we should relate these parameters to observable
quantities. In particular, we will see how the concentration $c$ is related to the magnetic field. One of the great
goals for ultracold atom gases is the realization of systems with well controlled microscopic parameters. This
is a central advantage as compared to solids or liquids for which usually the precise microscopic physics is only
poorly known. In order to achieve this goal the ``microphysical parameters'' should be fixed by the properties of the
individual atoms and molecules, e.g. by scattering in the vacuum or by the determination of binding energies. The
microscopic parameters are then a matter of atomic or molecular physics and do not involve collective effects of
many atoms or molecules.

Our functional integral approach relates the microscopic parameters to ``macroscopic observables''. It can be used for
arbitrary values of the density and temperature. In particular, it can be employed for the computation of properties
of excitations in the vacuum at zero density and temperature. Within one formalism we can therefore not only compute the
properties of many-body systems, but also the scattering behavior of individual atoms or molecules in the vacuum. This
enables us to relate our microscopic parameters (appearing in the action) to observables in atomic or molecular physics.
What is needed for this purpose is a computation of binding energies and scattering cross sections in the limit $T\to 0$,
$n\to 0$. This will be done in the present and next section.

We are interested in the limit where the density goes to zero while $\Tn =T/\epsilon_F$ is kept fixed at some value
$\Tn>\Tn_c$. Of course, then also $T$ goes to zero, but we do not have to bother with condensation phenomena or critical
phenomena (for $\Tn$ sufficiently larger than $\Tn_c$). In the limit $n\to 0, k_F\to 0$ also the concentration $c$ vanishes.
Actually, we have to consider two limits $|c|\to 0$ separately,
one for positive and the other for negative $c$. For positive $c$ the low density limit corresponds to a gas of molecules,
while for negative $c$ one obtains a gas of non-interacting fermionic atoms. If the scattering length is much smaller than
the average distance between two atoms or molecules, the properties of the dilute gas are directly related to the physics
of individual atoms or molecules.

\subsection{Atom and molecule phase}
The ``fermion gas limit'' or ``atom phase'' (negative $c$) is realized for positive $\bar{\nu}$ or negative scattering
length $a_R$. In this case
the molecular binding energy is positive such that the excitation of ``molecule resonances'' requires energy. Since for
fixed $\Tn$ and $k_F\to 0$ the temperature becomes very low, $T=\Tn\epsilon_F$, the excited states become strongly
suppressed and can be neglected. In this limit one finds $\sigex\to 1$ and therefore $\sigma \to \epsilon_F\to 0$ and
$\bar{a}\to a_R$. At fixed $\bar{\nu}$ and $\hpb$ the concentration vanishes $c=a_R k_F\to 0$.
Therefore the interaction effects are small. In particular, $\Tn_c$ vanishes in the limit $c\to 0$, cf. fig.
\ref{CrossoverTcAll}. Finally, the wave function renormalization $\ZpR$ diverges $\propto\hpb^2/\sqrt{\vec{q}^2}$ for a
vanishing momentum $\vec{q}^2\to 0$ in the propagator.

For a negative molecular binding energy or $\bar{\nu}<0$ the low density and low temperature state becomes a molecule gas.
Now the number density of the fermionic atoms is suppressed since for zero momentum their energy is higher as compared to
the molecules. We will show that this ``molecule phase'' is characterized by a nonzero negative value of $\sigma$.

Let us consider the quadratic term for $\bar{\phi}$ in the effective potential or the term linear in
$\rhob = \bar{\phi}^* \bar{\phi}$,
\begin{eqnarray}
\bar{m}_\phi^2 = \frac{\partial U}{\partial \rhob} (0) = \bar{\nu} - 2\sigma + \frac{\partial U_1^{(F)}}{\partial \rhob} (0)
+\frac{\partial U_1\hB}{\partial \rhob} (0).
\end{eqnarray}
According to eq. (\ref{BosNumber}) or \footnote{Eq. (\ref{BosNumber2}) becomes exact if the frequency dependence of the
correction to $\Apb$ can be neglected (cf. the appendix in \cite{Diehl:2005ae}) and $\bar{m}_\phi^2 + \Apb (q^2)q^2$ is used to
parameterize the exact inverse propagator at zero frequency.}
\begin{eqnarray}\label{BosNumber2}
\nRM&=& \int \frac{d^3q}{(2\pi)^3}\Big[\exp\Big(\frac{ \Apb(q)q^2 + \bar{m}_\phi^2}{\ZpR T}\Big) - 1\Big]^{-1}
\end{eqnarray}
the ``mass term'' $\bar{m}_\phi^2$ dominates the behavior for $T\to 0$ unless it vanishes. Our limit corresponds to
$\nRM\propto k_F^3\to 0$, $T\propto \Tn k_F^2\to 0$ or $\nRM\propto T^{3/2} \to 0$. This requires that $\bar{m}_\phi^2$
must vanish $\propto k_F^2$. In consequence, the chemical potential $\sigma$ reaches a nonzero negative value
\begin{eqnarray}
\lim\limits_{k_F\to 0}\sigma = \sigma_A = \frac{1}{2}\Big(\bar{\nu} + \frac{\partial U_1^{(F)}}{\partial \rhob} (0)
+\frac{\partial U_1\hB}{\partial \rhob} (0)\Big).
\end{eqnarray}
Here $\frac{\partial U_1^{(F)}}{\partial \rhob} (\rhob = 0)$ and $\frac{\partial U_1\hB}{\partial \rhob} (\rhob = 0)$ have to
be evaluated for $\sigma=\sigma_A$. We arrive at the important conclusion that for $\bar{\nu}<0$ the vacuum or zero density
limit does not correspond to a vanishing effective chemical potential but rather to negative $\sigma_A<0$.

The effective scattering length (\ref{InMediumScattLength}) becomes in the low density limit
\begin{eqnarray}\label{ScatterDens0}
\frac{1}{\bar{a}} &=& \frac{1}{a_R} + \frac{8\pi\sigma_A}{\hpb^2M} = -\frac{4\pi}{\hpb^2M} \big(\bar{\nu} -2 \sigma_A\big) \nonumber\\
&=& \frac{4\pi}{\hpb^2M} \frac{\partial U_1}{\partial \rhob}\big(\rhob=0).
\end{eqnarray}
We will see below that $(\partial U_1/\partial \rhob) (\rhob=0)$ does not vanish for $\sigma_A< 0$, $n\to 0$, $T\to 0$.
Therefore away from the resonance $\bar{a}(\sigma_A)$ remains finite for $k_F\to 0$ and the concentration parameter $c$
vanishes in the low
density limit. This is consistent with a noninteracting gas of molecules. It is not surprising that one encounters the
Bose-Einstein condensation of free bosons as the temperature is lowered below $T_c$.

In the molecule phase the nonzero $\sigma_A<0$ adapts the additive constant in the energy such that the molecular
energy level is a zero. In consequence, $-\sigma_A$ appears as a positive energy (gap or mass term) in the propagator
of the fermionic atoms. As it should be this suppresses the relative number of fermionic atoms in the limit $T\to 0$
and we end with a dilute gas of molecules. The two low density limits (with $\bar{\nu} \neq 0$) therefore both correspond
to the limit of vanishing concentration $|c|\to 0$. Positive and negative $c$ correspond to pure gases of molecules or
fermionic atoms, respectively.

\subsection{Binding energy}
In the molecule phase $\sigma_A$ has a simple interpretation in terms of the molecular binding energy in vacuum,
$\epsilon_M$, namely
\begin{eqnarray}
\epsilon_M = 2\sigma_A.
\end{eqnarray}
This can be seen by the computation of the scattering amplitudes in app. \ref{sec:scattvac}. Alternatively, we may find
this relation by comparing the densities of molecules and fermionic atoms in the low density limit. For small enough
density and temperature the interactions become unimportant and the ratio $\nRF/\nRM$ should only depend on the binding
energy in vacuum. Indeed, in our limit $\sigex$ is negative and diverges
\begin{eqnarray}
\lim\limits_{k_F\to 0} \sigex \to \frac{2M\sigma_A}{k_F^2}.
\end{eqnarray}

This simplifies the momentum integrals in the loops considerably. The wave function renormalization in the ``molecule
phase'' diverges for $\epsilon_M\to 0$
\begin{eqnarray}\label{ZpRlowdens}
\ZpR = 1 + \frac{M^{3/2}\hpb^2}{8\pi \sqrt{|\epsilon_M|}}.
\end{eqnarray}
Furthermore, for large negative $\sigex$ one finds $A_\phi= \Apn/\ZpR= 1/2$ \cite{Diehl:2005ae}. One therefore has
$\Apb/\ZpR =1/4M$ and $\nRM$ corresponds to a gas of dressed molecules with mass $2M$. On the
other hand, the small fraction of fermionic atoms obeys eq. (\ref{OmegF}) with $\sigma = \sigma_A$ resulting in
\begin{eqnarray}\label{OmegaXXA}
n_{F,0} \hspace{-0.1cm}= \Omega_{F,0} n = 2\hspace{-0.15cm}\int\hspace{-0.15cm} \frac{d^3q}{(2\pi)^3}\big[\exp\big\{\frac{1}{T}\big(\frac{q^2}{2M} - \sigma_A\big)\big\}
 + 1\big]^{-1}.\nonumber\\
\end{eqnarray}
In the low density limit the interactions become negligible and $n_{F,0}$ should reduce to the density of the free gas of
atoms with mass $M$. However,
even for zero momentum the energy of a single atom does not vanish - it is rather given by half the binding energy that is
necessary for the dissociation of a molecule. This corresponds precisely to eq. (\ref{OmegaXXA}), provided we identify
$\epsilon_M = 2\sigma_A$. We conclude that the binding energy is not simply given by $\bar{\nu}$ (which is defined for
$\sigma = 0$ and in the absence of molecule fluctuations) but rather obeys
\begin{eqnarray}\label{MicroPhysics}
\epsilon_M &=& \bar{\nu}  + \frac{\partial U_1}{\partial \rhob}\big(\rhob=0, \sigma = \sigma_A,T=0\big).
          % &=& -\Big(\bar{\nu}  + \frac{\hpb^2(2M)^{3/2}}{8\pi}\sqrt{\sigma_A}\Big)=-2\sigma_A.
\end{eqnarray}
A measurement of the binding energy of molecules in vacuum, $\epsilon_M$, can relate $\bar{\nu}$ to observation.

For this purpose we investigate the behavior of $\bar{m}_\phi^2 = \partial U/\partial\bar{\rho}(\bar{\rho}=0)$ as
$T\to 0$. First we note that the contribution
from the molecule fluctuations vanishes, since for $\bar{m}_\phi^2\geq 0$ one finds $(\partial U_1^{(B)}/\partial \rhob)
(\rhob=0)\propto \nRM\propto T^{3/2}$. For the contribution from the fermionic fluctuations one infers from eq.
(\ref{FermPot}) (now in dimensionful units)
\begin{eqnarray}
%\frac{\partial U_1\hF}{\partial\rhob} &=& -\hpb^2\int\frac{d^3q}{(2\pi)^3}\Big\{ \frac{\tanh \gamma_\phi}{4T\gamma_\phi} -\frac{M}{q^2}\Big\},\nonumber\\
\frac{\partial U_1\hF}{\partial\rhob}\Big|_{\rhob=0}\hspace{-0.25cm} &=& -\hpb^2M\hspace{-0.1cm}\int\hspace{-0.1cm}
\frac{d^3q}{(2\pi)^3}\Big\{\frac{1}{q^2\hspace{-0.1cm}-2M\sigma }
\tanh \frac{q^2-2M\sigma}{4MT}\nonumber\\
&&\qquad-\frac{1}{q^2}\Big\}
\end{eqnarray}
and therefore \footnote{For $\sigma >0$ the pole on the real positive axis leads to an additional factor i in front of the
r.h.s. of eq. (\ref{Eq100}). This has, however, no physical impact since $\sigma =0$ for the physical solution on the high
field side of the resonance $B>B_0$.} for $T\to 0$, $\sigma \leq 0$
\begin{eqnarray}\label{Eq100}
\frac{\partial U_1\hF}{\partial\rhob}(\rhob=0,\sigma)= \frac{\hpb^2}{8\pi} (2M)^{3/2} (-\sigma)^{1/2}.
\end{eqnarray}

For $\sigma <0$ the non-analytic behavior $\propto \sqrt{-\sigma}$ has an interesting consequence for the
threshold behavior near $\epsilon_M =0$. The equation determining $\sigma_A$ in the molecule phase (i.e. $\bar{m}_\phi^2
(\sigma_A)=0$) reads
\begin{eqnarray}\label{SigmaAEq}
\bar{\nu} + \frac{\hpb^2}{8\pi}(2M)^{3/2}(-\sigma_A)^{1/2} -2\sigma_A=0.
\end{eqnarray}
Independently of the value of $\sigma_A$ and $\hpb$ eq. (\ref{SigmaAEq}) can be transformed into
\begin{eqnarray}\label{EpsPropNu2}
\epsilon_M = 2\sigma_A = -\frac{16\pi^2(\bar{\nu} - 2\sigma_A)^2}{\hpb^4M^3} = -\frac{1}{M \bar{a}^2(\sigma_A)}.
\end{eqnarray}
We recover the well known universal result for the scattering length $\bar{a}(\sigma_A)$ \cite{WWStoofBos,XXStoofBos}
provided $\bar{a}$ can be identified with the physical scattering length $a$. This result is independent of the
value of the Feshbach coupling. One concludes $\epm(\bar{\nu}\to 0)\to 0$. Close to threshold ($\sigma_A \to 0$), the last
term in eq. (\ref{SigmaAEq}) becomes subdominant, implying a quadratic behavior $\epsilon_M \propto \bar{\nu}^2$. In contrast,
far away from threshold the binding energy equals the mean field binding energy, $\epsilon_M = \bar{\nu}$. For broad
resonances, the linear regime will only be approached far away from the resonance when $\sqrt{-\sigma_A} \gg
(2M)^{3/2}\hpb^2/(16\pi)$.

In the ``atom phase'' we define (cf. app. \ref{app:dispersion})
\begin{eqnarray}\label{EpsNuR}
\epsilon_M = \bar{m}_\phi^2(\sigma=0)/\ZpR = \bar{\nu}/\ZpR.
\end{eqnarray}
In our approximation one finds in the atom phase $\ZpR =1$. Since $\epm$ vanishes for $\bar{\nu}=0$ we find for the
transition from the molecule to the atom phase that $\epsilon_M$ is a continuous function of $\bar{\nu}$ and therefore of
$B$. The derivative $\partial \epsilon_M/\partial \bar{\nu}$ is discontinuous at the transition point while $\partial
\sigma/\partial\bar{\nu}$ remains continuous.

For both phases, we may write
\begin{eqnarray}
\epsilon_M = \frac{\bar{m}_\phi^2}{\ZpR} +2\sigma.
\end{eqnarray}
For negative $\epsilon_M$ the ground state value $\sigma
=\sigma_A$ is determined by $\bar{m}_\phi^2(\sigma_A) = 0$. In contrast, the other limit of a gas of fermionic atoms
corresponds to $\sigma = 0$, $\epsilon_M =  \bar{m}_\phi^2(\sigma =0)/\ZpR\geq 0$.

The value of $\epsilon_M$ could, in principle, differ between the two limits $n\to 0$ with negative or positive
$\epsilon_M$, due to the different ground state values of $\sigma$. It is interesting to address in more detail
the question of continuity as the magnetic field switches between the two
situations. In fact, we may view this qualitative change as a ``phase transition'' in vacuum as a function of $B$. The
corresponding ``order parameter'' is the value of $\sigma$ at $T=0$. For $B> B_0$ the ``atom phase'' is characterized by
$\sigma = 0$ and the mass term $\bar{m}_\phi^2 = \partial U /\partial\rhob (\rhob =0,\sigma=0)$ is positive, corresponding to
$\epsilon_M >0$. The ``molecule phase'' for $B<B_0$ shows a nonvanishing order parameter $\sigma = \sigma_A$. Now the mass
term vanishes, $\bar{m}_\phi^2=0$, whereas the atoms experience a type of gap $|\sigma_A|$. The binding energy is negative,
$\epsilon_M <0$. We find a continuous (``second order'') transition where $-\sigma_A$ approaches zero as $B$ approaches
$B_0$. In this case one has $\epsilon_M(B_0)=0$ and $\bar{m}_\phi^2(B_0) =0$ for $B$ approaching $B_0$ either from above
or below. (As a logical alternative, a discontinuous first order transition would correspond to a discontinuous jump of
$\sigma$ at $B_0$. Then also $\epsilon_M$ may jump and not be equal to zero at $B_0$.) As common for second order phase
transitions we find a non-analytic behavior at the transition point $B=B_0$.

\section{Parameters as a function of magnetic field}
\label{sec:concmag}

So far we have described the universal crossover physics in terms of two parameters $c$ and $\hpn$. For comparison
with experiment these parameters have to be related to to the microphysical properties of a given atomic system and to the
external magnetic field $B$. More precisely, the three parameters in the microscopic action (\ref{YukawaAction}),
$\bar{\nu}_\Lambda, \bar{h}_{\phi,\Lambda}$ and $\bar{\lambda}_{\psi,\Lambda}$ have to be known as functions of $B$ and
the effective UV cutoff $\Lambda$. The dependence on $\Lambda$ can be eliminated by trading $\bar{\nu}_\Lambda$ and
$\bar{h}_{\phi,\Lambda}$ for appropriately renormalized quantities $\bar{\nu}(B)$ and $\bar{h}_{\phi,0}$
connected to the atomic physics of individual atoms, i.e. evaluated for vanishing density and temperature. Furthermore,
$\bar{\nu}$ can be replaced by the reduced scattering length
\begin{eqnarray}
a_R(B) = - \frac{ \bar{h}_{\phi,0}^2(B) M}{4\pi\bar{\nu}(B)}.
\end{eqnarray}

The residual microscopic pointlike fermion interaction $\bar{\lambda}_{\psi,\Lambda}$ may be related to the background
scattering length $a_{bg}$
\begin{eqnarray}
a_{bg} = \frac{M\bar{\lambda}_{\psi,\Lambda}}{4\pi}.
\end{eqnarray}
We will neglect here a possible (weak) dependence of $a_{bg}$ on the magnetic field. Then the
magnetic field enters through $a_R(B)$. The microscopic information is now encoded in $a_R(B), a_{bg}$ and
$\bar{h}_{\phi,0}$. We recall that in the broad resonance limit the background scattering length $a_{bg}$ can also be
incorporated into $\bar{h}_{\phi,0}^2(B)$. By a suitable Fierz transformation we may equivalently use a formulation with
$\bar{\lambda}_{\psi,\Lambda}=0$.

We need to relate these parameters to our universal dimensionless parameters
\begin{eqnarray}
c( \bar{\nu}, \hpb,k_F, \sigma) &\to& c(B, k_F, \sigma),\nonumber\\
\tilde{h}_{\phi,0}( \bar{\nu}, \hpb,k_F, \sigma) &\to& \tilde{h}_{\phi,0}(B, k_F, \sigma).
\end{eqnarray}
Here $c$ and $\tilde{h}_{\phi,0}$ are evaluated in the limit $n\to 0$, $T\to 0$. Furthermore, the effective Yukawa coupling
appearing in the relation between the gap $\rex$ and the squared field $\tilde{\rho}$ includes renormalization effects,
$\hpn \equiv \hpn(\tilde{h}_{\phi,0}, \sigex,\Tn,c)$. This is discussed in the appendix \ref{sec:YukRenorm}.

\subsection{Concentration}

We may use the measurements of $\epsilon_M(B)$ in order to gain information about our parameters. Consider $^6\mathrm{Li}$
in a setting where the ``open channel'' consists of two atoms in the lowest energy states with nuclear spin $m_I = 1$ and
$m_I=0$, respectively. (These two lowest hyperfine states correspond to our two component fermion $\psi$.) The binding
energy near threshold has been measured \cite{Bartenstein04} for four different values of $B$ and fits well in this range
with
\begin{eqnarray}\label{EpsQuant}
\epsilon_M &=& -\beta (B_0- B)^2,
\end{eqnarray}
where, for $\lit$,
\begin{eqnarray}\label{BetaValue}
\beta^{(\mathrm{Li})} &=&  (7.22\cdot 10^{13}\mathrm{G}^2/\mathrm{eV})^{-1}  = (27.6\mathrm{keV}^3)^{-1},
\nonumber\\
B^{(\mathrm{Li})}_0&=& 834.1 \mathrm{G} = 16.29\,\mathrm{eV}^2.
\end{eqnarray}
The corresponding values for $^{40}\mathrm{K}$ are \cite{Jin04}
\begin{eqnarray}
\beta^{(\mathrm{K})} &=&  (4.97\cdot 10^{9}\mathrm{G}^2/\mathrm{eV})^{-1}  = (1.89\cdot 10^{-3}\mathrm{keV}^3)^{-1},
\nonumber\\
B^{(\mathrm{K})}_0&=& 202.10 \mathrm{G} = 3.95\,\mathrm{eV}^2.
\end{eqnarray}
From  eq. (\ref{EpsPropNu2}) we infer $\bar{a}(\sigma_A) = (-\epsilon_MM)^{-1/2}$ and we can also extract $\beta$ from
scattering experiments as described in app. \ref{sec:scattvac}. Close to the resonance we can therefore
relate the concentration parameter $c= \bar{a}(\sigma_A)k_F$ with the magnetic field \footnote{More precisely, $c(B)$
stands for $c(B,k_F, \sigma_A)$.} by
\begin{eqnarray}\label{GA}
c^{-1} (B) = - \Big(\frac{\beta M }{k_F^2}\Big)^{1/2} (B-B_0) = - \tau_B(B-B_0).
\end{eqnarray}

The relation (\ref{GA}) remains actually also valid for the ``atom gas'', $B> B_0$, $\epsilon_M >0$. In this case one has
$\sigma=0$ and
\begin{eqnarray}\label{GZ}
\bar{\nu} = \bar{\mu} (B-B_0) = -\frac{\bar{h}_{\phi,0}^2M}{4\pi a_R}
\end{eqnarray}
implies
\begin{eqnarray}\label{GC}
k_F c^{-1}(B) = \frac{4\pi\bar{\mu}}{\bar{h}_{\phi,0}^4M} (B_0- B).
\end{eqnarray}
It is experimentally verified that the relation $a^{-1} = \sqrt{\beta M}(B_0 - B)$ holds with the same value of $\beta$
(\ref{BetaValue}) as extracted from $\epsilon_M$ and we will confirm this by a ``computation from microphysics'' in app.
\ref{sec:scattvac}. This establishes the relation
\begin{eqnarray}\label{GD}
\beta = \frac{16\pi^2 \bar{\mu}^2}{\bar{h}_{\phi,0}^4M^3}
\end{eqnarray}
such that eqs. (\ref{GA}) and (\ref{GC}) coincide.

The relation (\ref{GA}) permits direct experimental control
of our concentration parameter for $T=0$ and $n=0$. For $^6\mathrm{Li}$  and $^{40}\mathrm{K}$ one finds
in the (arbitrary) unit \footnote{Of course, $k_F$ is not related to $n$ in this case.} $k_F =1\mathrm{eV}$
\begin{eqnarray}\label{taus}
\tau_B^{\mathrm{Li}} &=& 0.0088\mathrm{G}^{-1} = (113.1\mathrm{G})^{-1}=0.45\mathrm{eV}^{-2},\\
\tau_B^{\mathrm{K}} &=& 2.745\mathrm{G}^{-1} = (0.36\mathrm{G})^{-1}=140.55\mathrm{eV}^{-2}.\nonumber
% las calc gives \tau_B^{\mathrm{K}} &=& 2.748\mathrm{G}^{-1} = (0.36\mathrm{G})^{-1}=140.71\mathrm{eV}^{-2}.\nonumber
\end{eqnarray}
This will be easily extended to $T\neq 0$ and arbitrary $n$ below. The microscopic relation that is independent of $T$
and $n$ involves $\beta$ according to
\begin{eqnarray}\label{A24A}
 \frac{4\pi \bar{\nu}}{M\bar{h}_{\phi,0}^2} = (\beta M)^{1/2} (B-B_0) = \tau_B k_F (B - B_0).
\end{eqnarray}
We emphasize that the determination of $\tau_B$ in eq. (\ref{GA}) does not require knowledge of $\hpb$ since we can
use directly the experimental determination of $\beta$ from eq. (\ref{EpsQuant}).

Away from the resonance the issue of the correct choice of $c$ is more subtle. The observed scattering length $a$ is well
approximated by a resonance term
\begin{eqnarray}
a_{res} = -\frac{1}{\sqrt{\beta M}}\frac{1}{B- B_0}
\end{eqnarray}
and a constant (or weakly $B$ - dependent) background scattering $a_{bg}$,
\begin{eqnarray}
a(B) = a_{res}(B) + a_{bg}.
\end{eqnarray}
The question arises how the background scattering
is properly incorporated. In principle, not only the exchange of the dressed molecules contributes to the scattering,
but also a residual atom interaction in vacuum which is parameterized by the effective coupling $\bar{\lambda}_{\psi,0}$.
In the broad resonance limit our results are consistent with a minor role of $\bar{\lambda}_{\psi,0}$ in the molecule
phase and we will neglect it.

Indeed, we will see below how the background scattering is incorporated into $\bar{a}(\sigma_A)$ for the molecule phase,
such that we can associate $\bar{a}(\sigma_A)$ with the observed scattering length, including the background term
\begin{eqnarray}\label{eq93}
\bar{a}(\sigma_A) = a(B).
\end{eqnarray}
For the concentration this implies a simple relation to $a$ and therefore a somewhat more involved dependence on $B$. Our
central relation between the concentration $c$ and the magnetic field is simply expressed in terms of the
measured scattering length $a(B)$,
\begin{eqnarray}\label{TrivialRel}
c = a(B) k_F.
\end{eqnarray}

This relation is quite satisfactory. Indeed, in the broad resonance limit the interaction is essentially pointlike, both
for $a_{bg}$ and $a_{res}$. Physical results can therefore only depend on $a = a_{res} + a_{bg}$. This holds independently
of the precise value of $\bar{\lambda}_{\psi,\Lambda}$. Indeed, by a Fierz transformation we can shift the interaction
from $\bar{\lambda}_{\psi,\Lambda}$ to the contribution from molecule exchange $\sim \hpb^2/\bar{\nu}$ and vice versa
\cite{Diehl:2005ae}. Since for pointlike interactions the Fierz transformations are exact identities the physical results should only
depend on the sum $\bar{h}_{\phi,\Lambda}^2/(\bar{\nu}_\Lambda - 2\sigma_A) - \bar{\lambda}_{\psi, \Lambda}$ and not on the
value of an individual piece like $\bar{\lambda}_{\psi,\Lambda}$. In terms of the renormalized quantities the results can
then only depend on $a$ and not on $a_{res}$ and $a_{bg}$ separately. This feature is precisely realized by the
relation (\ref{TrivialRel}): all results depend only on the concentration $c$, and this is in turn is given by the total scattering
length $a$, independent of a decomposition into $a_{res}$ and $a_{bg}$.

Based in the above argument we will use the simple relation (\ref{TrivialRel}) also for $B>B_0$. Within our approximations
this implies that we associate $a_R = a_{res} + a_{bg}$ for $B>B_0$ which amounts to $B$ - dependent
$\bar{h}_{\phi,0}^2(B)$. In fact, our approximation does not resolve the Fierz ambiguity for $B>B_0$. In app.
\ref{sec:scattvac} we discuss in detail the scattering of atoms in vacuum. There we comment on the issue of the Fierz
ambiguity more extensively.

In summary, the dependence of $c^{-1}$ on $B$ obeys the nonlinear relation
\begin{eqnarray}\label{cB1}
\frac{k_F}{c} &=& \frac{1}{ a_{res}}\Big(\frac{1}{1 + a_{bg}/a_{res}}\Big)\\\label{cB2}
&=& -\frac{\hat{\tau}_B (B- B_0)}{1-a_{bg} k_F \hat{\tau}_B(B- B_0)}\nonumber
%B - B_0 &=&  -\frac{1}{\tau_B}\frac{c^{-1}}{1-a_{bg} k_F c^{-1}}
\end{eqnarray}
where $\hat{\tau}_B = k_F \tau_B = \sqrt{\beta M} =0.45 (140.55)\mathrm{eV}$ for $\lit$ ($\kal$) is independent
of $k_F$. This nonlinearity in the relation between the magnetic field $B - B_0$ and the coupling strength $c^{-1}$
is an important effect. For $\lit$ this produces the strong increase of the bare molecule fraction on the low field side
of the resonance as seen in fig. \ref{BareExpPart}.
We note that for negative $a_{bg}$ (as for $\lit$) the scattering length reaches zero at some value of $B$ in the BEC
regime, $B - B_0 = -\Delta = a_{bg}^{-1}(\beta M)^{-1/2}$. At this point $c$ vanishes. For $B$ below $B_0 - \Delta$ the
system belongs to the weakly coupled BCS regime. The transition between the BCS and BEC regimes in the weak coupling
limit can therefore be realized by experiment. For $\lit$ one finds $\Delta = 300\mathrm{G}$, for $\kal$ one has
$\Delta = - 7.8\mathrm{G}$.

\subsection{Yukawa coupling}
\label{sec:MicroYuk}
The discussion of the relevant range for our second parameter, the dimensionless Yukawa coupling $\hpn$, is more involved.
Let us denote the effective Yukawa coupling in vacuum in the atom phase by $\bar{h}_{\phi,0}$ or $\tilde{h}_{\phi,0}$
for the dimensionless counterpart. In app. \ref{sec:YukRenorm} we have computed its relation to the microscopic coupling
$\bar{h}_{\phi,\Lambda}$ appearing in the action (\ref{YukawaAction}).
From eq. (\ref{GD}) and the experimental determination of $\beta$ in the atom phase we can extract the ratio
\begin{eqnarray}\label{G1}
\frac{\bar{h}_{\phi,0}^2}{\bar{\mu}}  &=& \frac{4\pi}{(M^3\beta)^{1/2}} \\
&=& 2.516\cdot 10^{-7}\mathrm{eV}^{-2}\mathrm{G}=4.915\cdot 10^{-9}.\nonumber
\end{eqnarray}
(Here the numerical value is given for $^6\mathrm{Li}$.) For the broad Feshbach resonance in the $^6\mathrm{Li}$ system
the molecule state belongs to a singlet of the electron spin, resulting in $\mu_M \approx 0$. The microscopic value of
$\bar{\mu} = \partial \bar{\nu}_\Lambda/\partial B$ obtains its essential contribution from the magnetic moment of the atoms
in the open channel, which is well approximated \cite{Stoof98} by the Bohr magneton $\mu_B = 5.788\cdot 10^{-11}
\mathrm{MeV/T} = 0.2963 \mathrm{MeV}^{-1}$. For $\kal$ we take $\bar{\mu} =1.57\mu_B$ \cite{Chwedenczuk04}. With
\begin{equation}\label{GH}
\bar{\mu}^{(\mathrm{Li})} = 2\mu_B,\quad \bar{\mu}^{(\mathrm{K})} = 1.57 \mu_B,
\end{equation}
and $k_F = 1\mathrm{eV}$ this yields high values
\begin{eqnarray}\label{G3}%609.7
\tilde{h}_{\phi, 0}^{\mathrm{Li}} &=& 610,\quad (\tilde{h}_{\phi, 0}^{\mathrm{Li}})^2 = 3.72\cdot 10^5, \nonumber\\
\tilde{h}_{\phi, 0}^{\mathrm{K}} &=& 79,\quad \,\,\, (\tilde{h}_{\phi, 0}^{\mathrm{K}})^2 = 6.1\cdot 10^3.%79.0
\end{eqnarray}
Using the broad resonance criterion $\hpn^2 > 100$, we see that indeed $\lit$ and $\kal$ belong to this class, cf. fig.
\ref{hpUniversality}.

In the molecule phase the vacuum value of $\hpb$ differs from $\bar{h}_{\phi,0}$, as we have shown in app.
\ref{sec:YukRenorm}. If we incorporate $a_{bg}$ in $\bar{\lambda}_{\psi,\Lambda}$ such that $\bar{h}_{\phi,\Lambda}$ is
independent of $B$ one obtains
\begin{eqnarray}\label{hphim}
\bar{h}_{\phi,0}^{(m)} = \frac{\bar{h}_{\phi,0}}{1 - a_{bg}/\bar{a}(\sigma_A)}.
\end{eqnarray}
It is interesting to see how the renormalization effects turn $\bar{a}(\sigma_A)$ into the full scattering length including
the background scattering. Indeed, assuming that $\bar{h}_{\phi,0}$ is independent of $B$ and $\bar{\nu} \propto B-B_0$ we
may identify the resonant contribution to the scattering length as
\begin{eqnarray}
a_{res} = -\frac{M\bar{h}_{\phi,0}}{4\pi \bar{\nu}} = -(\beta M)^{-1/2}\frac{1}{B- B_0}.
\end{eqnarray}
For $\bar{a}(\sigma_A)$ this yields
(\ref{A24A})
\begin{eqnarray}\label{QQA}
\bar{a}(\sigma_A) = -\frac{M}{4\pi} \frac{(\bar{h}_{\phi,0}^{(m)})^2}{\bar{\nu} - 2\sigma_A}
= a_{res}\Big(\frac{\bar{h}_{\phi,0}^{(m)}}{\bar{h}_{\phi,0}}\Big)^2 \frac{\bar{\nu}}{\bar{\nu} - 2\sigma_A}.
\end{eqnarray}
Precisely at the resonance the renormalization effects vanish.

Away from the resonance, however, we have to take into account two renormalization effects which appear in the
order $a_{bg}/a_{res}$. First, the ratio $(\bar{h}_{\phi,0}^{(m)}/\bar{h}_{\phi,0})^2$ yields a $B$ dependent factor
$( 1 - a_{bg}/\bar{a})^2$ (with $\bar{a} \equiv \bar{a}(\sigma_A)$). Second, the ``renormalization of the detuning'' by the
shift $\bar{\nu} \to \bar{\nu} - 2\sigma_A$ induces an additional $B$ - dependence by a factor $\bar{\nu}/(\bar{\nu} -
2\sigma_A) = a_{res}/\bar{a}$. Eq. (\ref{QQA}) yields
\begin{eqnarray}\label{QQB}
\bar{a} = \frac{a_{res}^2}{\bar{a} (1 - a_{bg}/\bar{a})^2} = \bar{a}\Big(\frac{a_{res}}{\bar{a} -
a_{bg}}\Big)^2\nonumber\\
\end{eqnarray}
and we conclude that $\bar{a}$ is indeed given by $a_{res}(B) + a_{bg}$. We observe that $a_{bg}$ is related here to the
microscopic ``background atom interaction'' $\bar{\lambda}_\psi$ via the renormalization of $\bar{h}_{\phi,0}$
discussed in the appendix \ref{sec:YukRenorm}. Thus eq. (\ref{QQB}) gives a microphysical explanation for the nonlinear
$B$ - dependence of $\bar{a}$.

This issue is different for the atom phase where the $B$ - dependent
renormalization effects for $\bar{h}_{\phi,0}$ and $\bar{\nu}$ are absent. In this case $\bar{a} = a_R(B)$ only
reflects the resonant contribution to the scattering if we keep $\bar{h}_{\phi,0}^2$ independent of $B$. The background
contribution has then to be accounted for by the direct coupling $\bar{\lambda}_{\psi,0}$. Alternatively, we may absorb
$\bar{\lambda}_{\psi,\Lambda}$ into the molecule exchange contribution by a Fierz transformation, including an effective
$B$ - dependence of $\bar{h}_{\phi,0}^2$ for $B>B_0$. Then $a_R$ has to reflect the full dependence on $B$, including the
background scattering. If $\bar{\lambda}_{\psi,0}$ can be neglected this requires $a_R = a(B)$ and leads to the relation
(\ref{TrivialRel}).

The determination of $\epsilon_M = 2\sigma_A$ amounts to the condition
\begin{eqnarray}\label{GI}
\frac{\bar{m}_\phi^2(\sigma_A)}{\hpb^2} = \frac{\bar{\nu} - 2\sigma_A}{\hpb^2} + \frac{\partial U_1}{\partial \bar{r} }
(\bar{r} = 0, \sigma_A) =0.
\end{eqnarray}
The bosonic fluctuation contribution vanishes for $T\to 0$ and $\partial U_1\hF/\partial \bar{r}$ only depends on $\sigma$,
but not on $\hpb$ or $\bar{\mu}$. This is the reason why the behavior for $B\to B_0$ can be used to determine $c$ and the
ratio (\ref{G1}), but not $\hpn$ separately. In principle, independent experimental
information on $\tilde{h}_{\phi, 0}$ can be gained from measurements of the binding energy outside the
resonance. Indeed, for $\epsilon_M <0$ the general relation between $\epsilon_M$ and $\bar{\nu}$ reads
\begin{eqnarray}
\epsilon_M &=& \bar{\nu} - \gamma (1 - \sqrt{1 - 2\bar{\nu} /\gamma}),\\\nonumber
\gamma &=& \frac{\bar{h}_{\phi,0}^4M^3}{32\pi^2}.
\end{eqnarray}
Measurement of $\epsilon_M (B)$ over a wide enough range allows for the extraction of both $\bar{h}_{\phi,0}^2$ and
$\bar{\nu}$ independently. Far away from the resonance one has $|\gamma/\bar{\nu}|\ll 1$ such that $\epm =\bar{\nu}$.
From this range we extract the $B$ - dependence of $\bar{\nu}$, $\partial\epm /\partial B =\partial\bar{\nu}/
\partial B = \bar{\mu}$. On the other hand, close to the resonance the opposite limit $|\gamma/\bar{\nu}|\gg 1$ applies.
Then $\epm = - \bar{\nu}^2/(2\gamma)$ involves the concentration $c$, and, using $\bar{\nu} = \bar{\mu}(B - B_0)$, one
can extract $\bar{h}_{\phi,0}$ from $\gamma = -\bar{\nu}^2/(2\epm)$. Alternatively, one may extract $\bar{\nu}(B)$
and $\bar{h}_{\phi,0}(B)$ from a detailed investigation of the atom scattering near the resonance, as discussed in app.
\ref{sec:scattvac}.

\subsection{General relation between $c^{-1}$ and $B$}
\label{sec:phasediagram}

In subsect. A we have related the mean field scattering length $a_R$ in vacuum with the detuning magnetic field
$B-B_0$. This can be used directly for establishing the relation between the concentration
$c$ and $B-B_0$ for nonzero $T$ and $n$, using the definition (\ref{InMediumScattLength})
\begin{eqnarray}\label{crelation}
c^{-1} = k_F^{-1} \Big( a_R^{-1} + \frac{8\pi \sigma}{\hpb^2 M}\Big).
\end{eqnarray}
One finds
\begin{eqnarray}\label{bcrel}
\frac{1}{c} - \frac{16\pi \tilde{\sigma}}{\hpn^2} = -\tau_B (B-B_0) = -\tilde{b}.
\end{eqnarray}
Employing the relation
\begin{eqnarray}
\tilde{b} = \frac{8\pi \nun}{\hpn^2}
\end{eqnarray}
we may also write eq. (\ref{crelation}) in the form
\begin{eqnarray}\label{SigmaTildeB}
\tilde{\nu} = 2\tilde{\sigma} - \frac{\hpn^2}{8\pi c} = \frac{2M\bar{\mu} (B- B_0)}{k_F^2}.
\end{eqnarray}
We observe that the dimensionless Yukawa coupling $\hpn$ appears in eq. (\ref{bcrel}). It is related to $\bar{h}_{\phi,0}$
by eq. (\ref{hpRenormGen}) in app. \ref{sec:YukRenorm}. For practical calculations we use the relation (\ref{SigmaTildeB})
between $\tilde{\nu}$ and $B-B_0$.

In the broad resonance limit for large values of $\hpn^2$ of the same order of magnitude as the vacuum values (\ref{G3})
the terms $\sim \hpn^{-2}$ in eq. (\ref{bcrel}) can be neglected if $|\sigex|$ is not too large. This yields the simple
expression
\begin{eqnarray}
c^{-1} = -\tilde{b}.
\end{eqnarray}
We recall, however, that this formula should not be used far in the BEC regime where $\sigex$ takes large negative values
which diverge in the vacuum limit. For moderate or small values of $\hpn$ the term $\propto\sigex$ in eq. (\ref{bcrel})
should always be included.

\section{Dressed and bare Molecules: Comparison to experiments}
\label{CompExp}

\subsection{Bare molecules}
\label{sec:BareMolecs}
\begin{figure}
\begin{minipage}{\linewidth}
\begin{center}
\setlength{\unitlength}{1mm}
\begin{picture}(85,55)
      \put (0,0){
     \makebox(80,49){
 \begin{picture}(80,49)
       %\coordsyst{90}{54.9}{21}{14}
      \put(0,0){\epsfxsize80mm \epsffile{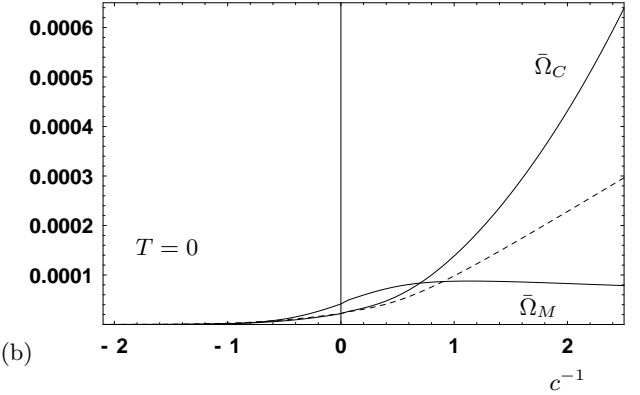}}
      \put(70,-3){$c^{-1}$}
      \put(-3,1){(b)}
      \put(68,39){$\OmC$ }
      \put(66,7){$\OmM$ }
      \put(15,15){$T =0$ }
     \end{picture}
      }}
\end{picture}
\end{center}
\vspace*{-1.25ex} \caption{Contributions $\OmC, \OmM$ to the fraction of closed channel
molecules $\OmB = \OmC + \OmM$. We plot the result for $\lit$ ($\tilde{h}_{\phi,0} =610$) for $T=0$. The condensed part
dominates $\OmB$ in the BEC regime, while in the vicinity of the resonance, the contributions are comparable. The dashed
line for $\OmC$ omits for $c^{-1}>0$ the background contribution to $\bar{a}$ and reflects the Fierz ambiguity.}
\label{BareExp2}
\end{minipage}
\end{figure}

The fraction of closed channel molecules is given by $\OmB = \OmM + \OmC$. It has been measured by a laser probe that
induces a transition to an excited molecular level - the quantity $Z$ in \cite{Partridge05} equals $\OmB$. Indeed, the laser
probe couples directly to the total number of ``bare molecules'' $\langle \hat{\phi}^*\hat{\phi}\rangle - \hat{n}_B$.
Importantly, we have seen in sect. \ref{DressedAndBareI} and \ref{EvalBeyondMFT} that this quantity involves the Yukawa
$\hpn$ coupling explicitly, cf. eq. (\ref{EqOmegaC}). In fig. \ref{BareExpPart} we compare our
results with the measured value for Li, using $T = 0$, $k_F= 0.493 \mathrm{eV}\hat{=}250\mathrm{nK}$. The solid line uses
$c= k_F(a_{res}(B) + a_{bg})$ (\ref{eq93}) for the whole range of $B$. The agreement with the measured value is very
convincing. In order to give and idea of the Fierz ambiguity
for $B> B_0$ we also display by the long-dashed line the result of the choice $c = k_F a_{res}(B)$. For $B<B_0$ the Fierz
ambiguity is removed by the renormalization effects as discussed in sect. \ref{sec:MicroYuk}. We emphasize that the
inclusion of the renormalization (\ref{hphiRen2}) for the Yukawa coupling is crucial. Omitting this effect (dashed line)
for $B<B_0$ in fig. \ref{BareExpPart} results in a clear discrepancy from the observations sufficiently far away from the
resonance.

At this place we may give a few more details of our computation of $\bar{\Omega}_B$,
\begin{eqnarray}
\OmB = \OmM + \OmC = 3\pi^2\big(\frac{n_M}{\ZpR} + 2\frac{\rex}{\tilde{h}_\phi^2}\big).
\end{eqnarray}
We use the Yukawa coupling as defined in (\ref{hpRenormGen}) in the place where it appears explicitly. All other
quantities are taken from the solution of the equations determining the crossover problem in the superfluid phase, eqs.
(\ref{Cross1SSB},\ref{Cross2SSB}). The formulae for $\bar{\Omega}_C$,
\begin{eqnarray}
\bar{\Omega}_C = \frac{6\pi^2}{k_F^3} |\bar{\phi}_0|^2 = 6\pi^2(1 - a_{bg}/a)^2 \frac{\rex_0}{\tilde{h}_{\phi,0}^2}
\end{eqnarray}
relates the closed channel condensate fraction to the superfluid order parameter $\phi_0$. In that sense Partrigde
\emph{et al.} indirectly measure the superfluid order parameter!
%The last expression becomes valid in the extreme BEC limit when the density equation is dominated by $n_{F,0}$.
%Using eq. (\ref{EpsPropNu2}), this formula relates $\OmB$ to the binding energy of the molecules, independently
%of $k_F$. It is quite plausible that this simplified description in terms of microscopic quantities breaks down in the
%vicinity of the resonance, where many-body effects become increasingly important.
From $\OmC$ we can infer $\OmB = \OmC(1 + \OmM/\OmC)$ by extracting the ratio $\OmM/\OmC$ from fig. \ref{BareExp2}. In
the BEC limit $\OmM/\OmC$ becomes negligible, while at resonance ($c^{-1}=0$) we find $\OmM/\OmC = 1.89$.

Let us briefly comment on the density dependence of our result. This is particularly simple in our dimensionless
formulation. Since $\rex_0$ is taken from the solution of the crossover problem it does not depend on $k_F$ if $c$ is kept
fixed. However, in order to find the right dimensionless value $\hpn = 2M\hpb/k_F^{1/2}$, $k_F$ must be inserted.
For fixed $c$ this yields $\bar{\Omega}_C\propto k_F$. The ratio $\OmM/\OmC$ depends also on $c$. The dependence of
$\OmB$ on $c$ introduces an additional density dependence of $\OmB$ if the scattering length $a$ is kept fixed. (Only for
fixed $a k_F$ one always has $\OmB \propto k_F$ in the broad resonance limit.)

In the scaling limit $c^{-1}=0$
the value $\tilde{r}_0$ and the ratio $\OmM/\OmC$ are independent of $k_F$ if we assume the broad resonance limit $\hpn\to
\infty$. In this limit we thus can confirm the simple scaling law at
resonance advocated by Ho \cite{Ho104}, $\bar{\Omega}_B \propto k_F$. A similar
result has been obtained by Levin \emph{et al.} \cite{IIChen05}, though neglecting the contribution $\bar{\Omega}_M$ which
is of the same order of magnitude at the resonance, cf. fig. \ref{BareExp2}.

In the BEC limit $\OmM/\OmC$ becomes negligible. Furthermore, since $\rex_0 \propto c^{-1}\propto k_F^{-1}$, one finds
that $\OmB$ becomes independent of $k_F$. Note that this result is only valid for $k_F$ which respect the BEC regime
condition, $c^{-1} =(a k_F)^{-1} >1$ or $k_F < a^{-1}$. %This can be seen in fig. [????????] where we plot $\OmB$ as a function of
%$k_F$ for three values of $B$, corresponding to the BEC, resonance and BCS regimes.

In the deep BCS regime $c^{-1} =|a k_F|^{-1} \gg 1$, we also find numerically that $\OmM/\OmC\ll 1$. Using the standard
BCS result relating the superfluid order parameter at $T=0$ and $c^{-1}$, our scaling form yields $\OmB \propto
k_F\exp(-\pi/(a k_F))$. Again, in order to stay in the desired regime, $k_F$ must be restricted to values
$k_F\ll |a|^{-1}$. In sum, we find the following scaling behaviors with $k_F$
\begin{eqnarray}\label{ScalingBeh}
\OmB \propto\left\{
\begin{array}{l}
  {\mathrm{const.} \qquad\qquad  \text{BEC}}  \\
  {k_F\qquad\qquad\quad\,\, \,\text{Resonance}}  \\
  {k_F\mathrm{e}^{-\pi/(ak_F)}\quad\,\,  \text{BCS}}
\end{array}\right. .
\end{eqnarray}

\begin{figure}
\begin{minipage}{\linewidth}
\begin{center}
\setlength{\unitlength}{1mm}
\begin{picture}(85,110)
      \put (0,0){
     \makebox(80,49){
     \begin{picture}(80,49)
       %\coordsyst{90}{54.9}{21}{14}
      \put(0,0){\epsfxsize80mm \epsffile{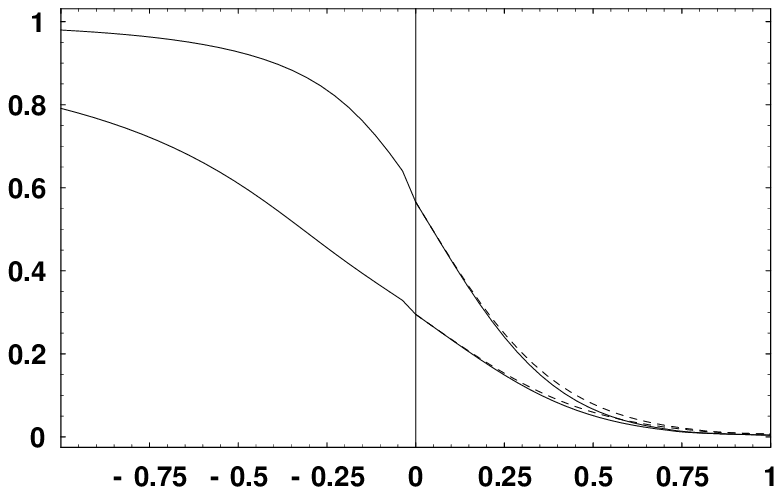}}
      \put(62,-2){$B- B_0$[G]}
      \put(10,43){$\Omega_B$}
      \put(65,40){$\kal$}
      \put(10,34){$\Omega_C$}
      \put(-2,-1){(b)}
      \end{picture}
      }}
       \put (0,54){
     \makebox(80,49){
     \begin{picture}(80,49)
       %\coordsyst{90}{54.9}{21}{14}
      \put(0,0){\epsfxsize80mm \epsffile{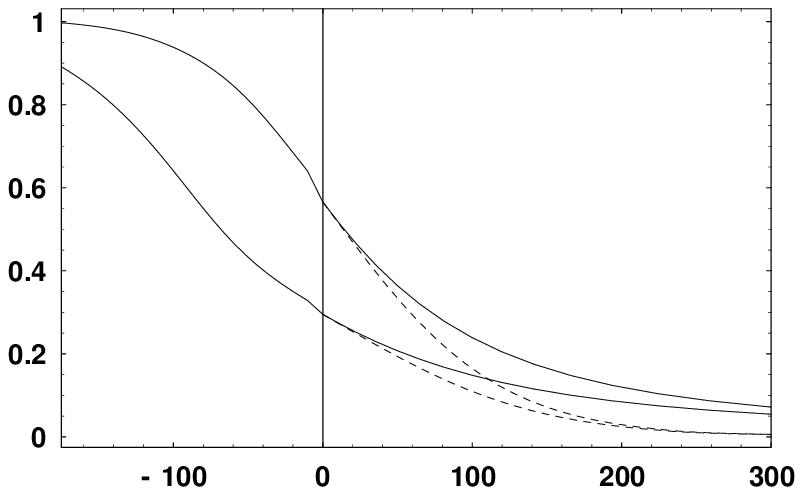}}
      \put(62,-2){$B- B_0$[G]}
      \put(10,44){$\Omega_B$}
      \put(65,40){$\lit$}
      \put(10,34){$\Omega_C$}
      \put(-2,-1){(a)}
      \end{picture}
      }}
\end{picture}
\end{center}
\vspace*{-1.25ex} \caption{Condensate fraction $\Omega_C$ and total fraction of dressed molecules $\Omega_B =
\Omega_M + \Omega_C$ at $T=0$, for $\lit$ (a) and $\kal$ (b). Due to the nonlinear relation of $c^{-1}$
and $B-B_0$ which is particularly strong for $\lit$ due to the large background scattering value, the curves are twisted
compared to the $\Omega(c^{-1})$ plots. We adjust $k_F$ to the values found in \cite{Jin04} ($k_F =1.50\mathrm{eV}
\hat{=} (2500 a_B)^{-1}$) and \cite{Ketterle04} ($k_F =1.39\mathrm{eV}\hat{=} (2700 a_B)^{-1}$). The dashed line reflects
the Fierz ambiguity as in figs. \ref{BareExpPart},\ref{BareExp2}.}
\label{DressedExp}
\end{minipage}
\end{figure}

\subsection{Dressed Molecules}
The condensate fraction measured in \cite{Jin04,Ketterle04} qualitatively refers to the condensation of ``dressed
molecules'' or di-atom states. At the present stage, however, the precise relation between the measured observables and the
condensate fraction $\Omega_C$ has not yet been established with sufficient quantitative accuracy. As argued in sect.
\ref{DressedAndBareI}, $\Omega_C$ is given by the expectation value of the renormalized field
\begin{eqnarray}
\Omega_C &=& 2\frac{\langle\hat{\phi}^*_R\rangle\langle\hat{\phi}_R\rangle}{n}= 6\pi^2\ZpR\rhon = 6\pi^2\rho ,\\\nonumber
\hat{\phi}_R &=& \ZpR^{1/2} \hat{\phi}.
\end{eqnarray}
In fig. \ref{DressedExp} we plot our result for the condensate fraction of dressed molecules at zero temperature as a
function of magnetic field, for
both the $\lit$ (fig. \ref{DressedExp} (a)) and the $\kal$ system (fig. \ref{DressedExp} (b)). Unlike the condensate
fraction of bare molecules, the contribution from the dressed condensate is an $\mathcal{O} (1)$ quantity.

We find qualitative agreement with the observations \cite{Jin04,Ketterle04} whereas for a quantitative comparison
one would need a more accurate relation between the measured observables and the condensate fraction $\Omega_C$ as
defined in our setting. First, the scales of the magnetic field $B$ for which $\Omega_C$ decreases from rather large
values to small values match the scales found in the experiments. This is another confirmation of our universal relation
between $B$, the particle density (or $k_F$) and the quantities $c^{-1},\hpn$. We find that in the BEC limit, the
condensate fraction approaches 1 very slowly. The observed condensate depletion is the effect of the
interaction between the dressed molecules, as expected for a weakly interacting Bogoliubov gas.

\begin{figure}
\begin{minipage}{\linewidth}
\begin{center}
\setlength{\unitlength}{1mm}
\begin{picture}(85,55)
      \put (0,0){
     \makebox(80,49){
 \begin{picture}(80,49)
       %\coordsyst{90}{54.9}{21}{14}
      \put(0,0){\epsfxsize80mm \epsffile{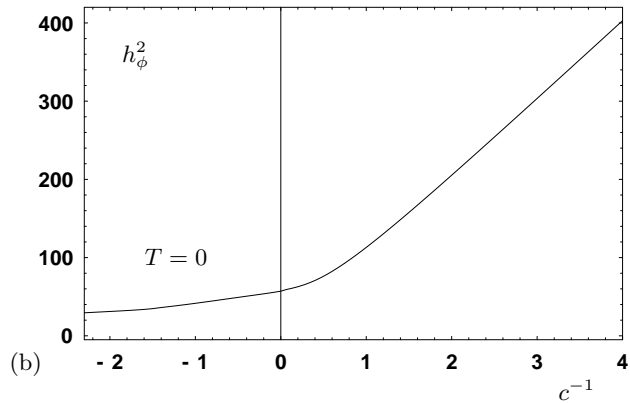}}
      \put(70,-3){$c^{-1}$}
      \put(-3,1){(b)}
      \put(12,42){$h_\phi^2$ }
     % \put(66,7){$\OmM$ }
      \put(15,15){$T =0$ }
     \end{picture}
      }}
\end{picture}
\end{center}
\vspace*{-1.25ex} \caption{Renormalized Yukawa coupling $h_\phi^2$ as a function of $c^{-1}$ at zero temperature in the
broad resonance limit.}
\label{HphiSqPlot}
\end{minipage}
\end{figure}

An interesting quantity is the value of the condensate fraction at the location of the resonance, $B =B_0$. For $T=0$ we
find a universal value $\Omega_C^0 = \Omega_C (T=c^{-1} =0) = 0.30$, which therefore should apply both \footnote{If
one would identify the ``condensate fraction'' quoted in refs. \cite{Jin04,Ketterle04} with our definition of $\Omega_C$
one obtains at the resonance $\Omega_C (\Tn=0.08, c^{-1}=0) = 0.13$ for $\kal$ \cite{Jin04} and $\Omega_C (\Tn=0.05, c^{-1}=0)
= 0.7$ for $\lit$ \cite{Ketterle04}.} for $\lit$ and $\kal$. In order to judge the reliability of this universal result we
may consider the relation
\begin{eqnarray}
\Omega_C = \frac{6\pi^2\rex_0}{h_\phi^2}
\end{eqnarray}
which involves the renormalized Yukawa coupling $h_\phi^2 = \hpn^2/\ZpR$. In the broad resonance limit and for $T=0$ the
value $\rex_0^0 = \rex_0 (T=c^{-1}=0) = 0.28$ is universal. The size of $\Omega_C^0$ is therefore determined by $h_\phi^2$,
$\Omega_C^0 = 16.85h_\phi^{-2}$. In the same limit we find from eq. (\ref{ZRFormula})
\begin{eqnarray}
h_\phi^{-2} = \frac{1}{8\pi^2} \int\limits_0^\infty d\qn \frac{\qn^2(\qn^2 - \sigex)}{[(\qn^2 - \sigex)^2 + \rex_0]^{3/2}}
\end{eqnarray}
where $\sigex^0 = \sigex (T=c^{-1}=0) = 0.50$. This yields $h_\phi^{-2} = 0.018$, $\Omega_C^0 =0.30$ as seen in fig.
\ref{DressedExp}. Away from the resonance both $\rex_0$ and $\sigex$ depend on $c$. In fig. \ref{HphiSqPlot} we show
$h_\phi^2$ as a function of $c^{-1}$ for $\Tn =0$.

In the broad resonance limit the value of the renormalized Yukawa coupling only depends on $c^{-1}$ and $\Tn$. This
universal value $h_\phi^2(c^{-1},\Tn)$ is reminiscent of the existence of a partial infrared fixed point
as some suitable infrared cutoff scale is lowered. The value of $h_\phi$ at the fixed point typically depends on $c^{-1}$ and $\Tn$,
but otherwise the memory of the initial conditions of the flow (e.g. $\bar{h}_{\phi,\Lambda}$) is lost. It is also plausible
that also $\tilde{\lambda}_\psi$ is governed by some fixed point. If this picture is true, the precise location of the fixed point may
change as the approximation method is improved, for example by taking the molecule fluctuations into account for the
computation of $\ZpR$ and the renormalization effects for $\hpn^2$. Nevertheless, if both the $\lit$ and the $\kal$
system are within the range of attraction of the fixed point, the value of $\Omega_C$ for $T=c^{-1} = 0$ must be the same!
Inversely, if the observations should establish different values of $\Omega_C^0$ this would imply that either no such
fixed point exists or that one of the systems is not yet close enough to the fixed point. Only in this case the system could
keep additional memory of the microscopic Yukawa coupling (e.g. $\bar{h}_{\phi,0}$) or the background scattering length
$a_{bg}$ such that $\Omega_C^0$ could depend on these parameters. Our findings so far, however, strongly suggest the
universal behavior predicting for broad resonances a universal curve $\Omega_C(c^{-1})$ for $T=0$.

\section{Conclusions}
\label{sec:conclusions}

The theoretical description of the crossover problem for ultracold fermionic atoms can be split into two steps. First,
the computation of universal properties in terms of the two relevant dimensionless couplings $c$ and $\hpn$. Second,
the relation of $c$ and $\hpn$ to the particular parameters of a given Feshbach resonance as the detuning of the
magnetic field $B-B_0$. Both steps can be performed within a common setting of non-relativistic quantum field theory,
described here in terms of a functional integral.

The first step can be done for all densities and temperatures and for arbitrary Feshbach resonances that can be
approximated by our system of one open and one closed channel. It reveals the universal aspects since all observables
are expressed in terms of only three parameters $c$, $\hpn$, $\Tn$. For broad Feshbach resonances (large $\hpn$) the
universality is further enhanced. The renormalized quantities relating to the ``macroscopic'' properties of dressed
molecules and atoms only involve a renormalized Yukawa coupling $h_\phi$ that takes a ``fixed point'' value
$h_\phi(c,\Tn)$ that is computable in terms of $c$ and $\Tn$. The ``dressed quantities'' therefore depend only on $c$ and
$\Tn$. In consequence, for $T=0$ or $T=T_c$ they turn to functions
depending only on the concentration $c$. Furthermore, in the scaling limit at resonance ($|c|\to \infty$) the
renormalized observables depend only on $\Tn$ and take predictable universal values for $T=0$ or $T=T_c$.

As a particular aspect of universality we find perfect equivalence between a purely fermionic description of the crossover
problem and a model that uses explicit bosonic degrees of freedom corresponding to the microscopic molecules in the
closed channel of the Feshbach problem. The microscopic molecules are a convenient tool to take into account the
non-locality of the effective fermion-interaction induced by the molecule exchange. This nonlocality is relevant for
narrow and intermediate Feshbach resonances. For broad Feshbach resonances, however, the nonlocality becomes subleading as
far as the properties of the dressed atoms and molecules are concerned. The broad resonance limit $\hpn\to \infty$
precisely corresponds to a pointlike microscopic fermion-interaction. As it should be, only one dimensionless parameter
(besides $\Tn$) characterizes the system, namely the scattering length $a$ appearing in the dimensionless concentration
$c=ak_F$.

For a test of our universality hypothesis we have computed several macroscopic observables as functions of $c,\hpn,\Tn$.
This concerns the fraction of microscopic molecules in the atom gas $\OmB$, the condensate fraction $\Omega_C$ and the
fraction of uncondensed dressed molecules $\Omega_M$. We also compute the correlation length for the molecules $\xi_M$
as well as their scattering length in medium $a_M$. The low momentum excitations in the superfluid phase are described by
a Bogoliubov theory for dressed molecules even if the microscopic molecules play no role. We have computed the corresponding
sound velocity $v_M$. All these quantities may be experimentally tested.

For these tests we need the second step, the relation of $c$, $\hpn$ and $\Tn$ to the physical properties of experiments
with a given Feshbach resonance at a given density $n$ and temperature $T$. Here $\Tn = T/\epsilon_F$ only involves $T$ and
$n$, $\epsilon_F =k_F^2/(2M)$, $n=k_F^3/(3\pi^2)$. In order to relate $c$ and $\hpn$ to observables in the two atom
system (at zero density and temperature) we compute the molecular binding energy $\epsilon_M$ and the cross section
for the atom-atom scattering $\sigma(s)$ as functions of $c$ and $\hpn$. (Here $s$ denotes the momentum of the scattering
atoms.) This is achieved by simply taking in our formalism the limit $n\to 0, T\to 0$. Comparing $\epm(c,\hpn),
\sigma(s,c,\hpn)$ to the measured values as functions of the magnetic field, $\epm(B), \sigma(s,B)$, we can extract
$c(B)$ and $\hpn(B)$. In particular, for the broad resonance limit $c(B)$ is simply related to the scattering length
$a(B)$ describing a pointlike interaction, $c(B) =a(B)k_F$. For a broad Feshbach resonance and ``dressed quantities'' the
only experimental input needed is therefore $a(B)$. Only for ``bare quantities'' (as the fraction of microscopic molecules)
one needs, in addition, the determination of $\hpn$.

The broad resonance limit leads to striking predictions that challenge experiment. For example, the condensate fraction
$\Omega_C$ should take the \emph{same} value at resonance, $B=B_0$, for all broad Feshbach resonances. (Here $B_0$
corresponds to a vanishing binding energy in vacuum, $\epm (B_0) =0$). For $T\to 0$ or $T\to T_c$ this should hold
for arbitrary density, whereas more generally $\Omega_C$ depends on $T/\epsilon_F$ and therefore on $n$ in a universal way.
In our approximation we find $\Omega_C=0.3$ for
$T=0$ but the precise value could change somewhat for improved approximations. If both the broad resonances for
$\lit$ and $\kal$ are well described by the broad resonance limit the experiments \cite{Jin04,Ketterle04} should find the
same $\Omega_C$! While the present experiments seem to indicate a substantially higher condensate fraction for $\lit$ as
compared to $\kal$, it remains to be seen if this tendency remains once the detailed relation between the observables
in both experiments and $\Omega_C$ is established.

The broad resonance limit poses the challenge of a strongly interacting quantum field theory. Several issues wait for
an improved treatment. The understanding of the renormalization flow for $h_\phi$ and $\tilde{\lambda}_\psi$ should reveal
the existence of an infrared stable (partial) fixed point and should allow to judge if both $\lit$ and $\kal$
are within the range of attraction of this fixed point. Indeed, the universal value of $\Omega_C(c^{-1}=T=0)$ requires that
for both systems the value of $h_\phi^2$ is sufficiently close to the fixed point value. The functional renormalization
group \cite{AAWetti,CWRG,Tetradis} seems to be an appropriate method to address these questions. This should also include
the effect of molecule fluctuations in the renormalization flow of $h_\phi,\ZpR$ and $\ApR$ as well as the computation
of the missing corrections to the wave function renormalization and gradient term for the fermionic atoms, $Z_\psi$ and
$A_\psi$. Another issue that is not yet settled in a satisfactory way in the present treatment concerns the behavior in the
superfluid phase for $T\to T_c$ and the universal critical behavior in the vicinity of $T_c$. These issues are mainly related
to the running of the molecule interaction $\lambda_\phi$.

Despite the present shortcomings we hope that our functional
integral approach opens the way for a systematic quantitative understanding of the relation between measured
microphysical parameters and macroscopic collective phenomena for ultracold fermionic atoms.
\\ \\

\textbf{Acknowledgement} \\\\
We would like to thank M. Bartenstein, T. Gasenzer, P.S. Julienne, B. Marcelis, H. Stoof and M. Zwierlein for useful
discussions.

\begin{appendix}

\section{Renormalization of Feshbach coupling}
\label{sec:YukRenorm}

So far we have not been very precise about the meaning of the Yukawa coupling $\hpb$. Just as $\bar{\nu}$ it has to
be renormalized, this time with multiplicative renormalization. In the action (\ref{YukawaAction}) appears the
microscopic Yukawa coupling $\bar{h}_{\phi,\Lambda}$ (and similar for $\bar{\lambda}_{\psi,\Lambda}$). In the vacuum
($T=0,n=0$) one can measure the Yukawa coupling by scattering experiments of single atoms. This involves a UV  - renormalized
coupling corresponding to the effective molecule - atom vertex at low momentum. We will use the vacuum value in the atom
phase, $\bar{h}_{\phi,0}$, in order to define a UV - renormalized Yukawa coupling. Finally, in an
atom gas at arbitrary $T,n$ the relation between the gap and $\phi_0$ involves a renormalized Yukawa coupling
$\hpb(T,\sigma)$. Corresponding dimensionless
quantities are denoted $\tilde{h}_{\phi,\Lambda},\tilde{h}_{\phi,0},\hpn (T,\sigma)$. The quantity $\hpn (T, \sigma)$
typically appears in the fermion loop through the gap $\rex$ and we use the shorthand $\hpn$. Actually, the situation is
even more complex since a momentum dependent $\hpn (q)$ appears in the momentum integrals of the Schwinger-Dyson
equations. We will neglect this in a simplified treatment.

\begin{figure}
\begin{minipage}{\linewidth}
\begin{center}
\setlength{\unitlength}{1mm}
\framebox{\begin{picture}(80,38)
       \put (0,-13){
     \makebox(80,38){
     \begin{picture}(80,38)
       %\coordsyst{90}{54.9}{21}{14}
      \put(0,0){\epsfxsize80mm \epsffile{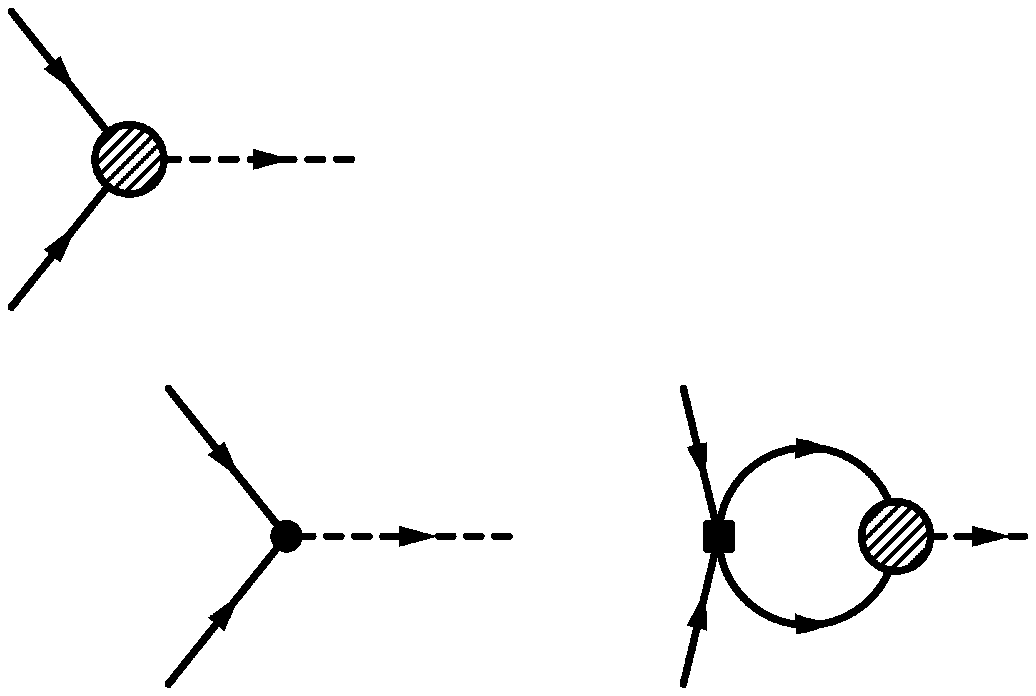}}%with the line vanishing at resonance: BareExp4.eps
   %   \put(20,2){FEYNMAN DIAGRAMS}
%      \put(-3,1){(b)}
      \put(43,21.5){$+$ }
      \put(12,21.5){$=$ }
      \end{picture}
      }}
\end{picture}}
\end{center}
\vspace*{-1.25ex} \caption{Graphical representation of the Schwinger-Dyson equation for the Yukawa or Feshbach coupling.
Fermions are represented by solid, molecules by dashed lines. The vacuum value of the Yukawa coupling, $\bar{h}_{\phi,0}$,
is signalled by the small blob, the full quantity $\hpb$ by the fat shaded blob. The vacuum value of the background
scattering vertex $\bar{\lambda}_\psi$ enters as a small square.}
\label{SDEYuk}
\end{minipage}
\end{figure}

We estimate the UV-renormalization effects for the Yukawa or Feshbach coupling by the use of an appropriate Schwinger-Dyson
equation shown in fig. \ref{SDEYuk}. This yields
\begin{eqnarray}
\hpb &=& \bar{h}_{\phi,\Lambda} - \hpb \bar{\lambda}_{\psi,\Lambda} \bar{J}_{h,\Lambda},\\\nonumber
\bar{J}_{h,\Lambda} &=& \frac{1}{4T}\int\frac{d^3q}{(2\pi^3)}\frac{\tanh\gamma_\phi}{\gamma_\phi}.
\end{eqnarray}
We first evaluate $\bar{J}_{h,\Lambda}$ for $T\to 0,\sigma\to 0$ and find that the integral is proportional to the ultraviolet cutoff
$\Lambda$
\begin{eqnarray}
\bar{J}_{h,\Lambda} (T=0,\sigma=0) = \frac{M\Lambda}{2\pi^2}.
\end{eqnarray}
Thus the relation \footnote{Since the loop integral is dominated by high momenta $q^2\approx \Lambda^2$ we use here
$\bar{\lambda}_{\psi,\Lambda}$ (instead of a more accurate description in terms of a momentum dependent
$\bar{\lambda}_\psi(q^2)$).} between $\bar{h}_{\phi,0}$ and $\bar{h}_{\phi,\Lambda}$ involves $\Lambda$
\begin{eqnarray}
\bar{h}_{\phi,0} = \bar{h}_{\phi,\Lambda} - \hpb \bar{\lambda}_{\psi,\Lambda}\frac{M\Lambda}{2\pi^2}.
\end{eqnarray}
Next we express $\hpb$ in terms of $\bar{h}_{\phi,0}$ instead of $\bar{h}_{\phi,\Lambda}$
\begin{eqnarray}
\hpb = \bar{h}_{\phi,0} - \hpb \bar{\lambda}_\psi \bar{J}_h
\end{eqnarray}
where $\bar{J}_h$ converges now in the ultraviolet
\begin{eqnarray}
\bar{J}_h  =  \frac{1}{4T}\int\frac{d^3q}{(2\pi^3)}\Big[\frac{\tanh\gamma_\phi}{\gamma_\phi}
- \frac{4TM}{q^2}\Big].
\end{eqnarray}
All this is completely analogous to the renormalization of $\bar{\nu}_\Lambda, \bar{\nu}$ discussed in detail in
\cite{Diehl:2005ae}, where the same momentum integral appears (up to a prefactor). The relevant momentum scale for
$\bar{\lambda}_\psi$ is now low, adapted to the momentum range dominating $\bar{J}_h$. We employ $\bar{\lambda}_\psi
= 4\pi a_{bg}/M$ and obtain for the dimensionless coupling ($a_{bg} =-0.38 \mathrm{eV}$ for $\lit$)
\begin{eqnarray}\label{hpRenormGen}
\hpn =  \frac{\tilde{h}_{\phi,0}}{1 + 8\pi a_{bg} k_F \tilde{J}_h}
\end{eqnarray}
with
\begin{eqnarray}\label{YukawaIntegral}
\tilde{J}_h = %- \frac{\tilde{h}_\sigma^2}{\tilde{m}_b^2}\sum_n \Tn \int\frac{d^3\qn}{(2\pi^3)}\frac{1}{P_F(\tilde{Q})P_F(-\tilde{Q})}.
\frac{\bar{J}_h}{2Mk_F} = \frac{1}{4\Tn}\int\frac{d^3\qn}{(2\pi^3)}\Big[\frac{\tanh\gamma_\phi}{\gamma_\phi}
- \frac{2\Tn}{\qn^2}\Big].
\end{eqnarray}

For the symmetric phase a first interesting limit concerns the BEC regime for $\sigex \to -\infty$ where
\begin{eqnarray}\label{hpBEC}
\lim\limits_{\sigex\to -\infty} \tilde{J}_h  = - \frac{\sqrt{-\sigex}}{8\pi}
\end{eqnarray}
such that
\begin{eqnarray}
\hpn = \tilde{h}_{\phi,0} \frac{1}{1 - a_{bg} k_F \sqrt{-\sigex}}.
\end{eqnarray}
A second limit relevant for the BCS regime takes $\Tn\to 0$ for $\sigex=1$,
\begin{eqnarray}
\lim\limits_{\Tn \to 0} \tilde{J}_h (\sigex  =1)  \propto \ln \Tn .
\end{eqnarray}
These two different limits have important consequences for the value of the renormalized Yukawa coupling in vacuum. For the
atom phase (positive binding energy $\epsilon_M$) we conclude that the limit $k_F\to 0, T\to 0$ discussed in sect.
\ref{sec:lowdens} implies a vanishing of the correction $\sim k_F \tilde{J}_h$ in eq. (\ref{hpRenormGen}) such that the
vacuum value of the Yukawa coupling is indeed given by $\tilde{h}_{\phi,0}$. In contrast, for the molecule phase ($\epsilon_M
<0$) the limit (\ref{hpBEC}) applies. With $\epsilon_M = 2 \sigma_A = - 1/(M\bar{a}^2(\sigma_A))$ (\ref{EpsPropNu2}) one
finds an enhancement of the vacuum Yukawa coupling for $a_{bg} >0$
\begin{eqnarray}\label{vaccoup}
\tilde{h}_{\phi,0}^{(m)} &=& \frac{\tilde{h}_{\phi,0}}{1-a_{bg} \sqrt{-\epsilon_M M}} =
\frac{\tilde{h}_{\phi,0}}{1-a_{bg} /\bar{a}(\sigma_A)}\nonumber\\
&=& \frac{\tilde{h}_{\phi,0}}{1-a_{bg}k_F c^{-1}}.
\end{eqnarray}
This correction vanishes at resonance.

We may also evaluate eq. (\ref{hpRenormGen}) in the superfluid phase. Using the standard BCS gap equation
((\ref{Cross1SSB}) with $\Omega_M =0$), one obtains
\begin{eqnarray}\label{hphiRen2}
\hpn =  \frac{\tilde{h}_{\phi,0}}{1 - a_{bg} k_F c^{-1}}.
\end{eqnarray}
This formula should be reliable for low $T$ where indeed $\Omega_M$ small. Eq. (\ref{hphiRen2}) smoothly connects
to the vacuum, eq. (\ref{vaccoup}). Despite several shortcomings in the derivation of (\ref{vaccoup}), mainly due to
the neglect of the momentum dependence, we believe that the relation (\ref{vaccoup}) should hold approximately
also for a more reliable treatment as, e.g., the functional renormalization group equations.

\section{Gap equations}
\label{app:appb}

\subsection{Normal phase}
\label{sec:NormalPhase}
In the normal or symmetric phase the superfluid gap vanishes, $\rhoR_0=0$, and therefore the mass matrix (\ref{Mass2})
becomes degenerate.
The value of $m_\phi^2$ is determined by a gap equation ($\Omega_M$ depends on $m_\phi^2$, but not on $\lambda_\phi$),
\begin{eqnarray}\label{Cross1SYM}
m_\phi^2 = m_\phi^{(F)\, 2} + \frac{1}{3\pi^2}\lambda_\phi\hF\Omega_M
\end{eqnarray}
with
\begin{eqnarray}
m_\phi^{(F)\, 2} &=& \nu + \Delta m_\phi^{(F)\, 2},\nonumber\\
\end{eqnarray}
evaluated at $\rho_0=0$. Here we have introduced a renormalized detuning $\nu=\nun/\ZpR$ and work in the broad resonance
limit such that the term $-2\sigma/\ZpR = \mathcal{O}(\hpn^{-2})$ in the classical contribution to the boson mass can be
neglected. Explicitly, we use
\begin{eqnarray}\label{someIntegrals}
\Delta m_\phi^{(F)\, 2} &=& \frac{\partial\tilde{u}_1\hF}{\partial\rhoR} \\\nonumber
&=&- \frac{h_\phi^2}{4\Tn} \int\frac{d^3\qn}{(2\pi)^3}\big[\gamma^{-1} \tanh\gamma -2\Tn\qn^{-2}\big],\label{DeltaMPhi}\nonumber\\
\lambda_\phi\hF &=& \frac{\partial^2\tilde{u}_1\hF}{\partial\rhoR^2}\nonumber\\
&=&  \frac{h_\phi^4}{32\Tn^3}\int\frac{d^3\qn}{(2\pi)^3}\gamma^{-3}\big[\tanh\gamma-\gamma\cosh^{-2}\gamma
\big].\nonumber \label{BosNumber}
\end{eqnarray}

The effective chemical potential $\sigex$ is determined by the density equation $\Omega_{F,0} + \Omega_M =1$ ($\Omega_{F,0}
=\Omega_F$ in the symmetric phase), with
\begin{eqnarray}
\Omega_{F,0} &=& 6\pi^2\int \frac{d^3\qn}{(2\pi)^3}\big[\exp 2\gamma + 1\big]^{-1},\label{OmegF}\\
\Omega_M &=& 6\pi^2\int\frac{d^3\qn}{(2\pi)^3} \Big(\mathrm{e}^{(\ApR \qn^2 + m_\phi^2)/\Tn} - 1\Big)^{-1}\nonumber\\
      &=& \frac{3\Gamma(3/2)}{2}\Big(\frac{\Tn}{\ApR}\Big)^{3/2}\mathrm{Li}_{3/2} \big(e^{-m_\phi^2/\Tn}\big)\label{OmegM}\nonumber.
\end{eqnarray}
The polylogarithmic function $\mathrm{Li}_k(z)$ obeys
\begin{eqnarray}
\mathrm{Li}_k(z) &=& \sum\limits_{n=1}^{\infty}\frac{z^n}{n^{k}}.
\end{eqnarray}
For the computation of $\Omega_M$ we finally need $\ApR$ which reads in the broad resonance limit
\begin{eqnarray}
\ApR &=& \frac{h_\phi^2}{48\Tn^3}\int \frac{d^3\qn}{(2\pi)^3}\,\,\qn^2\gamma^{-3}\big[\tanh\gamma-
\gamma\cosh^{-2}\gamma\big].\nonumber\\
\end{eqnarray}
All displayed expressions are effectively $\hpn$ - independent in the broad resonance limit since the renormalized
Yukawa coupling $h_\phi = \ZpR^{1/2} \hpn$ becomes independent of $\hpn$ for $\hpn\to \infty$. Using eq. (\ref{SYMDens}) we
can relate $\nu$ to the concentration $c$
\begin{eqnarray}\label{cInvariance}
\frac{\nu}{h_\phi^2} = \frac{\nun}{\hpn^2} = -\frac{1}{8\pi c}.
\end{eqnarray}
With the condition $\Omega_{F,0} + \Omega_M =1$ we can determine $\sigex$ and $m_\phi^2$ self-consistently.

A second Schwinger-Dyson equation is obtained for the self-consistent determination of the dimensionless renormalized
four-boson coupling $\lambda_\phi$ \cite{Diehl:2005ae}, namely
\begin{eqnarray}\label{LambdaPhi}
\lpR \hspace{-0.15cm}&=& \frac{\lpR\hF}{1 - I_\lambda},\\\nonumber
I_\lambda \hspace{-0.15cm}&=& - \frac{3\lpR\hF}{2\Tn}\hspace{-0.15cm}\int\hspace{-0.15cm}\frac{d^3\qn}{(2\pi)^3}\alpha^{-1}\big[(\exp 2\alpha - 1\big)^{-1}%\nonumber\\ &&\qquad
+ 2 \alpha \sinh^{-2}\alpha \big],\\\nonumber
\alpha \hspace{-0.15cm}&=& \frac{\ApR\qn^2 + m_\phi^2}{2\Tn}.
\end{eqnarray}
Approaching the critical temperature from above, $m_\phi^2(T\to T_c) \to 0$, this leads to an infrared divergence of the
fluctuation integral $I$ in the second line. In turn, this will lead to a behavior $\lambda_\phi(T\to T_c) \propto m_\phi
\to 0$. This ``infrared freedom'' at the phase transition is a generic feature of bosonic systems.

\subsection{Superfluid phase}
The superfluid phase is characterized by a nonvanishing superfluid order parameter $\rhoR_0$ or (squared) gap
$\tilde{r} = h_\phi^2\rhoR_0$. The minimum condition yields a gap equation in the broad resonance limit
\begin{eqnarray}\label{Cross1SSB}
\nu + \Delta m_\phi^{(F)\, 2} + \frac{1}{3\pi^2}\lambda_\phi\hF \Omega_M =0
\end{eqnarray}
which expresses the vanishing of the full mass matrix term $m_\phi^2 = 0$ (Goldstone theorem). The fluctuation integrals
entering (\ref{Cross1SSB}) are the same as in (\ref{someIntegrals}), with the only difference that they have to be evaluated
at $\rho_0\neq 0$. This makes these expressions more complicated. For their explicit form, we refer to \cite{Diehl:2005ae}.
Dividing eq. (\ref{Cross1SSB}) by $h_\phi^2$ we find
\begin{eqnarray}\label{Cross1SSB2}
0 &=& - \frac{1}{8\pi c}  +  \frac{1}{h_\phi^2}\Big(\Delta m_\phi^{(F)\, 2} \\\nonumber
&&+ 2\lambda_\phi\hF \int\frac{d^3\qn}{(2\pi)^3} \frac{\ApR\qn^2+ \lpR\rhooR/2}{\sqrt{\ApR\qn^2 +2\lpR\rhooR}}\\
&&\quad\times \big(\exp \sqrt{\ApR\qn^2 +2\lpR\rhooR}/\Tn -1 \big)^{-1}.\nonumber
\end{eqnarray}
Note that without the bosonic fluctuation contribution $\propto \Omega_M$, eq. (\ref{Cross1SSB2}) is just the usual BCS gap
equation as obtained from a purely fermionic approach, since the factor $h_\phi^{-2}$ cancels the corresponding factor in
(\ref{DeltaMPhi}). Due to the additional term $\propto \lambda_\phi\hF$ our approximation scheme goes beyond the
presently available purely fermionic approaches even in the broad resonance limit, since it offers the possibility to
include fluctuations of the dressed boson field as well! However, the quantitative treatment of the bosonic fluctuations
still involves substantial uncertainties, especially for $T$ near $T_c$ in the superfluid phase. In general, the boson
fluctuations are expected to affect our
quantitative results mainly close to the critical temperature. At very low $\Tn$, they should be negligible. Therefore,
our results should be reliable in this regime, which is confirmed by the matching of recent QMC results \cite{Carlson03}
with our approach \cite{Diehl:2005ae}.

The atom densities obey
\begin{eqnarray}\label{Cross2SSB}
\Omega_F + \Omega_M + \Omega_C = \Omega_{F,0} + \Omega_M + \bar{\Omega}_C = 1.
\end{eqnarray}
For a broad resonance the condensate fraction $\Omega_C = n_C/n = 6\pi^2\rhoR_0 = 6\pi^2 \rex/h_\phi^2$ becomes independent
of the the value of $\hpn$. For the actual computations we use
\begin{eqnarray}
\Omega_{F,0} &=&  - 3\pi^2 \frac{\partial \tilde{u}_1\hF}{\partial\sigex} \nonumber\\\nonumber
&=& -3\pi^2\int \frac{d^3\qn}{(2\pi)^3}\big(\frac{\gamma}{\gamma_\phi}\tanh\gamma_\phi -1\big),\\
\Omega_M &=& 3\pi^2  \int\frac{d^3\qn}{(2\pi)^3}\Big\{\frac{\ApR \qn^2 + \lambda_\phi\rhoR_0}{
\sqrt{\ApR \qn^2(\ApR \qn^2 + 2\lambda_\phi\rhoR_0)}}\nonumber\\
&&\times\coth\frac{\sqrt{\ApR \qn^2(\ApR \qn^2 + 2\lambda_\phi\rhoR_0)}}{2\Tn} - 1\Big\},\nonumber\\
\bar{\Omega}_C &=& 6\pi^2\frac{\rex}{\hpn^2},
\end{eqnarray}
where $\bar{\Omega}_C$ can be safely neglected for a broad resonance.
Since $\Omega_M$ depends on $\lambda_\phi$ in the superfluid phase, we need another gap equation for the four-boson
coupling $\lambda_\phi$. We find $\lambda_\phi =0$ for the infrared value in the superfluid phase, and we refer to
\cite{Diehl:2005ae} for a discussion of how we actually deal with this coupling.

%In the superfluid phase, the solution of the system (\ref{Cross1SSB},\ref{Cross2SSB}) amounts in the self-consistent
%determination of $\sigex, \rhoR_0$ resp. $\rex$.

\subsection{Phase boundary}
At the phase boundary, the system is governed by particularly simple equations since $m_\phi^2 = \lambda_\phi = \rhoR_0=0$.
The vanishing of $m_\phi^2$ implies the constraint
\begin{eqnarray}\label{Cross1PB}
-\frac{1}{8\pi c} + \frac{1}{h_\phi^2}\Big(\Delta m_\phi^{(F)\, 2} + \frac{1}{3\pi^2}\lambda_\phi\hF\Omega_M \Big) =0
\end{eqnarray}
and the density equation can be expressed as
\begin{eqnarray}\label{Cross2PB}
 \Omega_{F,0} + \Omega_M = 1.
\end{eqnarray}
The fermionic integrals for $\Delta m_\phi^{(F)\, 2}, \lambda_\phi\hF, \Delta \ApR$ are the same as in eq.
(\ref{someIntegrals}) with $\Omega_M$ simplified to
\begin{eqnarray}
\Omega_M = \frac{3\Gamma(3/2)\zeta(3/2)}{2}\Big(\frac{\Tn}{\ApR}\Big)^{3/2}\label{OmegMPB}.
\end{eqnarray}
At the phase boundary, the solution of the system (\ref{Cross1PB},\ref{Cross2PB}) amounts to the self-consistent
determination of $\sigex, \Tn_c$. The critical line therefore indeed depends only on $c^{-1}$ in the broad resonance limit
$\hpn \to \infty$.

The two lines representing the critical temperature in fig. \ref{CrossoverTcAll} are obtained by either including (short
dashed) or omitting (solid) the boson fluctuations in eq. (\ref{Cross1PB}). Other current approaches, e.g.
\cite{Stoof05,IIChen05,YPieri}, always work with the standard gap equation omitting the boson fluctuations.

\section{Dispersion relation for molecules in magnetic field}
\label{app:dispersion}

Due to the interactions between unbound atoms and molecules the relation between energy and momentum (or frequency and wave
vector) for
single molecules differs from the case of free molecules of mass $2M$ where $\omega = q^2/4M$. We can extract the
dispersion relation $\omega (\vec{q})$ from our approximation to the full molecule propagator $\bar{\mathcal{P}}_\phi(Q)=
\bar{\mathcal{P}}_\phi(q_0,\vec{q})$ in vacuum, where $n=T=0$. In the absence of spontaneous symmetry breaking we work in the
$\phi^*, \phi$ basis where $\bar{\mathcal{P}}$ multiplies the unit matrix. The dispersion relation $\omega(q)$ follows from
\begin{eqnarray}\label{C1eq}
\bar{\mathcal{P}}_\phi(\mathrm{i}\omega, q) = 0.
\end{eqnarray}
For the ``molecule phase'' ($\epsilon_M < 0$) one has to use $\sigma = \sigma_A$ with $\epsilon_M = 2\sigma_A$ such that
$\bar{\mathcal{P}}_\phi(0,0) =0$, whereas for the atom phase ($\epsilon_M>0$) one employs $\sigma =0$ and $\bar{\mathcal{P}}_\phi(0,0)
= \bar{m}_\phi^2$. For the classical propagator
\begin{eqnarray}
\bar{\mathcal{P}}_\phi = \mathrm{i}q_0 + \frac{q^2}{4M}
\end{eqnarray}
we recover the classical dispersion relation $\omega = q^2/4M$.

In the broad resonance limit, the molecule propagator is, however, strongly dominated by fluctuation effects, generated by
atom exchange processes. In the molecule phase ($\epm<0$) care is needed for the
correct atom propagator $P_F^{-1}$ \footnote{We neglect possible modifications of $P_F$ due to a nontrivial wave function
renormalization $Z_\psi$ and gradient coefficient $A_\psi$.}. In a nonvanishing magnetic field the energy of the atoms as compared to the zero level
defined by the molecules is $E_A = -\sigma_A + \delta \tau_3/2+ q^2/2M$. Here $\delta$ accounts for the level splitting
between the two atom levels ($\tau_3$ is the Pauli matrix) and for a chemical potential related to the difference in
atom numbers in the two levels. For an equal mixture of atoms in the two levels one has $\delta=0$.
In the atom phase one has $\sigma_A=0$.

We can now write
$\bar{\mathcal{P}}_\phi(\mathrm{i}\omega , q) =
-\omega + q^2/4M + \Delta \bar{\mathcal{P}}_\phi + \bar{m}_\phi^2$. We concentrate first on the ``molecule phase'' $\epsilon_M < 0$
where $\bar{m}_\phi^2 =0$ and
\begin{eqnarray}
\Delta \bar{\mathcal{P}}_\phi \hspace{-0.15cm}&=&\hspace{-0.1cm} -\hpb^2\hspace{-0.1cm}\int\hspace{-0.1cm}\frac{d^4q'}{(2\pi)^4}
\Big\{\hspace{-0.08cm}\Big[\frac{(q' -q/2)^2}{2M}\hspace{-0.05cm} - \sigma_A +
\frac{\delta}{2} + \mathrm{i}q'_0 - \frac{\omega}{2}\Big]^{-1}\nonumber\\
&&\hspace{-0.25cm}\Big[\frac{(q' +q/2)^2}{2M} - \sigma_A\hspace{-0.1cm}- \frac{\delta}{2}- \mathrm{i}q'_0 -
\frac{\omega}{2}\Big]^{-1}\hspace{-0.35cm} \nonumber\\
&& - \Big(\Big(\frac{q'^2 }{2M}- \sigma_A\Big)^2 + q'^2_0  \Big)^{-1}\Big\}.
\end{eqnarray}
Here, the last piece subtracts the piece for $q=0$, $\omega=0$, $\delta=0$ which is already included in the definition of
$\bar{m}_\phi^2$. In particular, it contains the renormalization (\ref{Watwees}) and therefore provides for an UV -
regularization of the momentum integral. With $\hat{\omega} = M (\omega + 2 \sigma_A)$, $\hat{\delta}= M\delta$, $\hat{q}_0
= 2Mq'_0$ one finds the UV finite integral
\begin{eqnarray}\label{DelPphiInt}
\Delta \bar{\mathcal{P}}_\phi \hspace{-0.15cm}&=& \hspace{-0.15cm}-\frac{\hpb^2 M}{\pi}\hspace{-0.15cm}\int\hspace{-0.15cm}
\frac{d^3q'}{(2\pi)^3}\hspace{-0.15cm}\int \hspace{-0.15cm}d\hat{q}_0\big\{\big[(q' -q/2)^2 - \hat{\omega}
 +\hat{\delta} + \mathrm{i}\hat{q}_0\big]^{-1} \nonumber\\
&& \times\big[(q' + q/2)^2 - \hat{\omega} -\hat{\delta} - \mathrm{i}\hat{q}_0\big]^{-1}\nonumber\\
&&- \big[(q'^2 -2M\sigma_A)^2 + \hat{q}_0^2\big]^{-1}\big\}.
\end{eqnarray}
The $\hat{q}_0$ integration yields
\begin{eqnarray}\label{D5}
\Delta \bar{\mathcal{P}}_\phi\hspace{-0.15cm} &=&\hspace{-0.15cm} -\hpb^2M\hspace{-0.15cm}\int\hspace{-0.15cm}\frac{d^3q'}{(2\pi)^3}
\big\{(q'^2 + q^2/4 - \hat{\omega})^{-1}\hspace{-0.15cm} \nonumber\\
&&\qquad - (q'^2 - 2M\sigma_A))^{-1}\big\}\\\nonumber
\end{eqnarray}
and we obtain for $\omega + 2\sigma_A <q^2/(4M)$
\begin{eqnarray}\label{C5A}
\Delta \bar{\mathcal{P}}_\phi&=& \frac{\hpb^2 M^{3/2}}{4\pi}\big(\sqrt{q^2/4M -2\sigma_A - \omega} - \sqrt{-2\sigma_A}\big).\nonumber\\
\end{eqnarray}
For $\hat{\omega}=0$ the result is nonanalytic in $q^2$, $\Delta \bar{\mathcal{P}}_\phi \sim
\sqrt{q^2}$. For negative $\hat{\omega}$, however, the infrared divergence in $\partial\Delta \bar{\mathcal{P}}_\phi/\partial q^2$ is
regulated.

For $\omega = q^2/(4M)$ both $\Delta\bar{\mathcal{P}}_\phi$ and the classical contribution to
$\bar{\mathcal{P}}_\phi$ vanish. This zero of $\bar{\mathcal{P}}_\phi$ (\ref{C1eq}) then defines
the dispersion relation for the molecules
\begin{eqnarray}\label{ClassDisp}
\omega = \frac{q^2}{4M}.
\end{eqnarray}
We recover the dispersion relation for fundamental pointlike nonrelativistic molecules of mass $2M$. In the broad
resonance limit the composite molecules just behave as pointlike spin 0 bosons with mass $2M$, even if the microscopic
interaction between the atoms is completely pointlike and truly microscopic molecules play no role. (In this limit
$\bar{\mathcal{P}}_\phi$ is approximated by $\Delta\bar{\mathcal{P}}_\phi$.)

We can recover the relation (\ref{ClassDisp}) by an expansion of $\Delta\bar{\mathcal{P}}_\phi$ in $q^2$ and $\omega$ where $\bar{\mathcal{P}}_\phi
= - \ZpR \omega + \bar{A}_\phi q^2$ with
\begin{eqnarray}\label{ApbMik}
\Apb &=& \frac{\partial\bar{\mathcal{P}}_\phi}{\partial (q^2)}\Big|_{q^2=\omega=0} = \frac{1}{4M} + \frac{\hpb^2M}{32\pi}\frac{1}{\sqrt{-\hat{\omega}}}\Big|_{\omega=0}\\\nonumber
&=& \frac{1}{4M}\Big(1+\frac{\hpb^2}{8\pi}\frac{M^{3/2}}{\sqrt{-2\sigma_A}}\Big)
= \frac{1}{4M}\Big(1+\frac{1}{32\pi}\frac{\tilde{h}_\phi^2}{\sqrt{-\sigex_A}}\Big),\\\nonumber
\ZpR &=&  -\frac{\partial\bar{\mathcal{P}}_\phi}{\partial \omega}\Big|_{q^2=\omega=0}
= \Big(1+\frac{\hpb^2}{8\pi}\frac{M^{3/2}}{\sqrt{-2\sigma_A}}\Big) \nonumber\\
&=& 1 + \frac{1}{32\pi}\frac{\hpn^2}{\sqrt{-\sigex_A}}.
\end{eqnarray}
The dispersion relation reads now $\omega= (\Apb/\ZpR) q^2$ and coincides with eq. (\ref{ClassDisp}).

The dispersion relation is independent of the binding energy $\epsilon_M = 2\sigma_A$. Hence we can use the dispersion
relation even in the vicinity of the resonance, $\epsilon_M\to 0$, where both $\Apb$ and $\ZpR$ diverge. In our
approximation the dispersion relation is also independent of $\hpb$. We observe that
in eq. (\ref{ApbMik}) we have defined $Z_\phi$ by the term linear in $\omega$ in a Taylor expansion of $\bar{\mathcal{P}}_\phi$.
The result agrees with eqs. (\ref{FuncZphi},\ref{ZpRlowdens}) where $\ZpR$ is defined somewhat differently. On the other
hand, for the atom scattering discussed in sect. \ref{sec:scattvac}, eq. (\ref{LamEff}) we rather should employ a definition
\begin{eqnarray}\label{C9}
\ZpR(\omega) &=& -\frac{\mathcal{\bar{P}}_\phi}{\omega}\Big|_{\omega = -\epm, q^2=0}
= 1 + \frac{\hpb^2M^{3/2}}{4\pi |\epm|}
\end{eqnarray}
which differs by a factor two from eq. (\ref{ZpRlowdens}). Of course, the dispersion relation (\ref{ClassDisp}) is not
affected by the precise definition of $\ZpR$.

We next turn to the ``atom phase'' for $\epsilon_M>0$, where $\sigma_A =0$ and $\bar{m}_\phi^2 \geq 0$. As long as
$\omega < q^2/(4M)$ we can employ eq. (\ref{C5A}) with $\sigma_A=0$. For $\omega = q^2/(4M)$ the fluctuation correction
$\Delta \bar{\mathcal{P}}_\phi$ vanishes. For $\omega > q^2/(4M)$, however, the integral (\ref{DelPphiInt}) becomes purely
imaginary such that
\begin{eqnarray}
\bar{\mathcal{P}}_\phi &=& -\omega + \bar{m}_\phi^2 + \frac{q^2}{4M} + \mathrm{i} \bar{\Gamma} ,\nonumber\\
\bar{\Gamma} &=& \frac{\hpb^2M^{3/2}}{4\pi} \Big( \omega - \frac{q^2}{4M}\Big)^{1/2}.
\end{eqnarray}
Indeed, the correction $\Delta \bar{\mathcal{P}}_\phi$ accounts now for a nonzero decay width of the molecule - the resonant state
can decay into two atoms if its energy $\omega$ exceeds twice the kinetic energy of an unbound atom with momentum $q/2$,
i.e. $\omega >2\cdot (q/2)^2/(2M) = q^2/(4M)$. At the threshold $\omega = q^2/(4M)$ the decay width $\bar{\Gamma}$ vanishes.

Obviously, it is not possible to find a value $\omega (q^2)$ where $\bar{\mathcal{P}}_\phi$ vanishes identically. For
$\omega\leq q^2/(4M)$ one has $\Delta\bar{\mathcal{P}}_\phi \geq 0$ and therefore $\bar{\mathcal{P}}_\phi$ is strictly positive
for $\bar{m}_\phi^2 >0$. For $\omega >q^2/(4M)$ the nonzero $\bar{\Gamma}$ prevents a vanishing $\bar{\mathcal{P}}_\phi$. The
location of the resonance may be defined by the zero of the real part of $\bar{\mathcal{P}}_\phi$. Similarly, we may define
$\ZpR$ by the $\omega$ - dependence of the real part of $\bar{\mathcal{P}}_\phi$, i.e. $\ZpR (\omega) = -\mathrm{Re}
(\bar{\mathcal{P}}_\phi)/\omega$ or $\ZpR = -\partial\mathrm{Re}(\bar{\mathcal{P}}_\phi)/\partial\omega(\omega=0)$.
For the atom phase, our approximation yields $\ZpR = 1$ since $\Delta\bar{\mathcal{P}}_\phi$ is purely imaginary.
Associating the ``binding energy'' $\epm >0$ with the location of the resonance defined in this way yields
\begin{eqnarray}
\epm = \frac{\bar{m}_\phi^2}{\ZpR} = \bar{m}_\phi^2.
\end{eqnarray}
We note that for a resonance the precise definition of the location of the resonance remains somewhat arbitrary. For example,
our definition of $\epm$ does not coincide with the minimum of $|\bar{\mathcal{P}}_\phi|$. Due to the $\omega$ - dependence of
$\bar{\Gamma}$ this rather occurs for $\omega = \max (\omega_{min},0)$ where $\omega_{min}$ is defined by
\begin{eqnarray}
\frac{\partial}{\partial \omega}\,|\bar{\mathcal{P}}_\phi|^2 &=& 2\big(\omega_{min} - \frac{q^2}{4M} -\bar{m}_\phi^2\big) + \frac{\hpb^4M^3}{16\pi^2}
=0,\nonumber\\
\omega_{min} &=& \bar{m}_\phi^2 + \frac{q^2}{4M} - \frac{\hpb^4M^3}{32\pi^2}.
\end{eqnarray}
For $q^2 =0$, this yields $\omega_{min} = \epm - \hpb^4M^3/(32\pi^2)$. The maximum of the cross section is given by the
maximum of $\bar{\Gamma}(s)/|\bar{\mathcal{P}}_\phi|^2$ (cf. eq. (\ref{BreitWigner})) which occurs for $\omega = s_{max}$ given by
\begin{eqnarray}
&&\frac{\partial \bar{\Gamma}}{\partial \omega} \,|\bar{\mathcal{P}}_\phi|^2  = \frac{\partial|\bar{\mathcal{P}}_\phi|^2
\bar{\Gamma}}{\partial \omega},\nonumber\\
&&3\omega^2 - 2\epm \omega + \bar{\Gamma}^2 - \epm^2 =0.
\end{eqnarray}

\section{Atom Scattering in Vacuum}
\label{sec:scattvac}

In this appendix we discuss the scattering of atoms in vacuum \footnote{The ``vacuum'' means here the state with $n\to 0,
T\to 0$ and includes the presence of a homogeneous magnetic field $B$.} and extract the scattering length. Once the
effective action $\Gamma$ is computed it contains directly the information on the one particle irreducible Green functions
including corrections from fluctuations. Therefore the scattering amplitudes can be extracted without much further
calculation. Contact to the observable parameters can directly be established in the $T\to 0$, $n\to 0$ limit of our
approach.

The amplitude for the elastic scattering of two atoms corresponds to the effective four-fermion amplitude. In presence
of a field for the molecules we have two contributions: from the tree exchange of molecules and from the one-particle irreducible
four-fermion vertex. The tree exchange contribution can be obtained formally by solving the field equation for
$\phi$ for nonvanishing $\psi\epsilon\psi$. This solution becomes then a functional of
$\psi$. Reinserting this solution into $\Gamma$ yields the ``tree contributions'' to the four-fermion amplitude.
\begin{figure}
\begin{minipage}{\linewidth}
\begin{center}
\setlength{\unitlength}{1mm}
\begin{picture}(85,54)
      \put (0,0){
    \makebox(80,49){
     \begin{picture}(80,49)
       %\coordsyst{90}{54.9}{21}{14}
      \put(0,0){\epsfxsize80mm \epsffile{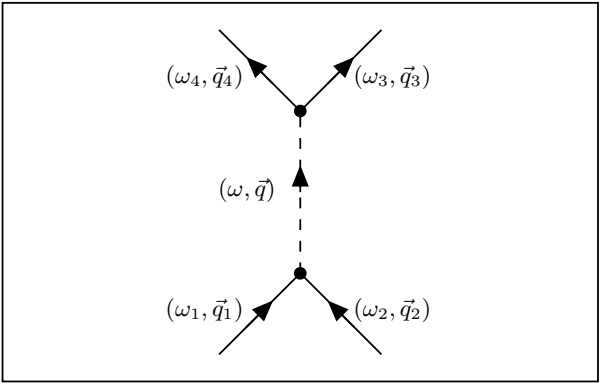}}
      \put(22,9){$(\omega_1,\vec{q}_1)$ }
      \put(47,9){$(\omega_2,\vec{q}_2)$ }
      \put(29,25){$(\omega,\vec{q})$ }
      \put(22,40){$(\omega_4,\vec{q}_4)$ }
      \put(47,40){$(\omega_3,\vec{q}_3)$ }
     \end{picture}
      }}
   \end{picture}
\end{center}
\vspace*{-1.25ex} \caption{Scattering with tree exchange of a molecule.}
\label{TreeExchange}
\end{minipage}
\end{figure}

Near the resonance the dominant piece arises from the exchange of molecules, as depicted in fig. \ref{TreeExchange}. With
$\vec{q}_1 =-\vec{q}_2$, $q_1^2/2M = q_2^2/2M = s/2$, $\vec{q}=\vec{q}_1 + \vec{q}_2 = 0$, $\omega= \omega_1 + \omega_2$
one obtains an effective interaction
\begin{eqnarray}\label{LamEff}
\bar{\lambda}_{eff}(s) = -\frac{\hpb^2}{\bar{\mathcal{P}}_\phi} +\bar{\lambda}_\psi = \frac{\hpb^2}{\ZpR( s -
\mathrm{i}\Gamma(s)-\epsilon_M) }
+\bar{\lambda}_\psi. \nonumber\\
\end{eqnarray}
This formula is valid both for positive and negative $\epsilon_M$. The first piece involves the inverse Minkowski -
propagator $\bar{\mathcal{P}}_\phi^{-1}(\mathrm{i}\omega,q)$ which is discussed in detail in app. \ref{app:dispersion}.
For positive $\epsilon_M$ the molecule state corresponds to a resonance with inverse propagator $\bar{\mathcal{P}}_\phi = -\ZpR\omega +
\Apb q^2 + \bar{m}_\phi^2 +\mathrm{i}\bar{\Gamma}(s)$ and $\omega_{1,2} =s/2$, $\omega =s$. In this case one has
(\ref{EpsNuR}) $\bar{m}_\phi^2= \bar{\nu} = \ZpR\epsilon_M $ and we use $\Gamma(s) = \bar{\Gamma}(s) /\ZpR$.
For $\epsilon_M <0$ the vacuum corresponds to $\sigma= \sigma_A$ and the inverse molecule propagator becomes
$\bar{\mathcal{P}}_\phi = -\ZpR\omega + \Apb q^2 +\mathrm{i}\bar{\Gamma}(s)$. On the other hand for the atoms on shell
(with $\sigma = \sigma_A$) one has now $\omega_{1,2} = (s -\epsilon_M)/2$, $\omega = s -\epsilon_M$,
resulting in eq. (\ref{LamEff}).

For low momentum scattering one can neglect $s$. For the molecule phase with $\epm<0$ and for large $\hpb$ we neglect
the ``direct part'' from $\bar{\lambda}_{\psi,0}$ and find
\begin{eqnarray}
a = - \frac{M}{4\pi} \frac{(\bar{h}_{\phi,0}^{(m)})^2}{\ZpR\epm}.
\end{eqnarray}
Comparison with the definition of $\bar{a}$ in eq. (\ref{CoupScLength}) yields
\begin{eqnarray}
\frac{a}{\bar{a}} = \frac{\bar{\nu}- \epm}{\ZpR\epm}.
\end{eqnarray}
Here we have to evaluate $\ZpR$ at $\omega=-\epm, \vec{q}=0$. Using the result (\ref{C9}) in app. \ref{app:dispersion} one
finds
\begin{eqnarray}
\ZpR (\omega=- \epm) = 1 + \frac{\bar{h}_{\phi,0}^2}{4\pi}\, M^{3/2}\epm^{-1/2}.
\end{eqnarray}
This can be compared to $\bar{\nu} - \epm$ extracted from eq. (\ref{SigmaAEq})
\begin{eqnarray}
\frac{\bar{\nu}- \epm}{\epm}= \frac{\bar{h}_{\phi,0}^2}{4\pi}\, M^{3/2}\epm^{-1/2}.
\end{eqnarray}
We conclude that for large $\bar{h}_{\phi,0}^2M^{3/2}\epm^{-1/2}$ the physical scattering length
$a$ coincides with $\bar{a}$
\begin{eqnarray}\label{120}
\bar{a} = a(1 +  4\pi \bar{h}_{\phi,0}^{-2}M^{-3/2} \epm^{1/2}).
\end{eqnarray}
In the broad resonance limit we therefore identify $\bar{a}$ with the physical scattering length.

We find from eqs. (\ref{EpsPropNu2}), (\ref{GZ}), (\ref{GD}) the familiar approximate form
\begin{eqnarray}\label{PhysScatt}
\bar{a}(\sigma_A)  &=& a_{bg}\Big(1 + \frac{\Delta}{B - B_0}\Big),\\\nonumber
\Delta &=& -a_{bg}^{-1}(M\beta)^{-1/2}.
\end{eqnarray}
For $\lit$ this yields $\Delta = 300 \mathrm{G} = 5.86 \mathrm{eV}^2$, with $a_{bg}=-1405 a_B =-0.38\mathrm{eV}^{-1}$.

We next turn to the atom phase where $\epm >0$. For a first discussion we also omit the decay width $\Gamma$ which
vanishes $\propto \sqrt{s}$ for $s\to 0$. The effective interaction becomes pointlike
\begin{eqnarray}
\bar{\lambda}_{eff} = -\frac{\hpb^2}{\ZpR\epsilon_M} + \bar{\lambda}_{\psi,0}
\end{eqnarray}
and we see that $\ZpR\epsilon_M$ replaces $(\bar{\nu}_\Lambda-2\mu)$ in the ``microscopic vertex'' (\ref{BosonCond2}).
In the atom phase the physical scattering length reads
\begin{eqnarray}
a(B) = \frac{\bar{\lambda}_{eff}M}{4\pi} = -\frac{\bar{h}_{\phi,0}^2M}{4\pi \ZpR\epsilon_M(B)} +
\frac{\bar{\lambda}_{\psi,0}M}{4\pi}.
\end{eqnarray}
The first part corresponds to $a_{res}$ (recall $\ZpR \epm = \bar{\nu} = \bar{m}_\phi^2$) and it is suggestive that the
second may now correspond to $a_{bg}$.

The situation is less satisfactory as for the molecule phase. In fact, $\bar{a} = -\bar{h}_{\phi,0}^2M/(4\pi\bar{\nu})$
accounts now only for the resonant part of the scattering amplitude, resulting in the relation
\begin{eqnarray}\label{eq120}
\bar{a} = (M\beta)^{-1/2} (B_0 - B)^{-1}.
\end{eqnarray}
In the broad resonance limit the interaction is pointlike and we may absorb the background scattering
$\bar{\lambda}_{\psi,\Lambda}$ into a more complicated $B$ - dependence of $\hpb^2$. This would result in the relation
(\ref{PhysScatt}) also for $\bar{a}$ in the atom phase. The $B$ - dependence of $\bar{a}$ or $c$ therefore depends on the splitting
of the microscopic interaction into the two pieces $\bar{\lambda}_{\psi,\Lambda}$ and $- \hpb^2/\bar{\nu}_\Lambda$. Since
all our results depend only on $c$ the $B$ - dependence of the observables is now affected by an unphysical ``Fierz
ambiguity'' \cite{Jaeckel02}. This is clearly a shortcoming of the approximation, in particular of our neglection of the
molecule fluctuations for the computation of $\ZpR$ and the renormalization effects for $\hpb$. The uncertainty arising
from the Fierz ambiguity is reflected by the two curves in fig. \ref{BareExpPart}, where the solid line reflects the choice
(\ref{PhysScatt}) whereas the long-dashed line corresponds to eq. (\ref{eq120}). Presumably the solid line reflects better the
true answer since at the end all macroscopic results can only depend on the combination $a_{res} + a_{bg}$. Close to the
resonance the Fierz ambiguity becomes irrelevant since $|a_{bg}|$ can be neglected as compared to $|a_{res}|$.

Close to the resonance the scattering at zero momentum provides the same information on the relation between $c$ and $B$ as
the binding energy $\epm (B)$ in the vicinity of $B_0$. Indeed, the dependence of the resonant scattering length $a_{res}$
or the cross section on the magnetic field involves the same combination of parameters. The resonant scattering length
$a_{res} = (M\beta)^{-1/2}(B_0 - B)^{-1}$ involves $\beta$, just as the binding energy (\ref{EpsQuant})
in the ``molecule phase'' for $\epm <0$. From the experimental parameterization of the scattering length
\begin{eqnarray}\label{PhysScattExp}
a(B) &=& a_{bg}\Big(1 + \frac{\Delta}{B - B_0}\Big),\\\nonumber
\end{eqnarray}
we can extract the parameter $\tau_B$ as
\begin{eqnarray}\label{tauB}
\tau_B = -(a_{bg} \Delta k_F)^{-1}.
\end{eqnarray}
This does not involve the Yukawa coupling. Experimental comparison of $\epm (B)$ and $a(B)$ on the respective sides of the
Feshbach resonance therefore amounts to a test of the computation of the vacuum properties.

For a realistic scattering of atoms we have to take into account the momentum dependence of  the atoms and the
nonvanishing decay width $\bar{\Gamma}$. From app. \ref{app:dispersion} we infer (for $q^2 =0$ and $\omega =
s - \epm\theta(-\epm)$)
\begin{eqnarray}\label{barGamma}
\bar{\Gamma}(s) = \frac{\hpb^2M^{3/2} s^{1/2}}{4\pi}.
\end{eqnarray}
On the atom side we find a resonant scattering for $s=\epm$ with a Breit-Wigner form.

With $\ZpR =1$ and
\begin{eqnarray}\label{BreitWigner}
\mathrm{Im} (\bar{\lambda}_{eff}) = \frac{\hpb^2\bar{\Gamma}(s)}{(s-\epm)^2 +\bar{\Gamma}^2(s)}
\end{eqnarray}
we may compute the total cross section $\sigma_{tot}$ as a function of $s$. Using the optical theorem $\sigma_{tot}
= |v|^{-1} \mathrm{Im}(\bar{\lambda}_{eff})$ one obtains ($|v| = |q|/M = \sqrt{s/M}$)
\begin{eqnarray}
\sigma_{tot} &=& \sqrt{\frac{M}{s}}\mathrm{Im} (\bar{\lambda}_{eff}(s)) \\\nonumber
&=& 4\pi(M s)^{-1}\Big(1+\frac{16\pi^2(s-\epsilon_M)^2}{\hpb^4M^3s}\Big)^{-1}.
\end{eqnarray}
For $s\to 0$, $\epsilon_M \neq 0$ this yields the expected result
\begin{eqnarray}\label{sigtot}
\sigma_{tot}^{(0)} = \frac{M^2\hpb^4}{4\pi \epm^2 } = 4\pi \bar{a}^2.
\end{eqnarray}

Independent experimental information on $\hpb$ can be gained from scattering in a momentum range where the pointlike limit
does not apply anymore. In particular, the peak height of the resonance at $s=\epm$ depends \footnote{More accurately, the
contribution from the background scattering $a_{bg}$ should be added.} on $\hpb$,
\begin{eqnarray}\label{hpMeasure}
\sigma_{tot}^{max} &=& \frac{4\pi}{M\epm} = \frac{4\pi}{M\bar{\nu}} = -\frac{16\pi^2 a_R}{M^2\hpb^2}\nonumber\\
&=& 16\pi^2\hpb^{-2}M^{-5/2}\beta^{-1/2}(B-B_0)^{-1}.
\end{eqnarray}
The value $\tilde{h}_{\phi, 0}^{res}$ extracted from resonant scattering (\ref{hpMeasure}) will again be a renormalized
coupling in vacuum, but now evaluated for the particular resonance frequency. It is therefore conceivable that it differs
from $\tilde{h}_{\phi, 0}$, eq. (\ref{G3}). We note that the cross section remains finite for all $|\epm| \neq 0$
($B\neq B_0$). On the other hand, one finds for $\epm =0$ and $s \to 0$ a divergence $\sigma_{tot} \propto s^{-1}$.

It is interesting to look at the behavior exactly on the Feshbach resonance ($\epsilon_M =0, B=B_0$) both from the picture
in the atom and the molecule phase. With
\begin{eqnarray}
\bar{\lambda}_{eff} (s) = -\frac{\hpb^2}{\bar{\mathcal{P}}_\phi(s)},\quad \bar{\mathcal{P}}_\phi(s) = -\ZpR (s) s +
\mathrm{i} \bar{\Gamma}(s)
\end{eqnarray}
one has
\begin{eqnarray}
|\bar{\mathcal{P}_\phi}(s)|^2 = \ZpR^2 (s) s^2 + \bar{\Gamma}^2.
\end{eqnarray}
In the molecule phase $\bar{\Gamma}(s)$ vanishes and
\begin{eqnarray}
\ZpR (s) s = s + \frac{\hpb^2}{4\pi}\,M^{3/2} s^{1/2}.
\end{eqnarray}
On the other hand, the atom phase is characterized by $\ZpR = 1$ and $\bar{\Gamma} (s)$ given by eq. (\ref{barGamma}).
While $|\bar{\mathcal{P}}_\phi (s)|$ is the same, the phase of $\bar{\lambda}_{eff}(s)$ jumps between the molecule and the
atom phase. This is a typical shortcoming of our approximation where the molecule fluctuations  are neglected in the
computation of $\ZpR$ and $\bar{\Gamma}$. In particular, the molecule fluctuations will provide for an imaginary part of
$\bar{\lambda}_{eff}(s)$ also for the molecule phase.

Our functional integral approach reproduces several results obtained previously by quantum mechanical two channel
computations in the vicinity of the Feshbach resonance.
The fluctuation correction to the mean field binding energy coincides exactly with the result from a quantum
mechanical coupled two-channel approach \cite{Koehler03}. The estimate for the Yukawa coupling given in \cite{YYStoof}
coincides with our ``vacuum coupling'' $\tilde{h}_{\phi,0}$ (\ref{G1}).

\end{appendix}

\bibliographystyle{apsrev}
\bibliography{Citations}

\end{document}